\documentclass{article}
\usepackage{amssymb}
\usepackage{rotating}
\usepackage{amsfonts}
\usepackage{amsmath}
\usepackage{amsthm}
\usepackage{caption}
\usepackage{float}
\usepackage{graphicx}
\usepackage{wrapfig}
\usepackage{enumerate}
\usepackage{multicol}
\usepackage{srcltx}
\usepackage{subcaption}
\usepackage{float}
\usepackage{array}
\usepackage{multirow}
\usepackage{amssymb}

\newtheorem{theorem}{Theorem}[section]
\newtheorem{proposition}{Proposition}[section]
\newtheorem{lemma}{Lemma}[section]
\newtheorem{corollary}{Corollary}[section]

\newtheorem{remark}{Remark}
\newcommand{\wh}{\widehat}

\newcommand{\wt}{\widetilde}

\newcommand\cG{{\cal G}}
\newcommand\cF{{\cal F}}

\newcommand\cX{{\cal X}}
\newcommand\cD{{\cal D}}

\newcommand\cS{{\cal S}}

\DeclareMathOperator*{\cWD}{\raisebox{-0.6ex}{\scalebox{1.5}{$\mathbb  W$}}}

\DeclareMathOperator*{\esssup}{ess\,sup}

\def\bbg{{\mathbb G}}
\def\bbf{{\mathbb F}}

\def\bbr{{\mathbb R}}

\def\text#1{\hbox{#1}}
\def\proof{{\noindent \bf Proof. }}

\def\endproof{\mbox{\ $\qed$}}

\def\E{{\bf E}}

\def\P{{\bf P}}

\def\Q{{\bf Q}}

\def\X{{\bf X}}

\def\d{\mathrm{d}}
\def\build #1_#2{\mathrel{\mathop{\kern 0pt #1}\limits_\zs{#2}}}
\newcommand{\zs}[1]{{\mathchoice{#1}{#1}{\lower.25ex\hbox{$\scriptstyle#1$}}
{\lower0.25ex\hbox{$\scriptscriptstyle#1$}}}}
\def\proof{{\noindent \bf Proof. }}
\def\endproof{\mbox{\ $\qed$}}
\usepackage[top=1in, bottom=1.5in, left=1.1in, right=1in]{geometry}

\title{Non-concave optimal investment with Value-at-Risk constraint: an application to life insurance contracts
}
\author{ Thai Nguyen  \and Mitja Stadje \thanks{Institute of Insurance Science and Institute of Financial Mathematics, University of Ulm, Germany. Email: thai.nguyen@uni-ulm.de,mitja.stadje@uni-ulm.de} \thanks{Institute of Insurance Science and Institute of Financial Mathematics, University of Ulm, Germany. Email: mitja.stadje@uni-ulm.de} }

\vspace{2mm}

\begin{document}

\maketitle
\begin{abstract}
This paper studies a Value-at-Risk (VaR)-regulated optimal portfolio problem of the equity holders of a participating life insurance contract. In a setting with unhedgeable mortality risk and complete financial market, the optimal solution is given explicitly for contracts with mortality risk using a martingale approach for constrained non-concave optimization problems. We show that regulatory VaR constraints for participating insurance contracts lead to more prudent investment than in the case of no regulation. This result is contrary to the situation where the insurer maximizes the utility of the total wealth of the company (without distinguishing between contributions of equity holders and policyholders), in which case a VaR constraint may induce the insurer to take excessive risks leading to higher losses than in the case of no regulation, see \cite{BasakShapiro01}. Compared to the unregulated problem, the VaR-constrained strategy leads to a higher expected utility for the policyholders, highlighting the potential usefulness of a VaR-regulation in the context of insurance. 
The prudent investment behavior is more significant if a VaR-type regulation is replaced by a portfolio insurance (PI)-type regulation. Furthermore, a stricter regulation (a smaller allowed default probability in the VaR problem or a higher minimum guarantee level in the PI problem) enhances the benefit of the policyholder but deteriorates that of the insurer. For both types of regulation, the gains in terms of expected utility are greater for higher participation rates, while being smaller for higher bonus rates. 
We also extend our analysis to frameworks where dividend and premature death benefit payments are made at an intermediate time date. 


\end{abstract}



\section{Introduction} \label{sec:int}
This paper investigates the equity holders' optimal investment problem of a participating contract under financial regulation. This problem is particularly relevant for insurance companies that operate under Solvency II with a Value-at-Risk (VaR) constraint. Participating contracts are life insurance products which provide the policyholders at least a guaranteed amount in downside market situations and a shared profit in good market scenarios. To participate in such a contract, the policyholders pay a premium (participation fee) which is collected together with the (equity holders') insurer's participation amount in an investment pool. At maturity, the policyholders receive a payoff which is linked to the investment performance. The equity holders' payoff is determined as the residual amount. Our derivation of the optimal solutions relies on the combination of a martingale approach for non-concave and non-differentiable objective functions and a point-wise optimization technique with constraints. 

Earlier studies on equity-linked life insurance contracts usually analyze pricing or optimal design problems. Some focus on the policyholders' perspective assuming specific investment strategies like constant proportion portfolio insurance (CPPI) or generalized constant-mix see, e.g., \cite{branger2010,mahayni2015,pezier2013}. Recently, \cite{Chen2017} considers participating life insurance contracts from the policyholders' perspective under a fair pricing constraint and tax privileges.

In this paper, we consider two common contract designs. In the first design, we assume that the equity holders have only limited liability, i.e., the policyholders' payoff is less than the guaranteed amount in case of a default of the insurance company. In the second design, we assume that the policyholders are fully protected against an insolvency of the insurance company (i.e., the final payoff to the policyholders always exceeds or equals the guaranteed amount). Note that in the case of full protection, the equity holders may suffer a negative payoff in case of insolvency. To describe the behavior of the equity holders in the loss domain, we use an $S$-shaped utility function adopted from prospect theory, see \cite{Tversky1992,HeKou16,Lin2017}.

Risk management and regulations based on a terminal VaR constraint are well-known in banking and insurance regulations. Banks are allowed to develop their own internal VaR models which are subject to supervisory approval based on standardized back-testing procedures controlled by the Basel Committee on Banking Supervision. In insurance contexts, to ensure that pension funds fulfill  their  obligations  to  pensioners,  pension regulators usually impose a VaR-type constraint on nominal funding ratios. The new regulatory framework for insurance companies in Europe, Solvency II, advocates
risk-based regulation which focuses on downside risk and suggests using measures
such as the ruin probability or VaR. 
The problem of utility maximization/optimal asset allocation under VaR-type constraints has been studied extensively in the literature, see e.g.  \cite{BasakShapiro01,boyle2007,CNSmult,Cuoco08, miller2017,wong2017,CNS17,wei2018}. 

This paper solves the equity holders' problem of utility maximization under a regulatory constraint imposed at maturity. We obtain closed-form solutions for various kinds of constraints. We first explicitly solve the problem for the two kinds of contracts mentioned above under a VaR regulation extending the martingale approach to non-concave utility maximization problems with constraints. Second, motivated by the fact that regulators
usually affect insurance contract designs by imposing a minimum capital requirement which is used to control adverse events, we consider a portfolio insurance (PI) constraint to enhance the protection for the policyholders. 
 Finally, we extend the result to a two-period optimization framework where premature death benefits and dividends can be paid at an intermediate date. This analysis is particularly useful for insurers which need to set a long-term investment while having to follow short-term liabilities with periodic VaR-type regulations. A further extension to multiple periods along these lines is possible but is omitted.

Our theoretical and numerical results show that already in the case of no regulation there is a moral hazard problem since the insurer does not have an incentive to ensure that there is any capital in the loss states where the terminal wealth falls below the minimal guarantee. The reason is that any terminal wealth in those states only benefits the policyholders and comes at the expense of a lower terminal wealth in the more prosperous states where the equity holders receive a positive residual. On the other hand, introducing a VaR constraint as in Solvency II forces the equity holders to enlarge the proportion of hedged loss states, leading to a genuine improvement for the policyholders. This result is contrary to the situation where the insurer maximizes the utility of the total wealth of the company without distinguishing between equity holders and policyholders, in which case a VaR constraint may induce the insurer to take excessive risk leading to higher losses in the worst 1\% of scenarios than in the case of no regulation, see e.g. \cite{BasakShapiro01,CNS17,kraft2013,CNSmult,wei2018}. Compared to the unregulated problem, the VaR-constrained strategy leads to a higher expected utility for the policyholders, implying an increase in the attractiveness of participating contracts. This more prudent investment behavior described above is more pronounced if a VaR based regulation is replaced by a PI-based regulation. Furthermore, a stricter regulation (a smaller allowed default probability in the VaR problem or a higher minimum guarantee level in the PI problem) will enhance the benefit of the policyholder but deteriorate that of the insurer. For both (VaR and PI) types of regulation while being stronger for higher participation rates, this regulatory effect becomes less for smaller bonus rates (given the same participation rate). Our results also show that in certain situations, for instance in the cases of relatively high participation rates and relatively low bonus rates, policyholders overall may prefer a VaR regulation over a PI regulation with a relatively low minimum guarantee level in terms of expected utility. Finally, the introduction of a full protection makes the equity holders' investment also generally more prudent in bad market scenarios but changes the investment behavior relatively little in good market scenarios.


The problem of {\it non-concave utility} maximization without constraints has been considered by many authors e.g., \cite{carpenter2000,ross2004,karoui2005,carassus2009, reichlin2013, bichuch2014,larsen2005}, using concavification techniques. In a framework without constraints, \cite{reichlin2013} proves the existence of an optimal terminal wealth. However, no specific payoff is provided. We remark that the technique developed here to deal with VaR constraints can be applied for the principle-agent problem with option-like compensations considered e.g. in \cite{carpenter2000,ross2004}. Note that such a risk-sharing problem may require some kind of fairness condition that guarantees that neither party can systematically benefit from the contract. In insurance contexts, when such a condition (being referred to as fair pricing constraint or participation constraint see, e.g., \cite{pitacco2011,Chen2017}) is satisfied, both the parties are willing to participate in the contract initially. While the optimization developed here can be applied to such a framework we mainly focus on regulatory aspects of VaR constraints motivated from the Solvency II regulations rather than on the problem of optimal product design. 
On the other hand, our framework allows for a pool of multiple contracts of heterogeneous policyholders with mortality risk for which the individual fairness is not the main point from the insurer's perspective as it might be too challenging to take it into account for every policyholder. It seems in our setting more reasonable to assume that only those policyholders enter the contract in the first place, for whom a fair pricing constraint holds without the need for the insurer to readjust its strategy. In addition, it seems very challenging to specify even for one policyholder the individual fair participating constraint in the two-period framework with death and dividend payments studied above due to the market incompleteness and possible intermediate money withdrawals.

When finishing this paper we noticed that \cite{Lin2017} has independently investigated participating insurance contract problem with a power utility function without regulation or mortality, doing extensive, interesting numerical comparisons between various hedging portfolio insurance strategies, and also numerically analyzing possible restrictions on the portfolio weights. We on the other hand consider general utility functions with mortality risk focusing on the impact of regulation for a pool of multiple contracts of heterogeneous policyholders. To the best of our knowledge, non-concave utility maximization problems together with (non-concave) constraints have not been considered before. Moreover, we extend the results to a two-period framework where premature death benefits and dividend payments can be considered at an intermediate date, which shows the flexibility of our method.


The paper is organized as follows: In Section~\ref{sec:mod}, we introduce the asset model and the parametric family of contract payoffs. We then solve the unregulated problems with mortality in Section \ref{sec:Uncon}. In Section~\ref{Se:VaR0}, we investigate the constrained problems. The results are numerically illustrated in Section \ref{sec:Num}. An extension to a two-period framework is presented in Section \ref{sec:Twoperiod}. All
technical proofs are reported in the Appendix.

\section{The financial market and participating contracts} \label{sec:mod}
\subsection{The financial market}
We fix a probability space $(\Omega,\cF, \P)$ equipped with a $d$-dimensional Brownian motion $W$, where $\bbf=(\cF_t)_\zs{t\in[0,T]}$ is the natural filtration of $W$ and $T>0$ is the time horizon. All equations and inequalities are assumed to hold almost surely. We consider a {\it complete} financial market without transaction costs consisting of one risk free asset (the bank account) $S^0$ and $d\ge 1$ traded risky assets whose price dynamics $S^1,\cdots,S^d$  are $\bbf$-adapted processes. Let $\Q\sim \P$ be the unique equivalent martingale measure such that the discounted risky assets $S_t^i/S_t^0$, $i=1,\cdots,d$ are $(\bbf,\Q)$-local martingales. We remark that the Radon-Nikodym derivative $\frac{\d \Q}{\d \P}$ is $\cF_T$-measurable and can be easily be extended to larger  filtrations by setting $\Q(A)=\E[\frac{\d \Q}{\d \P}{\bf 1}_A]$.

To unify notation, we assume in the sequel that $\bbg:=(\cG_t)_\zs{t\in[0,T]}$ is the filtration which is identical to $\bbf$ in the case of no mortality and in the case of additional mortality risk is an enlarged filtration (containing $\bbf$ and additional mortality events to be specified later) such that $S$ is a $(\bbg,\Q)$ local martingale. 

A trading strategy is defined as a $\bbg$-adapted predictable process $\nu=(\nu_t)_\zs{t\in[0,T]}$, $\nu_t=(\nu_t^0,\cdots,\nu_t^d)$, representing the vector of holdings in the assets. We interpret $\nu_t$ as a self-financing strategy, so that its portfolio process $X_t^{\nu,X_0}$ with initial value $X_0$ can be represented as 
$$
X_t^{\nu,X_0}=X_0+ \sum_{i=0}^d\int_0^t\nu_s^i \d S_s^i=X_0+\int_0^t \nu_s \d S_s,\quad  S=(S^0,S^1,\dots,S^d), \quad t\in[0,T].
$$
To exclude arbitrage opportunities, we only consider strategies $\nu$ such that $X^{\nu,X_0}$ is bounded from below and the above stochastic differentiable equation admits a unique strong solution. Hence, the discounted portfolio $X^{\nu,X_0}/S^0$ is a $(\bbg,\Q)$ supermartingale. Choosing a self-financing portfolio strategy $\nu$ is equivalent to choosing a terminal wealth $X_T$ which can be financed by $X_0$.

The set of attainable terminal wealth values is defined by
\begin{equation}
\cX:=\{ X\; \mbox{is} \; \cG_T\mbox{-measurable, replicable, bounded from below and}\; \E[\xi_T X]\le X_0\},
\label{eq:Adm}\notag
\end{equation}
where $\xi_T$ is the pricing kernel defined by $\xi_T:=\frac{S_0^{0}}{S_T^{0}}\frac{\d \Q}{\d \P}$ with $\frac{\d \Q}{\d \P}$ being the Radon-Nikodym derivative. Throughout this paper we will assume that $\xi_T$ does not have any atom. Note that $\xi_T (\omega)$ can be interpreted as the (Arrow-Debreu) value per probability (or likelihood) unit, of a security which pays out \$1 at time $T$ if the scenario $\omega$ happens, and $0$ else. As this value is high in a recession and low in prosperous times, $\xi_T(\omega)$ has the property of directly reflecting the overall state of the economy at time $T$. We remark that in a consumption based pricing model $\xi_T$ in equilibrium corresponds to a constant times the marginal utility of consumption and is also called {\it pricing kernel} or {\it stochastic discount factor}.

Examples of complete markets are the Black-Scholes model, local volatility models, stochastic interest rate models with a zero-coupon bond or the Heston stochastic volatility model with a financial derivative as an additional hedging instrument.

\subsection{Participating contracts and payoffs}
Participating life insurance contracts are contracts that provide a return guarantee  and  participation  in  the  surplus  of  the insurance company's performance. These products have been widely considered in major European life insurance markets, see Table 1 in \cite{GIMR2015}. Valuation and hedging of participating life insurance contracts  have been well documented in the literature e.g. \cite{bryis1997,Bacinellorita2002,bernard2005,Gatzert2007,bernard2012,Schmeiser15,Mirza2018}.
\begin{figure}[h]%
\centering\includegraphics[width=0.7\columnwidth]{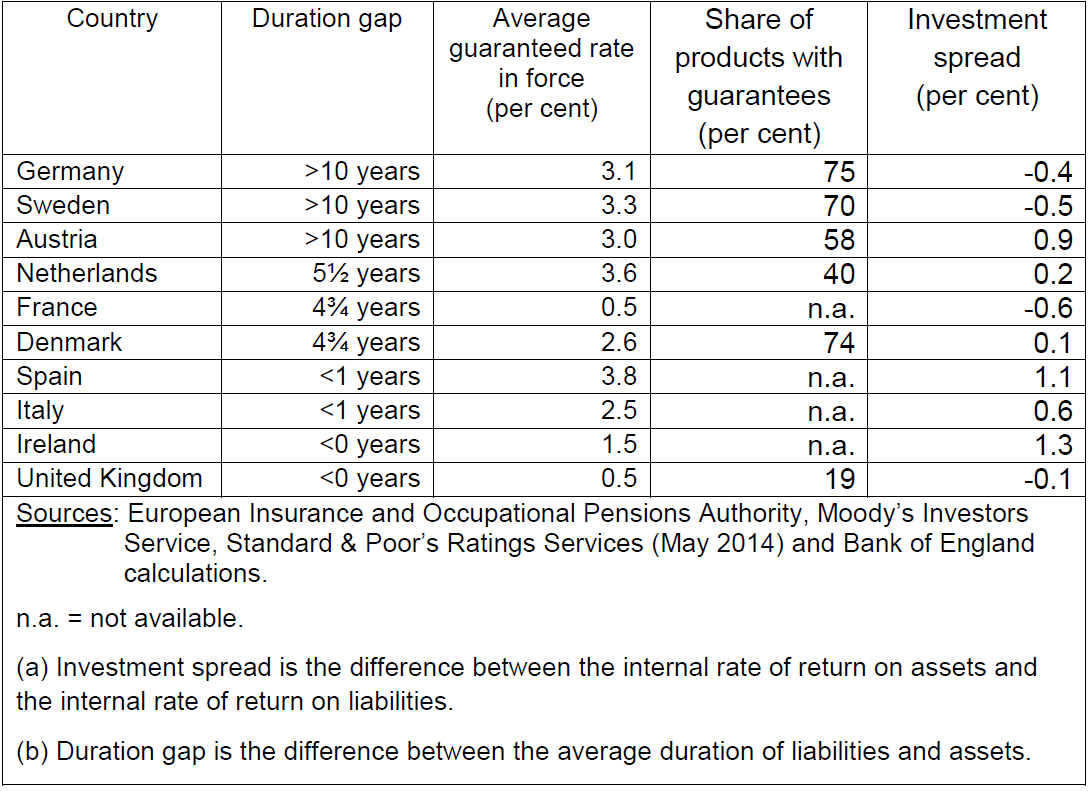}%
\caption*{\small Selected properties of major EU life insurance markets. Source: Table 1 in \cite{GIMR2015}.}%
\label{Tab1}%
\end{figure}

In this paper, we look at the insurer's investment problem related to a pool of multiple participating contracts having the same maturity. In particular, we assume that there are $n$ ($n\ge 1$) policyholders who each invest a single premium into an equity-linked life insurance contract with a maturity of $T$ years with $T<\infty$. At the initiation of the contract, each policyholder $i$ ($i\in\{1,\cdots,n\}$) invests an amount $L_0^i$. The total premium is given by $L_0=L_0^1+L_0^2+\cdots L_0^n$. The shareholders (equity holders) provide an initial equity $E_0>0$. In the sequel we assume that the insurance company or its management acts on behalf of the group of equity holders by maximizing their total utility. The initial portfolio value is given by the sum of all contributions, i.e.,  $X_0:=L_0+E_0$. We denote by $\alpha:=L_0/X_0$ the total share of the policyholders' contribution (or equivalently the debt ratio of the insurance company). $\alpha $ is also called the policyholders' participation rate. 

We assume that the individual policyholder $i$ receives the guaranteed payment $L_T^i$ plus some bonus which is proportional to her relative participation rate 
$k_i:=L_0^i/L_0$. Note that $\sum_ik_i =1$. For simplicity, we assume that the guaranteed interest rates are the same for all policyholders, in particular, $k_i=L_0^i/L_0=L_T^i/L_T$, where $L_T=L_T^1+\cdots+L_T^n$ is the total guaranteed liability. For example, in a Black-Scholes setting with constant interest rate $r$, we can choose $L_T^i=L_0^i e^ {gT}$, $i=1\cdots, n$, where $g\le r$.

The aggregated payoff of the insurance company to the group of policyholders is then defined as follows: If the terminal portfolio value $X_T$ is less than the total guaranteed liability $L_T$, bankruptcy is declared (default event) and the policyholders take what is left, $X_T$. If the terminal portfolio value exceeds $\wt{L}_T:=L_T/\alpha$, i.e. $\alpha X_T>L_T$, the surplus $\alpha X_T-L_T$ will be shared between the equity holders (represented by the insurer) and policyholders. In particular, we assume that 
the policyholders receive a surplus equal to $\delta\, \big[\alpha X_T -L_T\big]$, where the surplus (bonus) rate $\delta$ is the percentage of surpluses that is credited to the policyholders. When the terminal asset value lies between $L_T$ and $\wt{L}_T$ (the threshold where the participation bonus kicks in), the insurer is only able to distribute its guaranteed commitment $L_T$. To summarize, the insurer at time $T$ pays out to the policyholders
 \begin{align}
V_L^p(X_T) := &\, \begin{cases}
X_T & \text{if }X_T\le {L_T}, \\
L_T & \text{if } L_T<X_T\le {L_T}/\alpha, \\
L_T+\delta\, \big[\alpha X_T -L_T\big] & \text{if}\,  X_T> {L_T}/\alpha,\label{PsiL}
\end{cases}
\end{align}
or in a more compact form\footnote{$[\,\cdot\,]^+$ denotes the maximum $\max\{\,\cdot\,,0\}$.}, $$V_L^p(X_T)  =\,L_T+\delta\,  \big[\alpha X_T -L_T\big]^+-[L_T-X_T]^+.$$ 
The terminal payoff $V_L^{p}(X_T)$ will be shared among the policyholders corresponding to their relative contributions $k_i$. Thus, the payoff of policyholder $i$ ($i=1,\cdots,n$) is given by $k_i V_L^p(X_T)$ with  $\sum_ik_i =1$.

To exclude unrealistic cases, we assume throughout the paper that $0<\alpha<1$ and $0\le \delta<1$.  Hence, the policyholders take a long position in the bonus option and a short position in a
 defaultable put and benefit from the potential upsides over the final maturity guarantee. This type of {\it defaultable contracts} is frequently used in the literature of insurance contracts, see, for example \cite{Briys1994,bryis1997,grosen2000,Ballota2006,Bernard2010,bernard2012,Lin2017,Chen2017}. The equity holders then receive the residual asset value
 \begin{align}
 \label{PsiE}
 \nonumber V_E^p(X_T):= X_T - V_L^p(X_T) = &\,  \begin{cases}
 0 & \text{if } X_T\le L_T, \\
 X_T-L_T & \text{if } L_T<X_T\le \wt{L}_T, \\
 X_T-L_T- \delta\,\big[\alpha X_T-L_T \big] & \text{if}\,  X_T> \wt{L}_T
 \end{cases}\\[0.2cm]
  =&\,[X_T-L_T]^+-\delta\,\big[\alpha X_T -L_T \big]^+.
 \end{align}
Agreeing on the contract, the insurer takes a long position of the call $[X_T-L_T]^+$ and ${\delta}\alpha$ short positions of the bonus call $\big[ X_T -\wt{L}_T \big]^+$ with strike price $\wt{L}_T$.

For our analysis, we introduce
\begin{equation}
 \tilde\delta := \alpha\delta;\quad \wh{L}_T:=\wt{L}_T-L_T\quad \mbox{and} \quad f(x)=(1-\wt{\delta})x-(1-\delta)L_T.
\label{eq:fd}
\end{equation}
Hence, $\wt{\delta}$ is the actual (achieved) bonus rate of the policyholders, $\wh{L}_T$ is the difference between the bonus threshold and the guarantee, and $f(x)$ is the payoff that the insurer receives in case that the wealth $x$ is greater than the bonus threshold $\wt{L}_T$.

Like any profit-seeking company, the life insurance company sets up its investment mix to primarily maximize the benefits of its shareholders. Hereby,
we assume that the shareholders value their benefits through a strictly concave utility function $U$ defined on the positive real line,
which is twice differentiable and satisfies the usual Inada and the asymptotic elasticity (AE) (see \cite{kramkov1999}) conditions
\begin{equation}
 \lim\limits_{x\downarrow 0 } U^{\prime}(x)=\infty;\quad \lim\limits_{x\uparrow \infty } U^{\prime}(x)=0;\quad  \lim_{x\uparrow \infty } \frac{xU^{\prime}(x)}{U(x)}<1.
\label{eq:InadaAE}
\end{equation}
We assume furthermore that $U(0)=\lim_{x\to 0}U(x)>-\infty$. The case where $U(0)=\lim_{x\to 0}U(x)=-\infty$ is easier and can be treated without the concavification procedure, see more in Corollary \ref{Re:infty}. As usual, we denote by $I$ the inverse of $U^{\prime}$, the first derivative of the utility function. 

\subsection{Full protection and $S$-shaped utility function}\label{sec:full}
The policyholders of the defaultable contract defined by \eqref{PsiL} have to bear the risk of insolvency. This contract design may not serve well policyholders who aim at purchasing default-free policies. Motivated by \cite{Bernard2010}, we assume that the contract at maturity gives the policyholders at least some guaranteed amount $L_T$ without the possibility of default. More precisely, at time $T$ the policyholders receive the following payoff 
\begin{equation}
V_L^{np}(X_T) := L_T+\delta\,  \big[\alpha X_T -L_T\big]^+.
\label{eq:np}\notag
\end{equation}
In case that the insurance company is unable to fund the guarantee, this assumes that the guaranteed amount is taken over by an insurance guarantee fund or by the state government. As in the case of defaultable contract, the individual policyholder $i$ ($i=1,\cdots,n$) payoff can be defined by $k_i V_L^{np}(X_T)$.


Fully protected contracts have also been considered in \cite{Lin2017,Chen2017}. The equity holders take the residual portfolio value
 \begin{equation}
 \label{PsiE2}
 V_E^{np}(X_T):= X_T - V_L^{np}(X_T) = \,  \begin{cases}
 X_T-L_T & \text{if } X_T\le \wt{L}_T, \\
 X_T-L_T- \delta\,\big[\alpha X_T -L_T \big]^+ & \text{if } X_T>\wt{L}_T.\,
 \end{cases}\notag
  \end{equation}
Hence, in this {\it full protection} case, the equity holders take more risk because they may have a negative payoff in the insolvency case where $X_T-L_T<0$. Since the insurance company may have to pay this amount back at a later point in time and faces a loss of its reputation, we assume that the equity holders use an $S$-shaped utility (convex on the loss domain left to a reference point and concave on the gain domain right to a reference point) suggested by prospect theory \cite{Tversky1992} which allows for negative payoff values. In particular, the insurance company utility takes the form $U^{\cS}( V_E^{np}(X_T))$, where
\begin{equation}
U^{\cS}(x):=
 \begin{cases}
-U_\zs{lo}(-x) & \text{if } x\le 0, \\
U(x) & \text{if } x> 0,
 \end{cases}
\label{eq:ShapU}
\end{equation}
where $U_\zs{lo}$ and $U$ are two strictly concave utility functions satisfying $-U_\zs{lo}(0)=U(0)$ to guarantee that $U^{\cS}$ is continuous at $0$. We assume furthermore that $U$ satisfies the Inada and (AE) conditions. For example, we can take $U_\zs{lo}(x)=\eta U(-x)$, for some loss aversion degree $\eta>1$, see \cite{Tversky1992}. Note that for such {\it fully protected contracts}, the guarantee level $L_T$ is considered as the reference point which is naturally used to distinguish gains and losses. 

\begin{center}
	\begin{figure}[h]
      \begin{subfigure}[b]{0.49\textwidth}
   \includegraphics[width=\columnwidth,height=6cm]{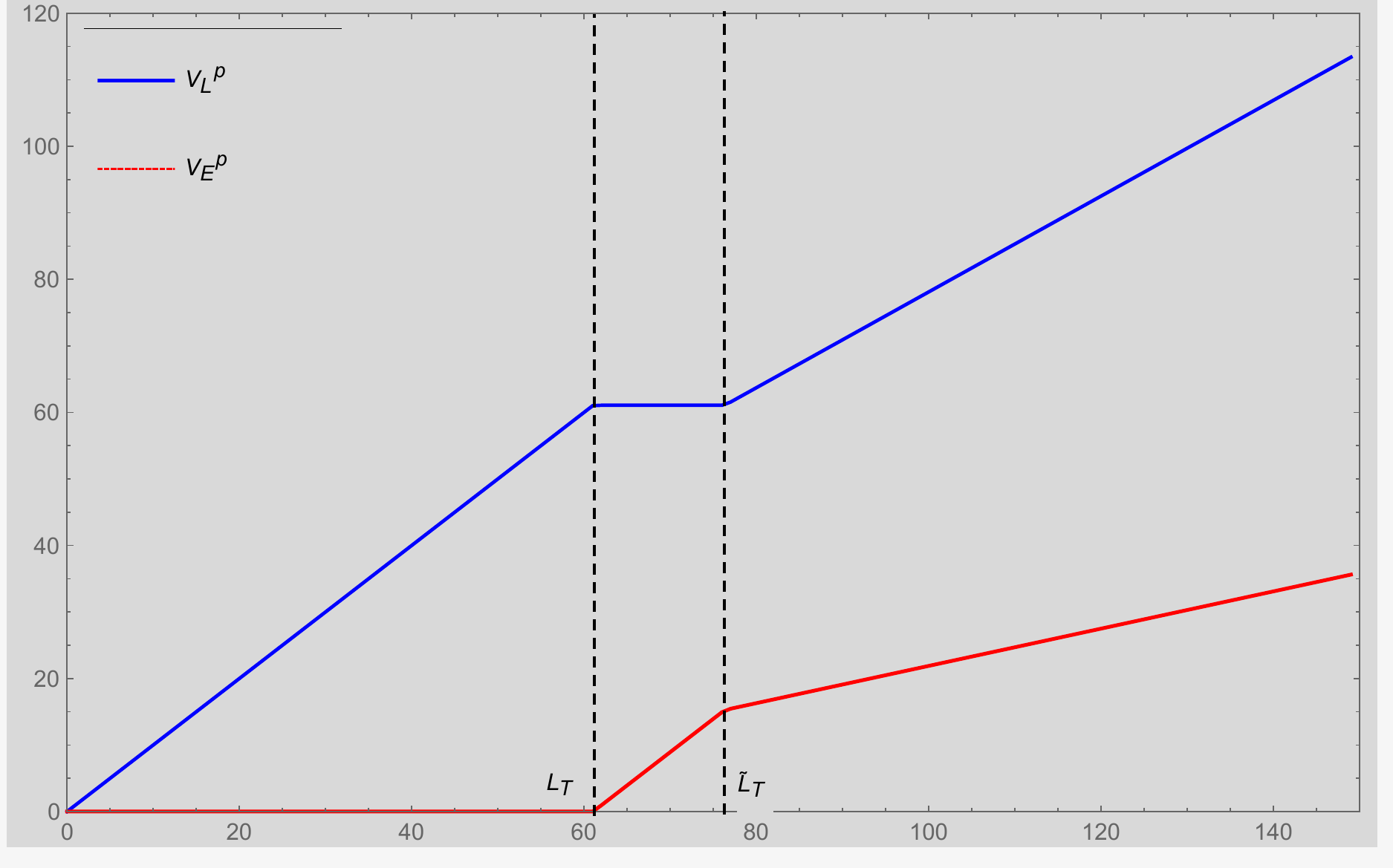}%
		 \caption{\small Defaultable contract payoffs}
		    \end{subfigure}
      \begin{subfigure}[b]{0.49\textwidth}
      \includegraphics[width=\columnwidth,height=6cm]{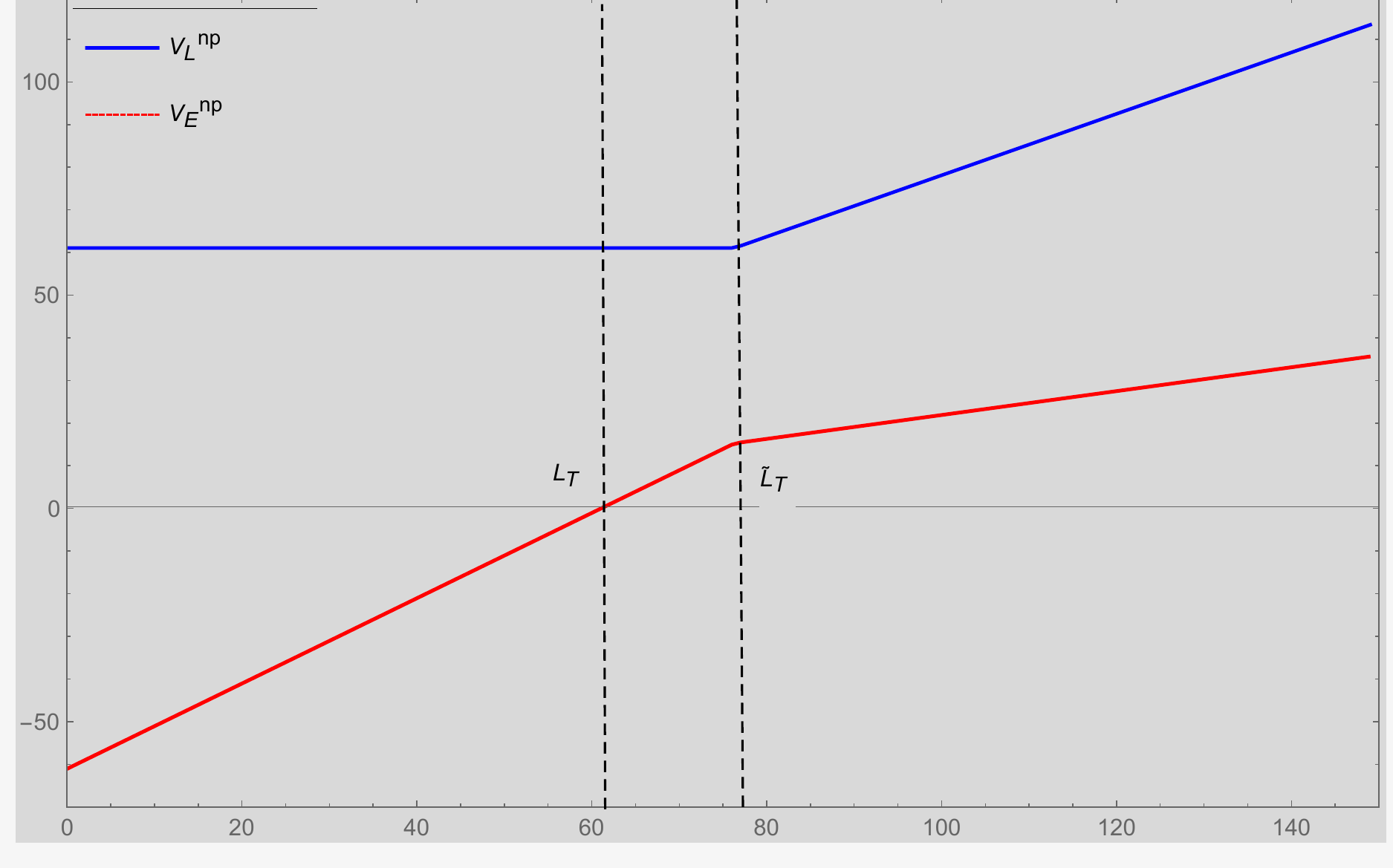}%
		\caption{\small Fully protected contract payoffs}
       \end{subfigure}
    \caption{\small Payoffs with $\alpha=0.8$, $\delta=0.9$, $L_T=50 e^{0.02\times 10}$}		
				 \end{figure}
	\end{center}
\begin{remark}
It has been suggested in \cite{Bernard2010} that the policyholders' payoff can take the following mixed form 
$L_T+\delta\,  \big[\alpha X_T -L_T\big]^+-(1-\rho)[L_T-X_T]^+$, for some $\rho\in[0,1]$. This means that the insurer ``sells back'' a portion of the default put to the policyholders. Note that $\rho=0$ leads to a defaultable contract while $\rho=1$ corresponds to a fully protected policy. In this paper we only focus on these extreme cases  ($\rho=0$ or $\rho=1$) and remark that the optimal investment problem for mixed contracts can be treated similarly.

\end{remark}

\subsection{Contracts with mortality risk}\label{subsec:mort}
Mortality is one of the most important risk factors in life insurance, strongly affecting pricing and premium principles. Note that when non-tradable stochastic mortality is considered, the market becomes incomplete. 
In this section we incorporate mortality risk into the optimal investment problem of the equity holders. We assume that the premature death of policyholder $i$ is modelled by the event $\{d_i=1\}$, where $d_i$ is a binomial random variable. We suppose that $d_1,\cdots, d_n$ are mutually independent and independent of the financial risk, see e.g. \cite{Bacinellorita2002,bernard2005}. 
We now work on an enlarged probability space $(\Omega,\cG, \P)$, where $\bbg=(\cG_t)_\zs{t\in[0,T]}$ is the enlargement of the financial filtration $\bbf$ that contains the information generated by the premature death variables $d_i,i=1,\cdots,n$. In other words, $\bbg$ represents the structure of the global information available to the insurer. Since $d_i,i=1,\cdots,n$ are discrete variables independent of $\bbf$, we obtain $\cG_t=\cF_t$ for all $0\le t<T$ and $\cG_T=\cF_T\vee\sigma(d_i,i=1,\cdots,n)$. In Section \ref{sec:Twoperiod} we will also allow deaths to be observed before maturity. 

First, consider the case of defaultable contracts. As discussed above, we suppose that policyholder $i$ receives the amount $k_i V_L^{p}(X_T)$ if 
she is alive at maturity. On the other hand, her family will receive a benefit in case of death which we for simplicity first assume to be the guarantee $L_T^i$ in case that the insurance company is solvent and $k_i X_T$ else. In Section \ref{sec:Twoperiod} we will also allow for arbitrary large death benefits and additionally take dividend payments into account. Overall, the payoff to policyholder $i$ will be equal to $k_i \min  (L_T,X_T)$. 
We note that we can also consider other death benefits than $L_T$ with our method.  
The policyholder $i$'s payoff is given by 
\begin{equation}
 k_i {\bf 1}_\zs{d_i=0}V_L^{p}(X_T)+k_i {\bf 1}_\zs{d_i=1}\min (L_T,X_T).\notag
\label{eq:VLtau}
\end{equation} 
Hence, the equity holders' payoff will be given by the difference between the terminal portfolio value and the total amount paid to the policyholders, i.e.,
\begin{equation}
V_E^{p,d}(X_T):=X_T-\sum_{i=1}^n k_i V_L^{p}(X_T){\bf 1}_\zs{d_i=0}- \sum_{i=1}^nk_i \min (L_T,X_T){\bf 1}_\zs{d_i=1}\notag
\end{equation}
or (noting that $\sum_{i=1}^n k_i=1$)
\begin{align}
V_E^{p,d}(X_T)&=(1-\kappa)\,V_E^{p}(X_T)+\kappa\,(X_T-L_T)^+, \label{eq:VEfirst}
 \end{align}
where $\kappa:=\sum_{i=1}^n k_i {\bf 1}_\zs{d_i=1}$. Economically, $\kappa$ represents the total proportion of the guaranteed payment $\min (L_T,X_T)$ that the insurer has to pay to those who have died before maturity.
\begin{remark} 
When all $k_i$ are identical, i.e., $k_i=1/n$, for all $i$, the set of possible outcomes of $\kappa$ is $\{0,1/n, 2/n,\cdots, 1\}$. Assuming furthermore that the death probabilities are the same for all policyholders, i.e., $\varepsilon:=\P(d_i=1)$ for all $i$, we can compute the probability densities of $\kappa$ as follows:
\begin{equation*}
\P(\kappa=m/n)= \binom{n}{m} \varepsilon^m (1-\varepsilon)^{n-m},\quad m\in\{0,\cdots,n\}.
\label{eq:density}
\end{equation*}
In general, the distribution of $\kappa$ is called generalized Poisson Binomial distribution (see e.g. \cite{zhang2018}) which can be computed using Panjer's recursion or the normal power approximation technique. Similar discussions in insurance contexts can be found in \cite{pitacco2007}. 
\end{remark}
The equity holders' payoff \eqref{eq:VEfirst} can be expressed as
\begin{align*}
V_E^{p,d}(X_T)&=\,  \begin{cases}
 0 & \text{if } X_T\le L_T, \\
 X_T-L_T & \text{if } L_T<X_T\le \wt{L}_T, \\
 \wt{f}(X_T,\kappa)& \text{if}\,  X_T>  \wt{L}_T
 \end{cases}
\end{align*}
or, 
\begin{equation}
V_E^{p,d}(X_T)=(X_T-L_T){\bf 1}_\zs{L_T<X_T\le \wt{L}_T}+\wt{f}(X_T,\kappa) {\bf1}_\zs{X_T>  \wt{L}_T},
\label{eq:VEtau}
\end{equation}
where
\begin{equation}
\wt{f}(X_T,\kappa):=\big((1-\wt{\delta})(1-\kappa)+\kappa\big)X_T - \big((1-\delta)(1-\kappa)+\kappa\big)L_T.
\label{eq:ftilde}
\end{equation}
Since $1-\wt{\delta}\le (1-\wt{\delta})(1-\nu)+\nu\le 1$ for all $\nu\in[0,1]$ we have for $X_T\ge \wt{L}_T$
\begin{equation}
(1-\wt{\delta})X_T - (1-\delta)L_T\le \wt{f}(X_T,\kappa)\le X_T-L_T.
\label{eq:comkappa}
\end{equation}

For the case of a full protection, we still assume that 
the guarantee $L_T^i=k_i L_T$ will be paid if policyholder $i$ dies before maturity.  
In other words, the policyholder $i$'s payoff is given by
\begin{equation}
k_i V_L^{np}(X_T){\bf 1}_\zs{d_i=0}+k_i L_T {\bf 1}_\zs{d_i=1}.
\label{eq:VLtaunp}\notag
\end{equation}
Thus, the equity holders' payoff in the case of full protection is given by
\begin{equation*}
V_E^{np,d}(X_T):=X_T-\sum_{i=1}^n k_i V_L^{np}(X_T){\bf 1}_\zs{d_i=0}-\sum_{i=1}^n k_i L_T {\bf 1}_\zs{d_i=1}=(1-\kappa) V_E^{np}(X_T)+\kappa (X_T-L_T).
\label{eq:VEtaunpfirst}
\end{equation*}
As before, the equity holders' payoff in this case can be expressed as
\begin{align}
V_E^{np,d}(X_T)=(X_T-L_T){\bf 1}_\zs{X_T\le \wt{L}_T}+\wt{f}(X_T,\kappa) {\bf1}_\zs{X_T>  \wt{L}_T}.\label{eq:VEtaunp}
\end{align}
We remark that for both cases mortality only affects the region $X_T> \wt{L}_T$, where the guarantee is met and profit is shared with the policyholders.  
\subsection{Insurer's optimization objective}
The insurer who is assumed to act on behalf of the equity holders wants to solve the following optimization problem 
 \begin{align} \label{eq:TPmort}
 \sup_{ X_T\in \cX} \E\left[{U}^{\cS}( V_E^{j,d}(X_T))\right], 
\end{align}
where ${U}^{\cS}$ is an $S$-shaped utility of the form \eqref{eq:ShapU} and the payoffs $V_L^{j,d}$ and $V_E^{j,d}$, $j\in\{p,np\}$ are defined in \eqref{eq:VEtau} and \eqref{eq:VEtaunp}. 
We remark that for defaultable contracts without mortality the insurer does not encounter negative values of the payoff. Therefore, without loss of generality, we can set ${U}^{\cS}\equiv U$ in this case. 

We furthermore assume the following integrability condition:

\vspace{2mm}
{\em
\noindent{\it Assumption (U)}: For any $\lambda\in(0,\infty)$, we have
$$
\E[U((1-\wt{\delta})^{-1}I(\lambda\xi_T))]<\infty \quad \mbox{and} \quad \E[\xi_T I(\lambda\xi_T)]<\infty,
$$
where $I:=(U')^{-1}$.
}

It is straightforward to check that Assumption (U) holds for commonly used utility functions like power or logarithmic. This condition is needed below to guarantee the existence of the Lagrangian multiplier $\lambda$.

Let $\Xi\subset [0,1]$ be the finite set of all possible outcomes of $\kappa$. Using the independence of $\kappa$ and $X_T$ we define
	\begin{align}
	U_\kappa(x)&:=\E_\kappa[U(\wt{f}(x,\kappa))]=\sum_{\nu\in\Xi} {U}\big(\wt{f}(x,\nu)\big)\P(\kappa=\nu)\nonumber\\
	&=\sum_{\nu\in\Xi} {U}\Big(\big((1-\wt{\delta})(1-\nu)+\nu\big)x - \big((1-\delta)(1-\nu)+\nu\big)L_T\Big) \P(\kappa=\nu),
	\label{eq:Uepsg}
	\end{align}
where $	\E_\kappa$ stands for the conditional expectation operator with respect to the distribution of $\kappa$. In other words, $U_\kappa(x)$ is a finite sum of concave functions weighted by the probability masses $\P(\kappa=\nu)$ with $\nu\in\Xi$. The following lemma is a direct consequence of the concavity of $U$ and the fact that	$\wt{f}(\wt{L}_T,\nu)=\wt{L}_T-L_T$. 
	\begin{lemma}\label{Le:Ukappa}
	The function $U_\kappa$ is strictly increasing and concave on $[\wt{L}_T,\infty)$ with derivative	given by	\begin{align}
	U'_\kappa(x)&:=\E_\kappa[\partial_{x} \wt{f}(x,\kappa) {U}'\big(\wt{f}(x,\kappa)]\nonumber\\
	&=\sum_{\nu\in\Xi} \big((1-\wt{\delta})(1-\nu)+\nu\big) {U}'\big(\wt{f}(x,\nu) \big)\P(\kappa=\nu).
	\label{eq:UepsgUprime}
	\end{align}
	Hence, its marginal inverse $I_\kappa=(U'_\kappa)^{-1}$ exists. Moreover, 
	$U'_\kappa(\wt{L}_T)\le U'(\wt{L}_T-L_T)$.
	\end{lemma}
	Note that $\kappa\equiv 1$ a.s corresponds to the extreme case where all policyholders almost certainly die before maturity. Similarly, the case where all policyholders survive at maturity with probability of 1 corresponds to $\kappa\equiv 0$ a.s. Below, we write $U_1$ and $U_0$ for the utility of these cases respectively. In particular,
	\begin{equation*}
	U_1(x):=U(x-L_T) \quad \mbox{and} \quad U_0(x)=U(f(x))={U}\big((1-\wt{\delta})x - (1-\delta)L_T\big).
	\label{eq:noname}
	\end{equation*}
	Thus, when all policyholders almost surely survive at maturity the equity holders' payoff takes the form \eqref{PsiE} (excluding mortality). However, when no policyholder survives until maturity, the insurer's payoff simply equals the difference between the portfolio value and the guaranteed payment, $X_T-L_T$.

Now, using the independence of the mortality and the reference portfolio we rewrite
the optimization problem \eqref{eq:TPmort} as
 \begin{align} \label{eq:TPmort2}
 \sup_{ X_T\in\cX}\E\left[\wt{U}^{\cS,j,\kappa}(X_T)\right],
\end{align}
where the objective utility is defined by 
\begin{equation}
 \label{Ueps}
\wt{U}^{\cS,j,\kappa}(x):=  \,  \begin{cases}
- U_{lo}(\epsilon_j(L_T-x))& \text{if } x\le L_T, \\ 
 U(x-L_T) & \text{if } L_T<x\le \wt{L}_T, \quad j\in\{p,np\},\\
 U_{\kappa} (x) & \text{if } x>\wt{L}_T, \,
 \end{cases}
  \end{equation}
with $\epsilon_{np}:=1$ and $\epsilon_p:=0$. 
It follows from Lemma \ref{Le:Ukappa} that $\wt{U}^{\cS,j,\kappa}$ is an $S$-shaped utility function. 
We remark that the derived utility $\wt{U}^{\cS,p,\kappa}$ (the case of defaultable contracts) coincides with $\wt{U}^{\cS,np,\epsilon}$ (the case of the full protected contracts) on $[L_T,\infty)$. For convenience, we introduce
\begin{equation}
q_p:=U_\zs{lo}(0 )<
q_{np}:=U_\zs{lo}(L_T )
\label{eq:q}
\end{equation}
and treat the case of defaultable contracts as a special case of an $S$-shaped utility.



\section{Unconstrained problem}\label{sec:Uncon}
As mentioned above,  
the derived utility function $\wt{U}^{\cS,j,\kappa}$ is neither concave nor differentiable. Motivated by non-linear compensation schemes for a fund manager, the maximization of the utility of terminal wealth for {\it non-concave utility} functions has been considered in \cite{carpenter2000, karoui2005, carassus2009,reichlin2013,bichuch2014}. Similar characterizations can be found in \cite{larsen2005,mahayni2015} where non-linear contract payoffs or changing preferences are considered. 

To obtain explicit solutions, below we use a Lagrangian approach to determine the optimal solution and point out the links to the concavification points of the derived utility function which are determined below.  
Recall that $U_\kappa$ is a strictly increasing and concave function on $[\wt{L}_T,\infty)$ with first derivative $U_\kappa'$ given by \eqref{eq:UepsgUprime}. Let $q\ge 0$ and define
	\begin{equation}
\Upsilon^{\kappa,q}(x):= U_\kappa(x)-xU'_\kappa(x) +q.
\label{eq:Psi2}
\end{equation}
	%
	%
The parameter $q$ represents the upper bound of the losses in case of an $S$-shaped utility function being used to deal with negative wealth. In particular, we set $q=U_\zs{lo}(L_T )$ in the case of full protection and $q= U_\zs{lo}(0 )$ in the case of defaultable contracts. 
%
 The optimal solution crucially depends on the sign of $\Upsilon^{\kappa,q}$ at the utility changing point $\wt{L}_T$. 
From Lemma \ref{Le:Ukappa} we observe that
\begin{align*}
\Upsilon^{\kappa,q}(\wt{L}_T)=U_\kappa(\wt{L}_T)-U_\kappa'(\wt{L}_T)\wt{L}_T+q\ge U(\wh{L}_T)-U'(\wh{L}_T)\wt{L}_T+q=\Upsilon^{1,q}(\wt{L}_T).
\end{align*}
The optimal solution will be characterized by $\wh{y}^{1,q}$ and $\wh{y}^{\kappa,q}$ given in the following lemma.
\begin{lemma}\label{Le:conf}
Let $q\ge 0$ and $\Upsilon^{\kappa,q}$ be defined as in \eqref{eq:Psi2}. 
If $\Upsilon^{1,q}(\wt{L}_T)>0$ then there exists a positive number $L_T<\wh{y}^{1,q}<\wt{L}_T$ satisfying $\Upsilon^{1,q}(\wh{y}^{1,q})=0$, i.e.,
\begin{equation}
U(\wh{y}^{1,q}-L_T )-U'(\wh{y}^{1,q}-L_T )\wh{y}^{1,q}+q\stackrel{}{=}0.
\label{eq:yPT}
\end{equation}
If $\Upsilon^{\kappa,q}(\wt{L}_T)<0$ then there exists a positive number $\wh{y}^{\kappa,q}>\wt{L}_T $ satisfying $ \Upsilon^{\kappa,q}(\wh{y}^{\kappa,q})=0$, i.e.,
\begin{equation}
U_\kappa(\wh{y}^{\kappa,q})-\wh{y}^{\kappa,q}U_\kappa'(\wh{y}^{\kappa,q})+q\stackrel{}{=}0.
\label{eq:yqPT}
\end{equation}
\end{lemma}
In fact, $\wh{y}^{1,q}$ is the tangency point of the straight line starting from the point $(0,-q)$ to the curve $U(x-L_T )$ in the first case and $\wh{y}^{\kappa,q} $ is the tangency point of the straight line starting from $(0,-q)$ to the curve $U_\kappa(x)$ in the second case. 

\vspace{2mm}
\proof We prove the first property. To this end, note that the left hand side of \eqref{eq:yPT}
is an increasing, continuous function in $\wh{y}^{1,q}$ due to the concavity of $U$. By Inada's condition, it takes values in $(-\infty,\Upsilon^{1,q}(\wt{L}_T)]$ and the conclusion follows from the intermediate value theorem. The second statement can be proved in the same way using the asymptotic elasticity condition of $U$. \endproof

\vspace{2mm}
Below we solve the optimization problem \eqref{eq:TPmort2} using a Lagrangian approach. The optimal terminal wealth will be expressed as a function of the price density price $\xi_T$ and a 
Lagrangian multiplier $\lambda$ defined via the budget equality. 
Recall that for both payoffs, the derived utility function $\wt{U}^{\cS,j,\kappa}$ defined in \eqref{Ueps} takes an $S$-shaped form. 

The optimal terminal wealth of Problem \eqref{eq:TPmort2} is characterized by the following theorem.
\begin{theorem}\label{Th:unconstPTeps} The optimal terminal wealth of Problem \eqref{eq:TPmort2} is given by
	\begin{equation*}\label{eq:uopteps}
	X_T^{j,\kappa,*}=
	\begin{cases}
	I_\kappa(\lambda\xi_T){\bf 1}_\zs{\xi_T<\xi_\zs{\wt{L}}^{\kappa}}
	+ \wt{L}_T{\bf 1}_\zs{\xi_\zs{\wt{L}}^{\kappa}\le \xi_T<\xi_\zs{\wh{L}}}
	+(L_T+I(\lambda \xi_T)){\bf 1}_\zs{\xi_\zs{\wh{L}}\le \xi_T<\wh{\xi}^{1,q_{j}}}\quad\mbox{if}\quad\,\Upsilon^{1,q_j}(\wt{L}_T)>0,	\\[2mm]
I_\kappa(\lambda\xi_T){\bf 1}_\zs{\xi_T<\xi_\zs{\wt{L}}^{\kappa}}
	+ \wt{L}_T{\bf 1}_\zs{\xi_\zs{\wt{L}}^{\kappa}\le \xi_T<\xi_\zs{U}^{q_{j}}} \quad \mbox{if}\quad \Upsilon^{\kappa,q_{j}}(\wt{L}_T)\ge 0\ge\Upsilon^{1,q_{j}}(\wt{L}_T),\\[2mm]
			I_\kappa(\lambda\xi_T){\bf 1}_\zs{\xi_T<\wh{\xi}^{\kappa,q_{j}}} \quad \mbox{if}\quad \Upsilon^{\kappa,q_{j}}(\wt{L}_T)<0,
	\end{cases}
	\end{equation*}
where $j\in\{p,np\}$ and
\begin{equation}
\wh{\xi}^{\kappa,q}:=\frac{U_\kappa'(\wh{y}^{\kappa,q})}{\lambda},\quad \xi_\zs{\wt{L}}^{\kappa}:=\frac{U_\zs{\kappa}'(\wt{L}_T)}{\lambda},\quad \wh{\xi}^{1,q}:=\frac{U'(\wh{y}^{1,q}-L_T)}{\lambda},\quad  \xi_\zs{U}^{q}:=\frac{U(\wh{L}_T)+q}{\wt{L}_T\lambda}, \quad\xi_\zs{\wh{L}}:=\frac{U'(\wh{L}_T)}{\lambda}.
\label{eq:xi11}
\end{equation}
The Lagrangian multiplier $\lambda$ is defined via the budget constraint. 
\end{theorem}
\proof See Appendix \ref{sec:unpro}. \endproof

Let us give some comments on Theorem \ref{Th:unconstPTeps}. First, when $\Upsilon^{1,q_j}(\wt{L}_T)>0$ or equivalently, the concavification point of $\wt{U}^{\cS,j,\kappa}$ lies in the interval $[L_T,\wt{L}_T]$, the optimal terminal wealth takes a four-region form determined by the utility changing points $\wh{y}^{\kappa,q_j}$ and $\wt{L}_T$. For $\Upsilon^{\kappa,q_j}(\wt{L}_T)<0$ or equivalently, for the concavification point being in the interval $[\wt{L}_T,\infty)$, the optimal terminal wealth takes a two-region form and is determined by $\wh{y}^{\kappa,q_j}$. When $\Upsilon^{1,q_j}(\wt{L}_T)\le 0\le\Upsilon^{\kappa,q_j}(\wt{L}_T)$, concavification is needed on the interval $[0,\wt{L}_T]$. The concave hull is then equal to the linear segment connecting zero with $\wt{L}_T$ and coincides with $U_\kappa(x)$ on $[\wt{L}_T,\infty)$. In this case the optimal terminal wealth takes a three-region form. In all cases, the optimal terminal wealth is decreasing in $\xi_T$ and ends up with zero from a certain value of the price density on, i.e., in the worst economic states. This reflects the moral hazard problem that the insurer does not have an incentive to ensure that there is any capital in case the terminal wealth falls below the minimal guarantee. The reason is that any terminal wealth in those states only benefits the policyholders and comes at the expense of a lower terminal wealth in the more prosperous states. We will later see that introducing a VaR constraint ameliorates this situation.


\vspace{2mm}
\noindent{\bf Optimal policy of wealth distribution}: 
Like the classical concave utility maximization problem, the optimal terminal wealth is given as a function of the state price density $\xi_T$ at maturity and the Lagrangian multiplier which is  determined via the budget equation. In particular, we observe from Theorem \ref{Th:unconst} that for $\Upsilon^{1,q_j}(\wt{L}_T)>0$, the four-region form solution is characterised by wealth levels defined by $[I_\kappa(\lambda\xi_T),\wt{L}_T,L_T+I(\lambda \xi_T),0]$ corresponding to the partition of the terminal market states with boundary points $[\xi_\zs{\wt{L}}^{\kappa},\xi_\zs{\wh{L}},\wh{\xi}^{1,q_{j}}]$. 
In other words, the wealth level at time $T$ is respectively assigned to $I_\kappa(\lambda\xi_T)$ on the sub-interval $(0,\xi_\zs{\wt{L}}^{\kappa})$, to $\wt{L}_T$ on $[\xi_\zs{\wt{L}}^{\kappa},
\xi_\zs{\wh{L}})$, to $L_T+I(\lambda \xi_T)$ on $[\xi_\zs{\wh{L}}, \wh{\xi}^{1,q_{j}})$ and finally to zero on $[\wh{\xi}^{1,q_{j}},\infty)$. Therefore, it is convenient to represent the terminal wealth in dependence of the state price  as 
\begin{equation*}
\cWD_\zs{[\xi_\zs{\wt{L}}^{\kappa},\xi_\zs{\wh{L}},\wh{\xi}^{1,q_{j}}]}^{[I_\kappa,\wt{L},I+L,0]}:=I_\kappa(\lambda\xi_T){\bf 1}_\zs{\xi_T<\xi_\zs{\wt{L}}^{\kappa}}
	+ \wt{L}_T{\bf 1}_\zs{\xi_\zs{\wt{L}}^{\kappa}\le \xi_T<\xi_\zs{\wh{L}}}
	+(L_T+I(\lambda \xi_T)){\bf 1}_\zs{\xi_\zs{\wh{L}}\le \xi_T<\wh{\xi}^{1,q_{j}}}.
\label{eq:WD4}
\end{equation*}
In the same spirit, the optimal wealth for the case $\Upsilon^{\kappa,q_{j}}(\wt{L}_T)\ge 0\ge\Upsilon^{1,q_{j}}(\wt{L}_T)$ and the case $\Upsilon^{\kappa,q_{j}}(\wt{L}_T)<0$ can be respectively represented by the three-region and the two-region wealth distributions as
\begin{equation*}
\cWD_\zs{[\xi_\zs{\wt{L}}^{\kappa},\xi_\zs{U}^{q_{j}}]}^{[I_\kappa,\wt{L},0]}:=I_\kappa(\lambda\xi_T){\bf 1}_\zs{\xi_T<\xi_\zs{\wt{L}}^{\kappa}}
	+ \wt{L}_T{\bf 1}_\zs{\xi_\zs{\wt{L}}^{\kappa}\le \xi_T<\xi_\zs{U}^{q_{j}}} \quad \mbox{and}\quad \cWD_\zs{[\wh{\xi}^{\kappa,q_{j}}]}^{[I_\kappa,0]}:=I_\kappa(\lambda\xi_T){\bf 1}_\zs{\xi_T<\wh{\xi}^{\kappa,q_{j}}}.
		\label{eq:WD3}
\end{equation*}
Thus, the optimal terminal wealth in Theorem \ref{Th:unconstPTeps}  can be expressed as
\begin{equation}
X_T^{j,\kappa,*}=\cWD_\zs{[\xi_\zs{\wt{L}}^{\kappa},\xi_\zs{\wh{L}},\wh{\xi}^{1,q_{j}}]}^{[I_\kappa,\wt{L},I+L,0]}{\bf 1}_\zs{\Upsilon^{1,q_{j}}(\wt{L}_T)>0}+ \cWD_\zs{[\xi_\zs{\wt{L}}^{\kappa},\xi_\zs{U}^{q_{j}}]}^{[I_\kappa,\wt{L},0]} {\bf 1}_\zs{\Upsilon^{\kappa,q_{j}}(\wt{L}_T)\ge 0\ge\Upsilon^{1,q_{j}}(\wt{L}_T)}+ \cWD_\zs{[\wh{\xi}^{\kappa,q_{j}}]}^{[I_\kappa,0]}	{\bf 1}_\zs{\Upsilon^{\kappa,q_{j}}(\wt{L}_T)<0}. 
\end{equation}
These representations of wealth distribution will be used in the rest of the paper for notational convenience.

\vspace{2mm}
Let us now consider the case where $\kappa\equiv 1$, i.e., all the policyholders die before the maturity of the contract. In this case, the optimal terminal wealth is given by the following all-or-nothing form.

\begin{corollary}
If $\kappa\equiv 1$, the optimal terminal wealth is given by $X_T^{j,\kappa,*}=\cWD_\zs{[\wh{\xi}^{1,q_j}]}^{[L_T+I,0]}$, $j\in\{p,np\}$, where $\lambda$ satisfies the budget constraint.
\end{corollary}

We conclude this section by presenting the optimal solution without mortality for both kinds of contracts, i.e., $\kappa=0$. 
In this case, the inverse marginal function $I_0=(U'_1)^{-1}$ in Lemma \ref{Le:Ukappa} can be computed explicitly as 
\begin{equation}
h(x):=I_0(x)= \frac{I\big(x/(1-\tilde\delta))  + (1-\delta)L_T}{1-\tilde\delta}\,.
\label{eq:h}
\end{equation}
Note that $h$ is a decreasing mapping from $\big(0,(1-\tilde\delta)U'(\wh{L}_T)\big]$ to $[\wt{L}_T, \infty)$. 

\begin{corollary}[Defaultable contract without mortality]\label{Th:unconst} The optimal solution to the unconstrained problem $\max_{X_T\in \chi}\E[U(V_E^p(X_T))]$ is given by
\begin{equation*}
	X^{p,*}_T=	\cWD_\zs{[\xi_\zs{\wt{L}}^0,\xi_\zs{\wh{L}},\wh{\xi}^{1,0}]}^{[h,\wt{L},I+L,0]}{\bf 1}_\zs{\Upsilon^{1,0}(\wt{L}_T)>0}+ \cWD_\zs{[\xi_\zs{\wt{L}}^0,\xi_\zs{U}^{0}]}^{[h,\wt{L},0]} {\bf 1}_\zs{\Upsilon^{0,0}(\wt{L}_T)\ge 0\ge\Upsilon^{1,0}(\wt{L}_T)}+ \cWD_\zs{[\wh{\xi}^ {0,0}]}^{[h,0]}{\bf 1}_\zs{ \Upsilon^{0,0}(\wt{L}_T)<0},
\label{eq:Xp}
\end{equation*}
where 
\begin{align*}
&\xi_\zs{\wt{L}}^0:=\frac{(1-\wt{\delta})U'(\wh{L}_T)}{\lambda},\quad \xi_\zs{U}^0:=\frac{U(\wh{L}_T)}{\wt{L}_T\lambda},\quad\wh{\xi}^{1,0}:=\frac{U'(\wh{y}^{1,0}-L_T)}{\lambda},\quad \wh{\xi}^{0,0}:=\frac{(1-\wt{\delta})U'(f(\wh{y}^{0,0}))}{\lambda},
\label{eq:xi1_0}
\end{align*}
and $\xi_\zs{\wh{L}}:=U'(\wh{L}_T)/{\lambda}$ and $\lambda$ is defined via the budget constraint $\E[\xi_T	X_T^{p,*}]=X_0$.
\end{corollary}
\begin{corollary}\label{Re:infty}
Concavification is not needed for the case where $U(0)=-\infty$. In this situation, it can be deduced directly from the proof of Theorem \ref{Th:unconst} in Appendix \ref{sec:unpro} that the optimal terminal wealth is given by $\cWD_\zs{[\xi_\zs{\wt{L}}^0,\xi_\zs{\wh{L}}]}^{[h,\wt{L},I+L]}(\lambda,\xi_T).$
\end{corollary}
The optimal terminal wealth for a fully protected contract without mortality takes an all-or-nothing form which is given in the following Corollary. 
\begin{corollary}[Fully protected contract without mortality]\label{Th:unconstPT} The optimal solution to the $S$-shaped utility unconstrained problem for the full-protection payoff $V_E^{np}$ without mortality is given by
	\begin{equation*}\label{eq:uoptPT}
	X^{np,*}_T=	\cWD_\zs{[\xi_\zs{\wt{L}}^0,\xi_\zs{\wh{L}},\wh{\xi}^{1,q_{np}}]}^{[h,\wt{L},I+L,0]}{\bf 1}_\zs{\Upsilon^{1,q_{np}}(\wt{L}_T)>0}+ \cWD_\zs{[\xi_\zs{\wt{L}}^0,\xi_\zs{U}^{q_\zs{np}}]}^{[h,\wt{L},0]} {\bf 1}_\zs{\Upsilon^{0,q_{np}}(\wt{L}_T)\ge 0\ge\Upsilon^{1,q_{np}}(\wt{L}_T)}+ \cWD_\zs{[\wh{\xi}^ {0,q_\zs{np}}]}^{[h,0]}{\bf 1}_\zs{ \Upsilon^{0,q_{np}}(\wt{L}_T)<0},
	\end{equation*}
where 
\begin{align*}
\xi_\zs{U}^{q_{np}}:=\frac{U(\wh{L}_T)+U_\zs{lo}(L_T)}{\wt{L}_T\lambda},
\quad\wh{\xi}^{1,q_\zs{np}}:=\frac{U'(\wh{y}^{1,q_\zs{np}}-L_T)}{\lambda}\quad\mbox{and} \quad \wh{\xi}^{0,q_\zs{np}}:=(1-\wt{\delta})\frac{U'(f(\wh{y}^{0,q_{np}}))}{\lambda}
\label{eq:xi1_betalo}
\end{align*}
and $\lambda$ is defined via the budget constraint $\E[\xi_T	X_T^{np,*}]=X_0$.
\end{corollary}

\begin{remark}\label{Re:Svs}


We observe that for any $x\ge \wt{L}_T$, $\Upsilon^{1,q_{np}}(x)>\Upsilon^{1,0}(x)$, which implies that $\wh{y}^{1,q_{np}}<\wh{y}^{1,0}$ whenever they both exist. Similarly, $\wh{y}^{0,q_{np}}<\wh{y}^{0,0}$ whenever they both exist. This means that the concavification area for the full protection case will be reduced in comparison to the case of a defaultable contract, assuming that the insurer uses the same concave utility function $U$ for the gain part.

\end{remark}
\vspace{2mm}

	\vspace{2mm}
\noindent{\bf Optimal strategy:} We have determined the optimal terminal wealth as a function of $\xi_T$ and a multiplier $\lambda$ that satisfies the budget constraint. For the reader's convenience, we briefly discuss how to deduce the optimal strategy by applying the martingale representation theorem in the case of a Brownian filtration. To simplify the presentation we consider a simple Black-Schole's model with one risky asset given by 
$$
\d S_t=S_t(\mu \d t+\sigma \d W_t),
$$
and $S_t^0=e^{rt }$ where $r, \mu, \sigma$ are positive constants. The pricing kernel $\xi_t$ admits the dynamics
$\d \xi_t=-\xi_t(r\d t+\theta\d W_t)$, $\xi_0=1$, where $\theta:=(\mu-r)/\sigma$ is the market price of risk. 
The wealth process starting with an initial wealth $X_0>0$
related to the strategy $\nu_t$ (the amount invested in the risky asset at time $t$) can be expressed as
\begin{equation}
\d X_t =  (r X_t+ \nu_t(\mu-r))  \d t + \nu_t \sigma \d W_t, \quad X_0>0.\label{eq:Mol.00}
\end{equation}
To guarantee the existence of the strong unique solution of \eqref{eq:Mol.00} we require $\nu_t$ to be a $\bbg$-progressively measurable process satisfying the square integrability condition $\int_0^T\nu_t ^2 \d t<\infty$ a.s. Recall that $\bbg\equiv \bbf$ in this case.

Now, let $Z=X(\lambda,\xi_T)$ be the optimal terminal wealth and define $Z_t:=\xi_t^{-1}\E[\xi_T Z\vert \cF_t]$. Then, the process $(\xi_t Z_t)_\zs{t\in[0,T]}$ is a martingale under $\P$. Therefore, by the martingale representation theorem, there exists an adapted process $(\varsigma_t)$ such that $\int_0^T \varsigma_t^2 \d t<\infty$ a.s. and
$$
\xi_t Z_t=X_0+\int_0^t \varsigma_s \d W_s,\quad t\in[0,T],
$$
or equivalently, $\d (\xi_t Z_t)=\varsigma_t \d W_t$. On the other hand, applying It\^{o}'s lemma we get
\begin{align*}
\d  Z_t= \d  (\xi_t Z_t \xi_t^{-1})&=\xi_t^{-1}\d  (\xi_t Z_t ) + \xi_t Z_t \d \xi_t^{-1}+ \d  (\xi_t Z_t )  \d \xi_t^{-1} \notag\\
&=( \xi_t^{-1} \varsigma_t+Z_t \theta) \d W_t+ \bigg((r+\theta^2)Z_t+\xi_t^{-1}\varsigma_t \theta \bigg ) \d t.
\end{align*}
Since $\int_0^T \varsigma_t^2 \d t<\infty$ a.s. we remark that the process $Z_t$ being the unique strong solution of the above stochastic differential equation is a continuous process in $[0,T]$. Now, identifying the dynamics of $\d  Z_t$ with equation \eqref{eq:Mol.00} we deduce that the optimal strategy $\nu_t^*$ is given by 
$$
\nu_t^*= \sigma^{-1}\xi_t^{-1}\varsigma_t+ \sigma^{-1}\theta Z_t ,\quad X_t^*=Z_t.$$
Using the inequality $(a+b)^2\le 2(a^2+b^2)$ and the continuity of the processes $\xi$ and $Z$ we obtain
$$
\int_0^T (\nu_t^{*})^2 \d t\le 2 \sigma^{-2}\left(\int_0^T \varsigma_t^2 \d t \,\sup_{t\in[0,T]}\xi_t^{-2}+\theta^2 T \sup_{t\in[0,T]} Z_t^2 \right)
<\infty, \quad a.s. $$
which implies that $\nu_t^*$ is admissible. 

The same argument can be applied for the constrained problems below. More explicit representations can be provided for power and logarithmic utility functions. 

\section{Optimal investment under regulations}\label{Se:VaR0}

\subsection{The VaR-constrained problem}\label{Se:VaR}
As seen in the previous section, under the absence of regulation the insurer on behalf of the equity holders will optimally choose a strategy which may lead to insolvency at maturity. In this case, the policyholders suffer severe losses since the terminal portfolio value may be zero for very bad market scenarios. In practice, an appropriate investment must take some regulatory constraints into account. According to Solvency II, the insurance company needs to ensure that a VaR-type regulation (i.e., a default probability constraint) shall be satisfied. In addition, the insurance company
is interested in achieving at least a target payment to serve the promised guaranteed amount to the policyholders. Therefore, the insurer has to choose an optimal dynamic portfolio under a VaR constraint, which can be stated as
\begin{align} \label{eq:const.1}
 \sup_{ X_T\in\cX}\E\left[\wt{U}^{\cS,j,\kappa}(X_T)\right],
 \quad\text{s.t.} \quad\P(X_T<L_T)<\beta,\quad j\in\{p,np\},
\end{align}
for some allowed default probability level $0\le \beta<1$. 

\begin{theorem}\label{Th:const} Let $\beta\in (0,1)$ and define $\bar{\xi}$ so that $\P(\xi_T>\bar{\xi})=\beta$. Suppose that $X_0\ge\E[\xi_T L_T {\bf 1}_\zs{\xi_T\le \bar{\xi}}]$. Then, the VaR-constrained problem \eqref{eq:const.1}, $j\in\{p,np\}$, admits the following optimal solution:
\begin{itemize}
	\item If $\Upsilon^{1,q_j}(\wt{L}_T)>0$ then
	\begin{equation*}
X^{VaR,j,\kappa,*}_T=\cWD_\zs{[\xi_\zs{\wt{L}}^{\kappa},\xi_\zs{\wh{L}},\bar{\xi}]}^{[I_\kappa,\wt{L},I+L,0]}{\bf 1}_\zs{ \bar{\xi}\ge \wh{\xi}^{1,q_j}}+\cWD_\zs{[\xi_\zs{\wt{L}}^{\kappa},\xi_\zs{\wh{L}},\wh{\xi}^{1,q_{j}}]}^{[I_\kappa,\wt{L},I+L,0]}{\bf 1}_\zs{ \bar{\xi}< \wh{\xi}^{1,q_{j}}}.
		\end{equation*}
		\item If $\Upsilon^{\kappa,q_j}(\wt{L}_T)\ge 0\ge\Upsilon^{1,q_j}(\wt{L}_T)$ then
\begin{equation*}
X^{VaR,j,\kappa,*}_T=\cWD_\zs{[\xi_\zs{\wt{L}}^{\kappa},\xi_\zs{\wh{L}},\bar{\xi}]}^{[I_\kappa,\wt{L},I+L,0]}{\bf 1}_\zs{ \bar{\xi}\ge \xi_\zs{\wh{L}}}+\cWD_\zs{[\xi_\zs{\wt{L}}^{\kappa},\bar{\xi}]}^{[I_\kappa,\wt{L},0]}{\bf 1}_\zs{ \xi_\zs{U}^{q_{j}}\le\bar{\xi}< \xi_\zs{\wh{L}}}
+\cWD_\zs{[\xi_\zs{\wt{L}}^{\kappa},\xi_\zs{U}^{q_{j}}]}^{[I_\kappa,\wt{L},0]}{\bf 1}_\zs{ \bar{\xi}< \xi_\zs{U}^{q_{j}}}.
		\end{equation*}
		\item If $\Upsilon^{\kappa,q_j}(\wt{L}_T)<0$ then
\begin{align*}
X^{VaR,j,\kappa,*}_T=\cWD_\zs{[\xi_\zs{\wt{L}}^{\kappa},\xi_\zs{\wh{L}},\bar{\xi}]}^{[I_\kappa,\wt{L},I+L,0]}{\bf 1}_\zs{ \bar{\xi}\ge \xi_\zs{\wh{L}}}+\cWD_\zs{[\xi_\zs{\wt{L}}^{\kappa},\bar{\xi}]}^{[I_\kappa,\wt{L},0]}{\bf 1}_\zs{ \xi_\zs{\wt{L}}^{\kappa}\le \bar{\xi}< \xi_\zs{\wh{L}}}
+
\cWD_\zs{[\bar{\xi}]}^{[I_\kappa,0]}{\bf 1}_\zs{ \wh{\xi}^{\kappa,q_j}\le \bar{\xi}<\xi_\zs{\wt{L}}^{\kappa}}
+
\cWD_\zs{[ \wh{\xi}^{\kappa,q_j}]}^{[I_\kappa,0]}{\bf 1}_\zs{\bar{\xi} <\wh{\xi}^{\kappa,q_j}}.
\end{align*}
\end{itemize}
In each case, the Lagrangian multiplier $\lambda$ is defined via the budget constraint.
\end{theorem}
\proof See Appendix \ref{sec:pro}. \endproof

\vspace{2mm}

We observe first that when $\bar{\xi}< \wh{\xi}^{1,q_j}$/$\bar{\xi}< \xi_U^{q_j}$/$\bar{\xi}<\wh{\xi}^{\kappa,q_j}$ in the first/second/third case, the VaR constraint is not binding and the corresponding unconstrained solution $X_T^{j,\kappa,*}$ given by \eqref{eq:uopteps} is still optimal for \eqref{eq:const.1}. On the other hand, when the VaR constraint is binding (active) the terminal wealth will be (partially) shifted to the right of the concavification point. We can observe that for $j\in\{n,np\}$, there exists $\xi^*_j>0$ such that $$X^{VaR,j,\kappa,*}_T\ge X^{j,\kappa,*}_T  \quad \mbox{for all}\quad \xi_T\ge \xi^*_j,$$
meaning that the VaR-terminal wealth dominates the unconstrained terminal wealth for the most negative loss states (due to the fact that $X^{VaR,j,\kappa,*}_T$ and  $X^{j,\kappa,*}_T $ are decreasing functions of $\xi_T$). Furthermore, the inequality is strict for some states. Hence, introducing a VaR-constraint forces the equity holders to enlarge the proportion of hedged loss states, leading to a genuine improvement for the policyholders. This result is contrary to the situation where the insurer maximizes the utility of the total wealth of the company without distinguishing between equity holders and policyholders, in which case a VaR constraint may induce the insurer to take excessive risk leading to higher losses than in the case of no-regulation, see \cite{BasakShapiro01}. 
However, there is still a region of market scenarios in which the optimal terminal wealth equals zero, which means that a VaR regulation does not lead to a full prevention of moral hazard. The intuitive reason is that under a VaR regulation, the insurance company is only required to keep the portfolio value above $L_T$ with a given probability $1-\beta$. Once the regulation is probabilistically fulfilled the insurer can push the remaining risk into the tail to seek a higher potential wealth level in good market states. This is consistent with the classical VaR-constrained asset allocation problem with concave utility functions \cite{BasakShapiro01}. Nevertheless, in our case with participating contracts, the use of a VaR constraint does not lead the insurance company to bigger losses than in the case of no regulations as in the classical VaR problem. On the contrary, our results show that a VaR constraint {\em strictly improves the risk management for the loss states}.

\subsection{The PI-constrained problem}

In this section, we try to better protect the policyholders from the equity holders' moral hazard by, instead of having a VaR constraint, assuming that the insurance company has to keep the portfolio value almost surely above some given level minimum capital requirement $l$. Hence the insurance company needs to solve
\begin{equation} \label{eq:PI1}
 \sup_{ X_T\in\cX}\E\left[\wt{U}^{\cS,j,\kappa}(X_T)\right],
 \quad\text{s.t.} \quad X_T\ge l\quad a.s.,\quad j\in\{p,np\}.
\end{equation}
For simplicity we assume that $l$ is deterministic. 

Let us first consider the case  $0\le l\le L_T$. As before, concavification is characterized by the following generalized version of Lemma \ref{Le:conf}.


\begin{lemma}[Tangency point for PI-problem]\label{Le:mini}
Let $q\ge 0$ and $\Upsilon^{\kappa,q}_l$ be defined by
	\begin{equation}
\Upsilon^{\kappa,q}_l(x):= U_\kappa(x)-U'_\kappa(x)(x-l) +q.\notag
\label{eq:Psi2l}
\end{equation}
If $\Upsilon^{1,q}_l(\wt{L}_T)>0$ then there exists a positive number $L_T<\wh{y}^{1,q}_l<\wt{L}_T$ satisfying $\Upsilon^{1,q}_l(\wh{y}_l^{1,q})=0$, i.e.,
\begin{equation}
U(\wh{y}^{1,q}_l-L_T )-U'(\wh{y}^{1,q}_l-L_T )(\wh{y}^{1,q}_l-l)+q\stackrel{}{=}0.\notag
\label{eq:yPTl1}
\end{equation}
If $\Upsilon^{\kappa,q}_l(\wt{L}_T)<0$ then there exists a positive number $\wh{y}^{\kappa,q}_l>\wt{L}_T $ satisfying $ \Upsilon^{\kappa,q}_l(\wh{y}^{\kappa,q}_l)=0$, i.e.,
\begin{equation}
U_\kappa(\wh{y}^{\kappa,q}_l)-U_\kappa'(\wh{y}^{\kappa,q}_l)(\wh{y}^{\kappa,q}_l-l)+q\stackrel{}{=}0.\notag
\label{eq:yqPTl2}
\end{equation}
\end{lemma}
As before, $\wh{y}^{1,q}_l$ is the tangency point of the straight line starting from the point $(l,-q)$ to the curve $U(x-L_T )$ in the first case, and $\wh{y}^{\kappa,q}_l $ is the tangency point of the straight line starting from $(l,-q)$  to the curve $U_\kappa(x)$ in the second case. For our problem with $l\in [0,L_T]$ the loss magnitude $q$ is equal to $q_p^l:=0$ for defaultable contracts or 
$q_{np}^l:=U_\zs{lo}(L_T -l)$ for fully protected contracts.


We remark that to hedge against the minimum capital level $l$, the investor must start with at least $\E[\xi_T l]$ as the initial capital. Next, we show in Theorem \ref{Th:l} that the optimal terminal wealth under a PI constraint takes a similar form as in the unconstrained case but with bounded losses. The moral hazard is restricted thanks to the additional guarantee $l$.
\begin{theorem}\label{Th:l} Assume that $0\le l< L_T$ and $X_0\ge \E[\xi_T l]=le^{-rT}$. Then, the optimal solution to the insurance portfolio problem \eqref{eq:PI1} is given by 
	\begin{equation*}\label{eq:lopt}
	X_T^{PI,j,\kappa,*}=\cWD_\zs{[\xi_\zs{\wt{L}}^{\kappa},\xi_\zs{\wh{L}},\wh{\xi}^{1,q_{j}^l}_l]}^{[I_\kappa,\wt{L},I+L,l]}{\bf 1}_\zs{\Upsilon^{1,q_{j}^l}_l(\wt{L}_T)>0}+ \cWD_\zs{[\xi_\zs{\wt{L}}^{\kappa},\xi_\zs{U}^l]}^{[I_\kappa,\wt{L},l]} {\bf 1}_\zs{\Upsilon^{\kappa,q_{j}^l}_l(\wt{L}_T)\ge0\ge\Upsilon^{1,q_{j}^l}_l(\wt{L}_T)}+ \cWD_\zs{[\wh{\xi}^{\kappa,q_{j}^l}_l]}^{[I_\kappa,l]}	{\bf 1}_\zs{\Upsilon^{\kappa,q_{j}^l}_l(\wt{L}_T)<0}, \quad j\in\{p,np\},
	\end{equation*}
where, similarly to \eqref{eq:xi11},
\begin{equation}
\wh{\xi}^{1,q^l_j}_l:=\frac{U'(\wh{y}^{1,q^l_j}_l-L_T)}{\lambda}, \quad 
\wh{\xi}^{\kappa,q^l_j}_l=\frac{U_\kappa'(\wh{y}^{\kappa,q^l_j}_l)}{\lambda}
\quad
\mbox{and}
\quad
\xi_\zs{U}^l:=\frac{U(\wh{L}_T)}{(\wt{L}_T-l)\lambda}.
\label{eq:xil}
\end{equation}
The Lagrangian multiplier $\lambda$ is defined via the budget constraint $\E[\xi_TX_T^{PI,j,\kappa,*}]=X_0.$
\end{theorem}
\proof See Appendix \ref{se:profPI}. \endproof

In this theorem the moral hazard problem shown in Theorem \ref{Th:unconst} is solved by introducing a minimal bound. However, this comes at the expense of lowering the wealth significantly in the prosperous states and is thus rather costly, see the numerical illustration given below.
 
Let us now turn to the case where the minimum amount lies between the utility changing points $L_T< l<\wt{L}_T$ (i.e., between the guarantee and the bonus threshold). In this case, the derived utility function $\wt{U}^{\cS,j,\kappa}$ is strictly increasing and globally concave in the considered domain $[l,\infty)$. However, $\wt{U}^{\cS,j,\kappa}$ is not smooth at $\wt{L}_T$ due to Lemma \ref{Le:Ukappa}, which makes the classical utility maximization result inapplicable. We remark that concavification is not needed because we have global concavity in the optimization domain. The case $l\ge \wt{L}_T$ is just the classical portfolio insurance problem with concave utility function $U_\kappa$ and can be dealt with similarly. The following result can be directly obtained using the same Lagrangian technique. 
\begin{proposition}\label{Th:l2} Assume that $X_0\ge \E[\xi_T l]=le^{-rT}$. Then, for $ L_T< l<\wt{L}_T$, the optimal solution to the portfolio insurance problem \eqref{eq:PI1} is given by
	\begin{equation}\label{eq:lopt2}
	\cWD_\zs{[\xi_\zs{\wt{L}}^{\kappa},\xi_\zs{\wh{L}},\xi_{\wh{l}}]}^{[I_\kappa,\wt{L},I+L,l]}=
	I_\kappa(\lambda\xi_T){\bf 1}_\zs{\xi_T<\xi_\zs{\wt{L}}^{\kappa}}	+ \wt{L}_T{\bf 1}_\zs{\xi_\zs{\wt{L}}^{\kappa}\le \xi_T<\xi_\zs{\wh{L}}}	+(L_T+I(\lambda \xi_T)){\bf 1}_\zs{\xi_\zs{\wh{L}}\le \xi_T<\xi_{\wh{l}}}+l {\bf 1}_\zs{\xi_T\ge\xi_{\wh{l}}},\notag
	\end{equation}
where $\xi_{\wh{l}}:=U'(l-L_T)/\lambda$. When $l=L_T$ the optimal terminal wealth is 
	\begin{equation}\label{eq:lopt22}
	\cWD_\zs{[\xi_\zs{\wt{L}}^{\kappa},\xi_\zs{\wh{L}}]}^{[I_\kappa,\wt{L},I+L]}=
	I_\kappa(\lambda\xi_T){\bf 1}_\zs{\xi_T<\xi_\zs{\wt{L}}^{\kappa}}	+ \wt{L}_T{\bf 1}_\zs{\xi_\zs{\wt{L}}^{\kappa}\le \xi_T<\xi_\zs{\wh{L}}}	+(L_T+I(\lambda \xi_T)){\bf 1}_\zs{\xi_\zs{\wh{L}}\le \xi_T},\notag
	\end{equation}
	which can be seen as a limiting case by sending $\xi_{\wh{l}}$ in \eqref{eq:lopt2} to infinity. 
For $\wt{L}_T\le l$, the optimal terminal wealth is given by $\cWD_\zs{[\xi_\zs{l}^{\kappa}]}^{[I_\kappa,l]}	=I_\kappa(\lambda\xi_T){\bf 1}_\zs{\xi_T<\xi_\zs{l}^{\kappa}}
	+l {\bf 1}_\zs{\xi_T\ge\xi_\zs{l}^{\kappa}},$
where $\xi_\zs{l}^{\kappa}:=U'_\kappa(l)/\lambda$. The Lagrangian multiplier $\lambda$ satisfies the budget constraint.
\end{proposition}

\section{Numerical examples}\label{sec:Num}
We assume that the insurer's utility function is given by $U(x):=x^{1-\gamma}/(1-\gamma)$ with $\gamma=0.5$. For the full protection case, we assume that $ U_{lo}(x)=\eta U(-x)$. The market coefficients are 
$
\mu = 0.05; r = 0.03; \sigma = 0.3.
$
The contract has a maturity $T = 10$ and the total contribution (i.e., the initial capital) is fixed with $X_0 = 100$. The guarantee is given by $L_T = \alpha X_0 e^{g T}$, where the guaranteed interest rate is $g=0.02$. Note that the threshold where the participation bonus kicks in $\wt{L}_T=L_T/\alpha=X_0 e^{g T}$ is now fixed and independent of the participation rate in the comparative analysis below. For simplicity we do not consider mortality risk. Below we look at the investment behavior of the reference portfolio decided by the insurer in order to maximize its expected terminal utility with and without regulations. Note that constant proportion portfolio insurance (CPPI) strategies can be considered as a possible benchmark for comparative analysis, see \cite{Lin2017}. We emphasize that CPPI strategies are not optimal and provide less expected utility for the insurer. For an analytic discussion, it may therefore be useful to compare the insurer's strategy with the Merton strategy which maximizes the insurer's utility of the total wealth of the company without distinguishing between the equity and policyholders starting with the same total initial endowment. 
\begin{figure}[h]
\begin{subfigure}[b]{0.49\textwidth}
       	 \includegraphics[width=\columnwidth,height=7cm]{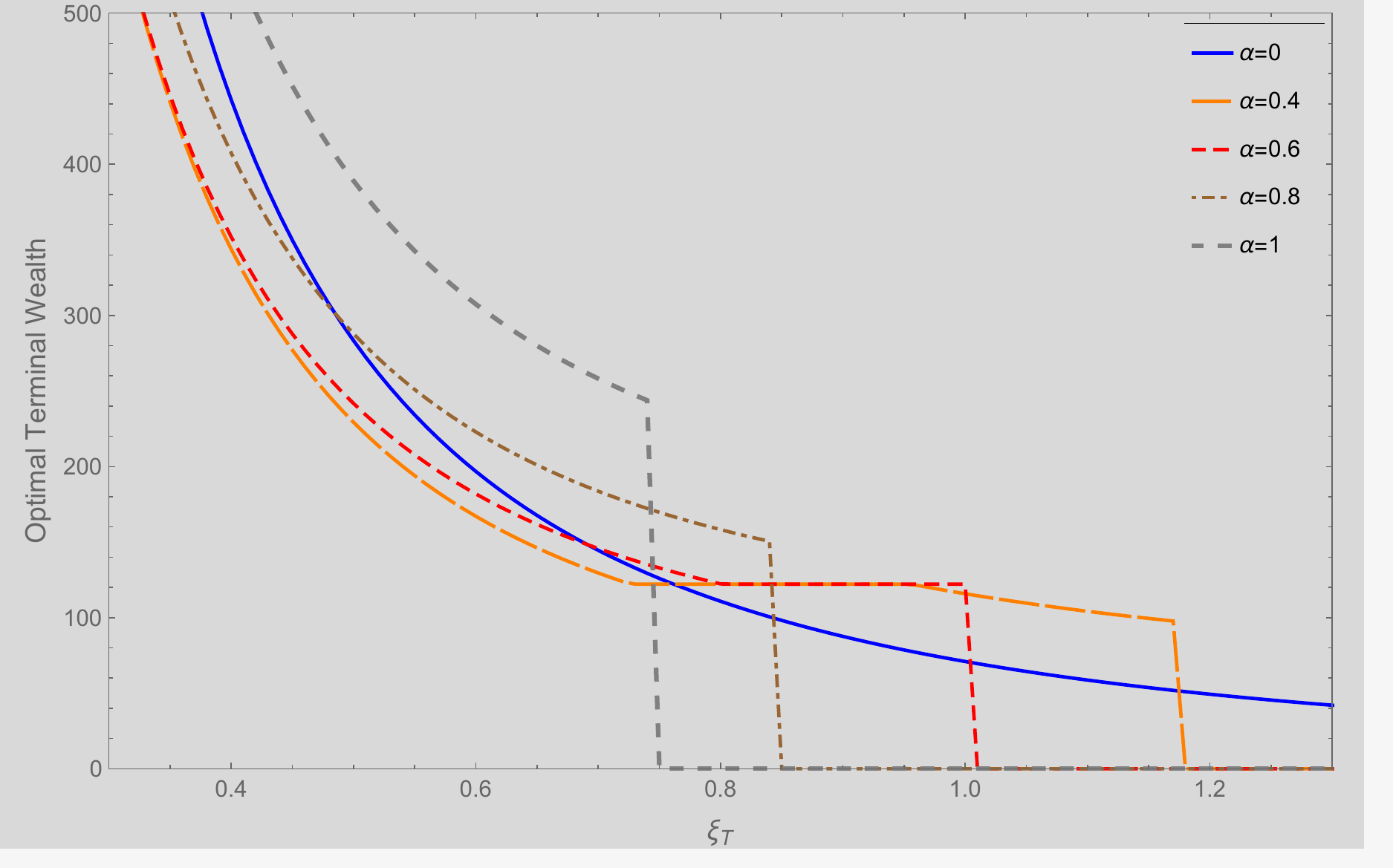}%
			\caption{\small Defaultable contract}
		 \end{subfigure}
		\begin{subfigure}[b]{0.49\textwidth}
         	 \includegraphics[width=\columnwidth,height=7cm]{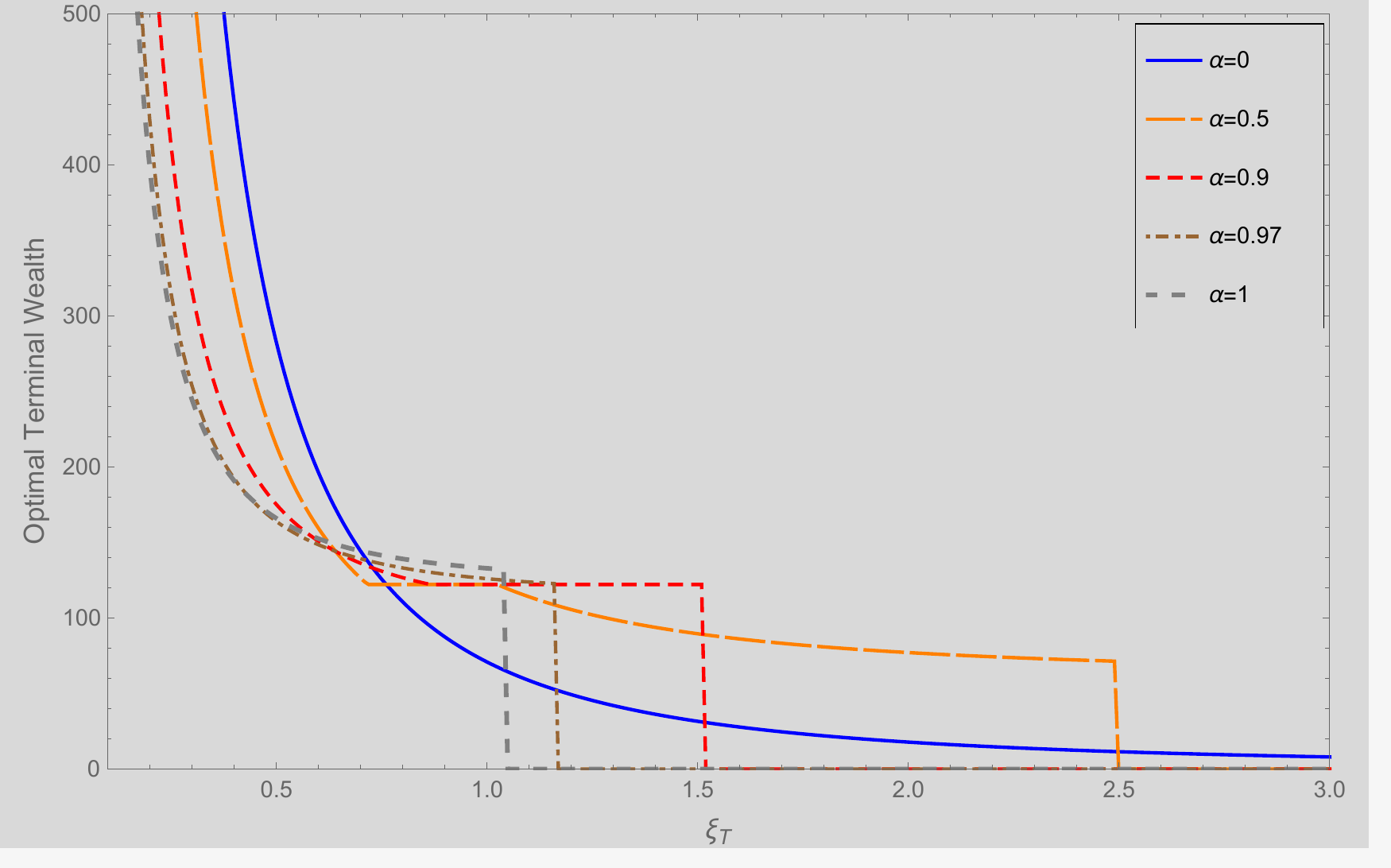}%
			\caption{\small Fully protected contract with $\eta=1.01$.}
		 \end{subfigure}
				\caption{Unconstrained optimal terminal wealth with $\delta=0.6$ without mortality}
				\label{Fig:OptXT}
		\end{figure}

Figure \ref{Fig:OptXT} plots the optimal terminal portfolio with different values of $\alpha$ for the case with a defaultable contract and with a fully protected contract. It can be observed that the optimal terminal wealth can take a two-, three- or four-region form as also shown in Theorem \ref{Th:unconst}. 
The first thing to note from the graph is that in both cases (with a defaultable or a fully protected contract) the investment riskiness admits a monotonicity in the overall participation rate $\alpha$. In particular for defaultable contracts, the insurer is most risk seeking for very large values of $\alpha$ which can be seen by the fact that 
these values in good states (i.e., for small values of $\xi_T$) yield a relatively high terminal wealth while in bad states the terminal wealth is lower, and the unhedged region where the terminal wealth is zero increases. The reason might be that as will be seen below very high participation rates mean that the insurer contributes very limitedly in the investment pool for which it tries to compensate (since the bonus rate is fixed) by taking on more risks in order to mimic a Merton like terminal wealth in the good states for itself. Since in this case the insurer gets a bonus without contributing much, this is a classical moral hazard induced behavior. 
Similar effects can be observed for fully protected contracts.
\begin{figure}[h]
      \begin{subfigure}[b]{0.49\textwidth}
      \includegraphics[width=\columnwidth,height=6cm]{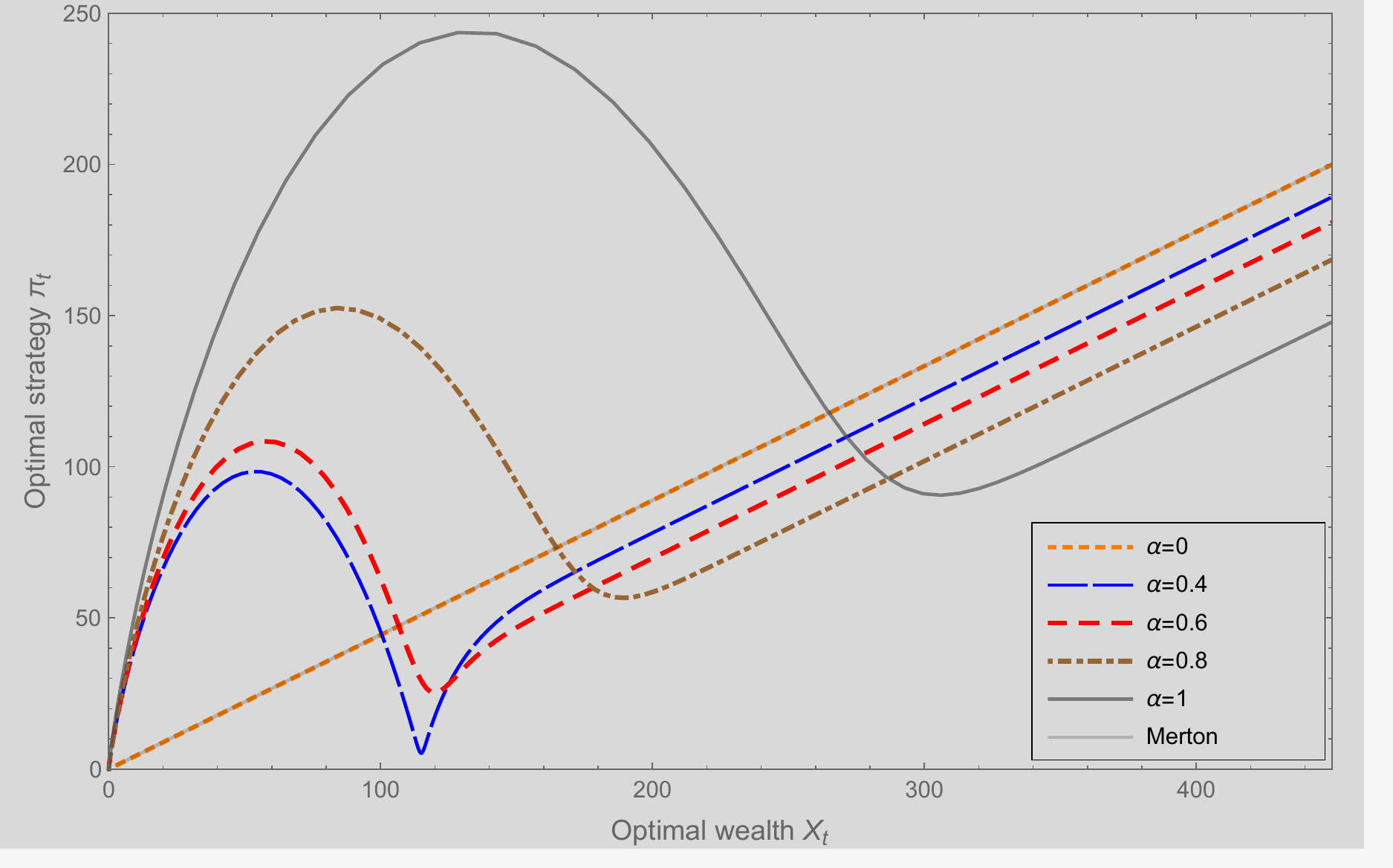}%
\caption{\small $\delta=0.6$ }%
    \end{subfigure}
      \begin{subfigure}[b]{0.49\textwidth}
      \includegraphics[width=\columnwidth,height=6cm]{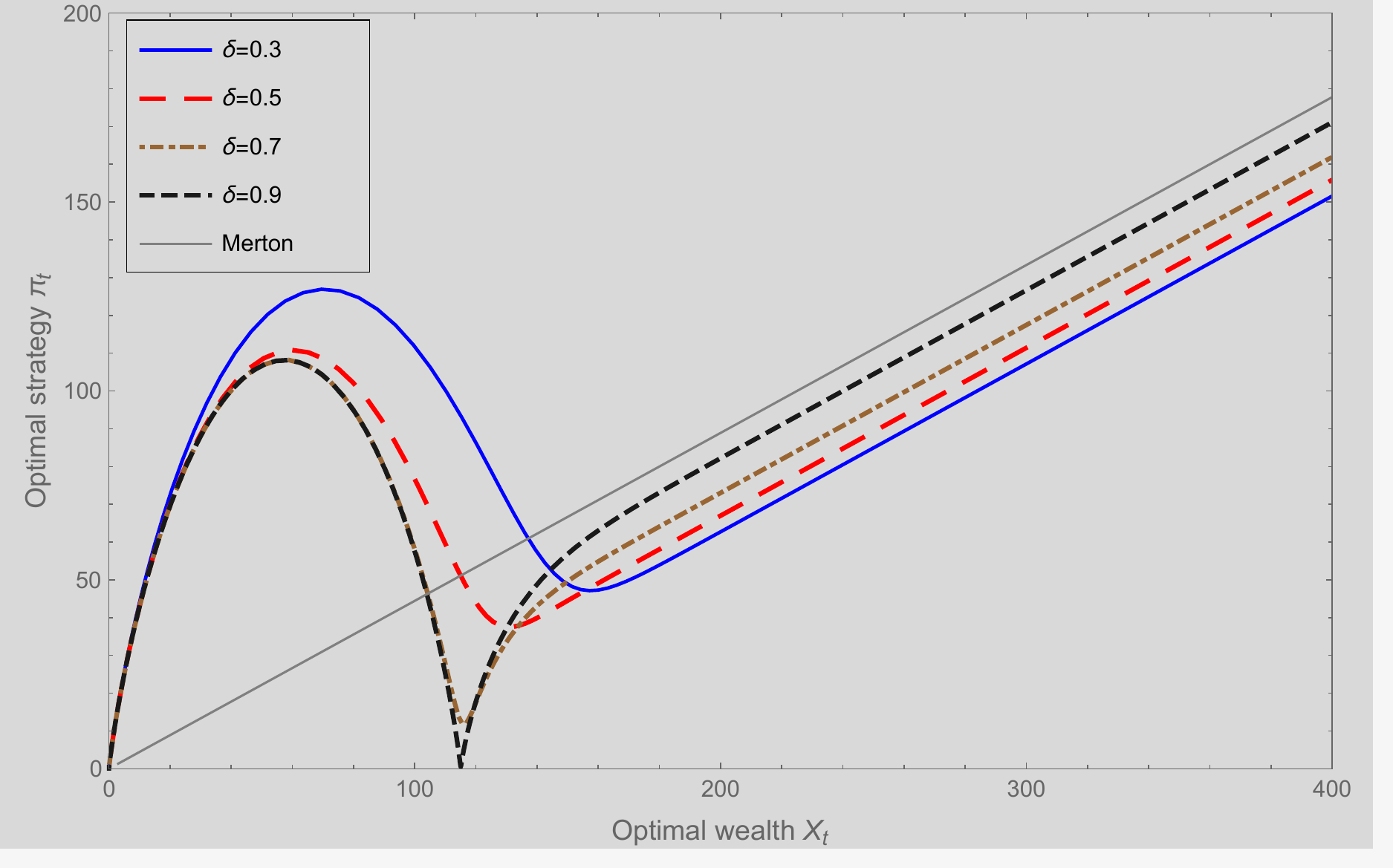}%
\caption{\small $\alpha=0.6$ }%
       \end{subfigure}
    \caption{\small Defaultable contract without mortality: effect of $\alpha$ and $\delta$ on the unconstrained optimal strategies at $t=8$.}
		\label{Fig:strategyuncon}
		 \end{figure}
		

Figure \ref{Fig:strategyuncon} depicts the effect of the participating ratio $\alpha$ and the bonus rate $\delta$ on the relationship between the optimal strategy and the wealth level at time $t=8$, two years before maturity. As revealed in the figure, the optimal amount invested in the risky asset is always non-negative and exhibits a peak-valley structure. The reason for the peak is that if the wealth becomes low the equity holder (insurer) is left with nothing or even with negative wealth and will try push the wealth back ``into the money'' by significantly investing into the risky asset. Once the wealth is above the guarantee level the investment behavior normalizes, i.e., becomes more like the Merton benchmark case. 

Consistent with what has been observed in Figure \ref{Fig:OptXT}, the investment riskiness for a fixed bonus rate increases if the policyholder contributes very much to the contract. The effect of the bonus rate $\delta$ on the optimal strategy is presented in the right panel of Figure \ref{Fig:strategyuncon}, showing that for a given participation rate the exposure to risk decreases for low terminal wealth and increases for a larger terminal wealth if the contract provides the policyholder a higher bonus rate $\delta$. Again this is due to two effects: the first effect is that with higher bonus rates the contract becomes less valuable for the equity holder, implicitly lowering her wealth and shifting the curve to the left. The second effect is that in order to mimic the Merton strategy for her personal account the equity holder needs to actually increase the riskiness of their position when higher bonus rates are paid to the policyholder. The graph shows that the first effect outweighs the second effect in bad economic scenarios while the second effect is stronger in good economic scenarios where the strategy overall is less sensitive to shifts in wealth.

\begin{center}
\begin{figure}[h]
      \begin{subfigure}[b]{0.32\textwidth}
		\includegraphics[width=\columnwidth,height=6.5cm]{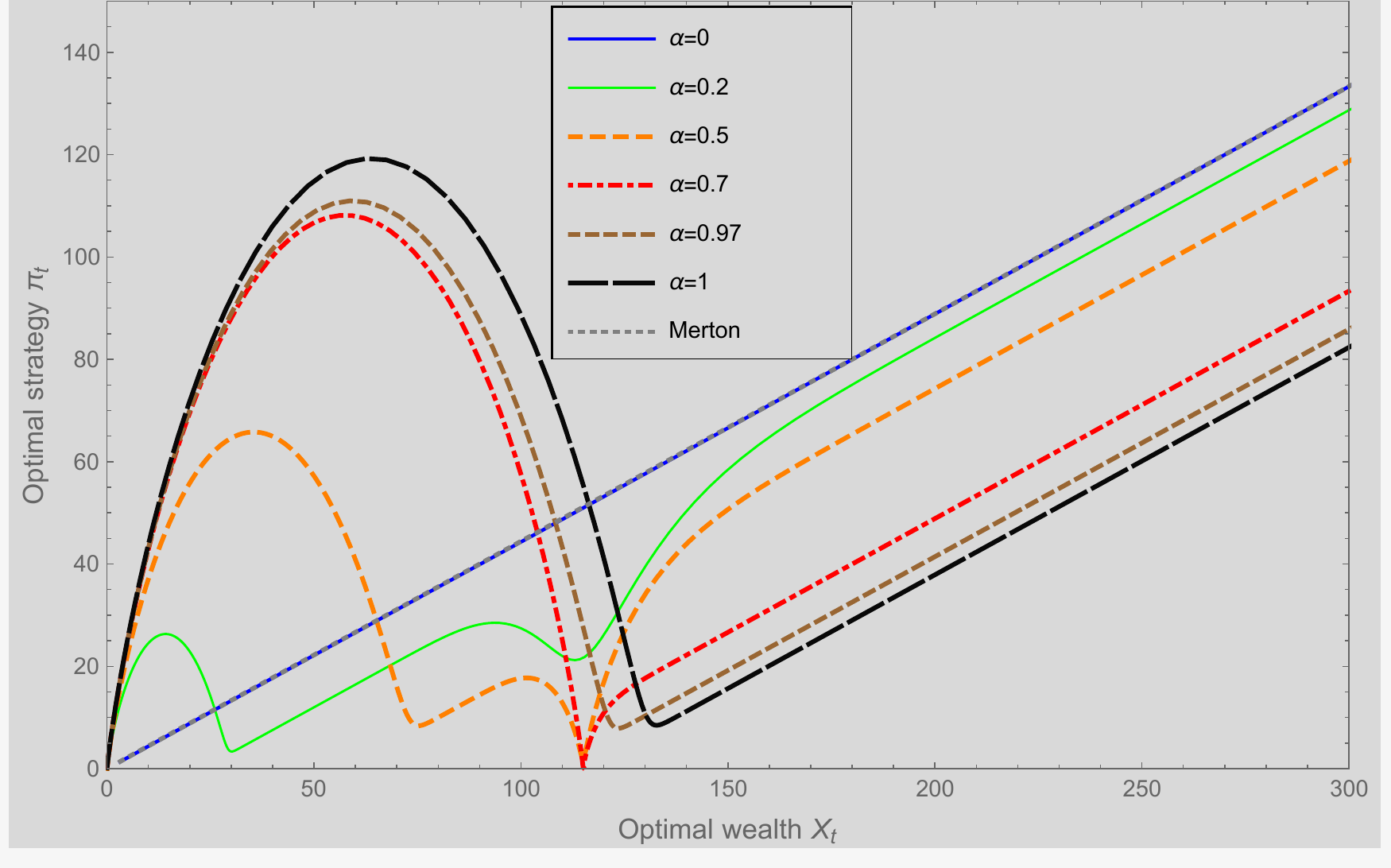}%
			\caption{\small $\delta=0.6$ and $\eta=1.01$}
    \end{subfigure}%
   \begin{subfigure}[b]{0.32\textwidth}
     	\includegraphics[width=\columnwidth,height=6.5cm]{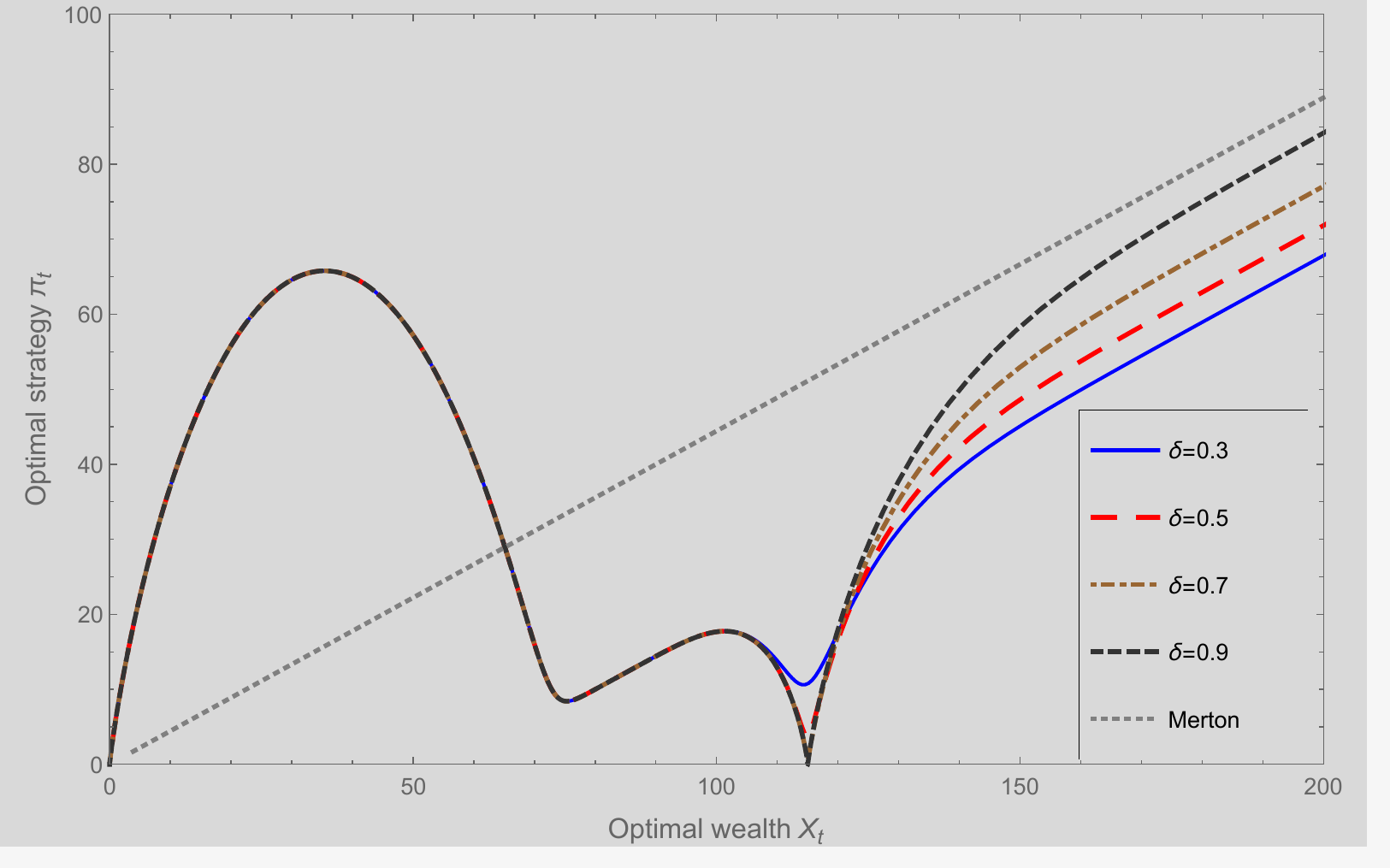}
		\caption{\small $\alpha=0.5$ and $\eta=1.01$ }
       \end{subfigure}%
			      \begin{subfigure}[b]{0.33\textwidth}
     	\includegraphics[width=\columnwidth,height=6.5cm]{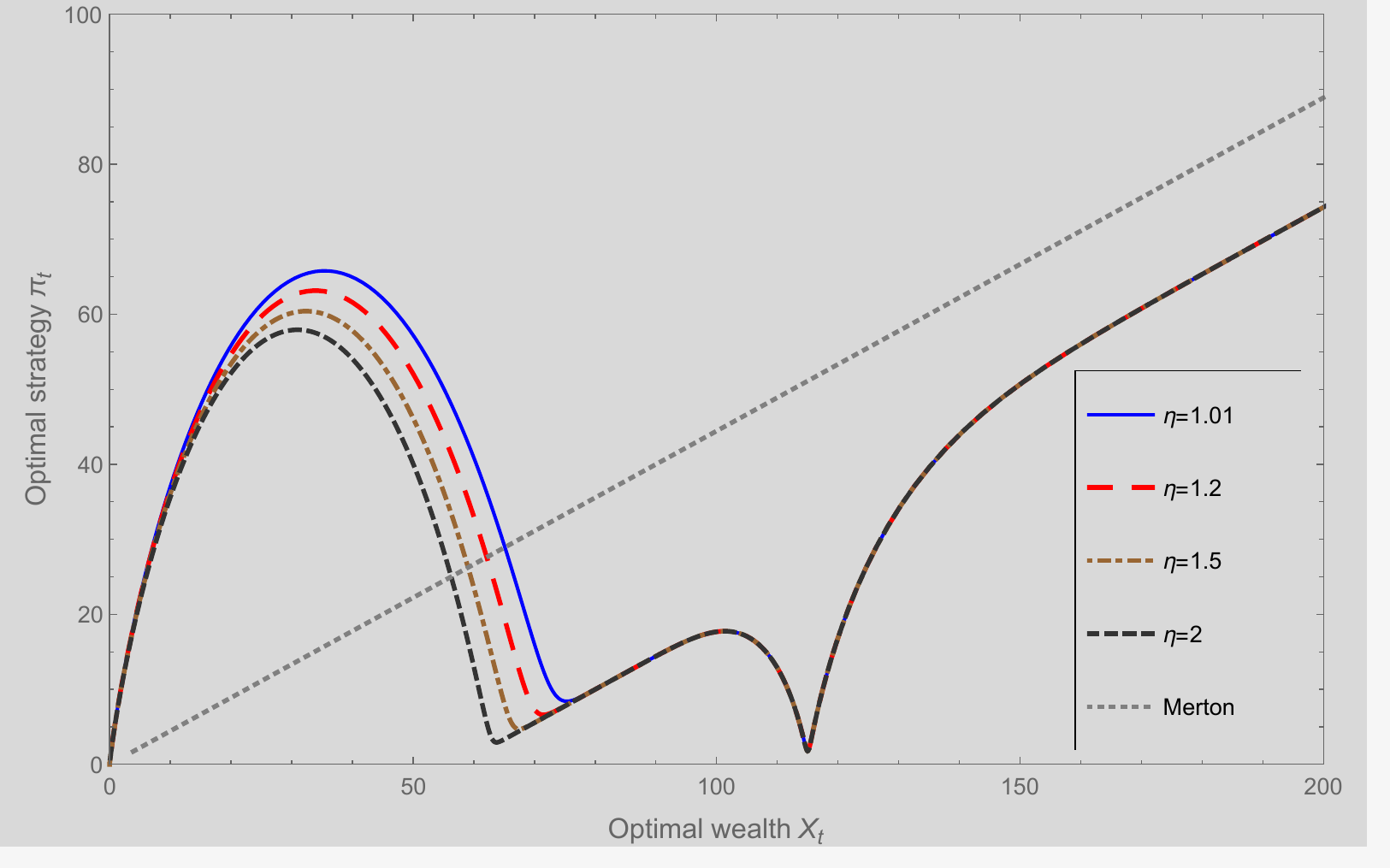}
		\caption{\small $\alpha=0.5$ and  $\delta=0.6$  }
       \end{subfigure}%
    \caption{\small Fully protected contract without mortality: effect of the participation rate $\alpha$, the bonus rate $\delta$ and the loss degree $\eta$ in the unconstrained strategy at $t=8$}
		\label{Fig:NP}
		 \end{figure}
		\end{center}
	
The effects of the participation rate $\alpha$, the bonus rate $\delta$ and the loss degree $\eta$ for a fully protected contract are illustrated in Figure \ref{Fig:NP}. It can be observed from the left panel that in line with what has been discussed in Figure \ref{Fig:strategyuncon} given a fixed bonus rate and a loss degree, changing $\alpha$ leads to monotone changes in the optimal strategy. In particular, high participation rates goes hand in hand with relatively risky investments of the insurance company in the loss states. Furthermore, the middle panel shows that given a participation rate, an increase in $\delta$ for fully protected contracts does not lead to a significant change in the exposure to risk in bad market scenarios. Hence, in these scenarios the investment of the insurer is rather robust with respect to the bonus rate and investments seem more motivated by the desire to avoid losses. Moreover, the higher $\delta$, the closer the strategy to the Merton strategy in case of good performance. 
The right panel also shows that increasing the loss aversion $\eta$ leads the insurer to more prudent investment strategies in case of bad performance. However, the risky investment is almost the same in case of good performance. To explain this, we remark that an increase in $\eta$ only leads to higher sensitivity with respect to losses while the gain part is not influenced. 
\begin{figure}[h]
      \begin{subfigure}[b]{0.49\textwidth}
    \includegraphics[width=\columnwidth,height=6.5cm]{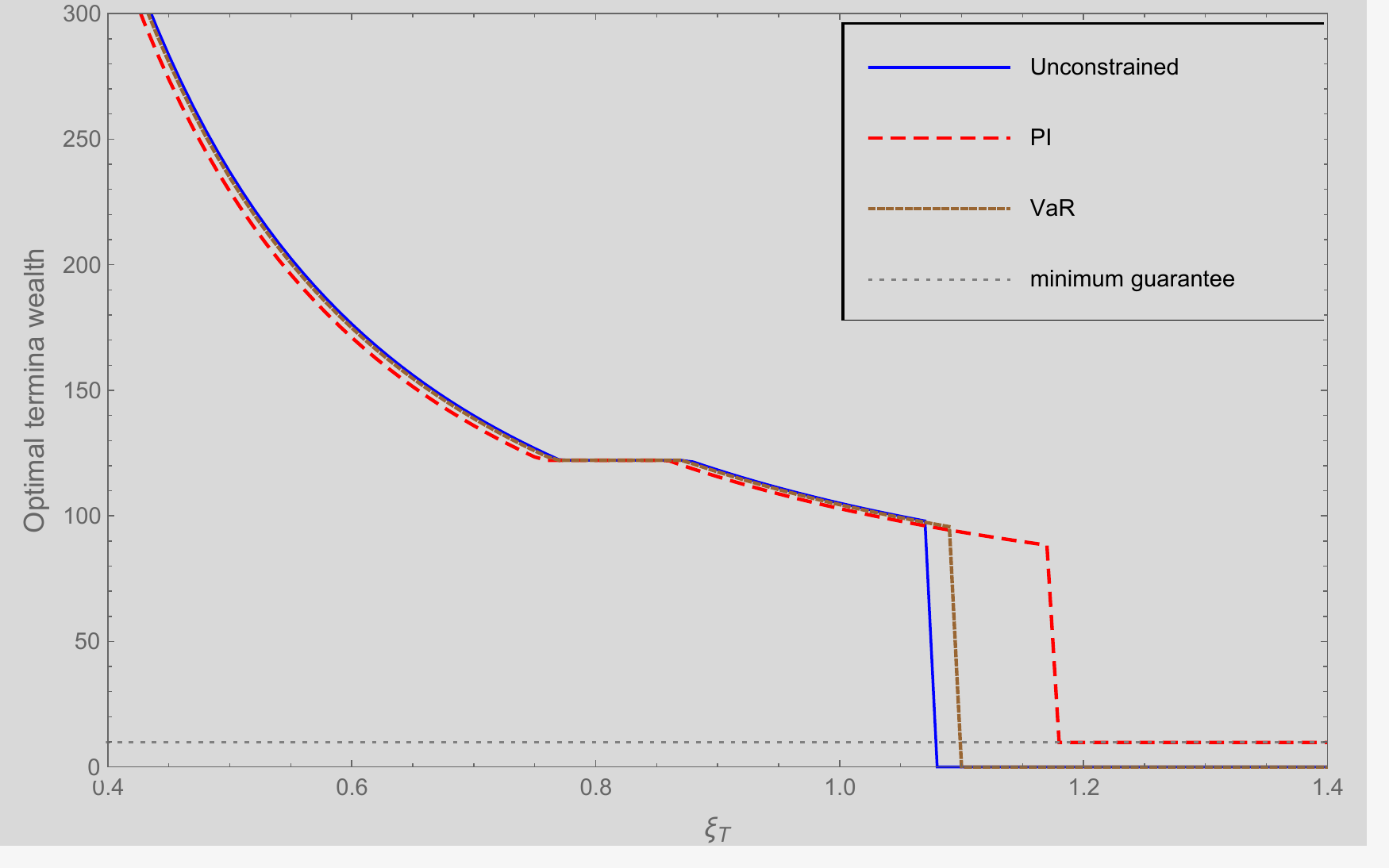}%
	\caption{\small Optimal terminal wealth. }
      \end{subfigure}
			      \begin{subfigure}[b]{0.49\textwidth}
    \includegraphics[width=\columnwidth,height=6.5cm]{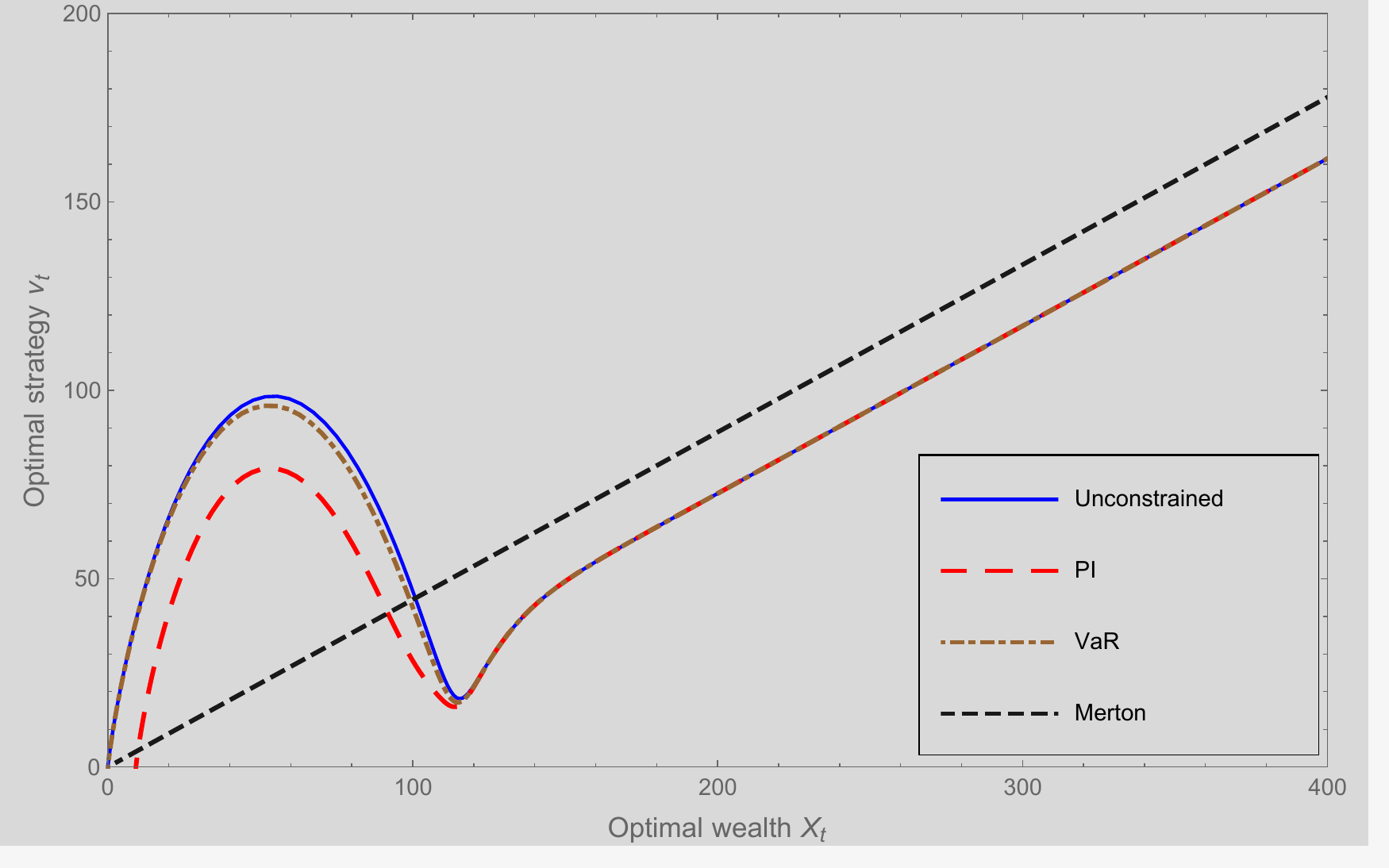}%
			\caption{\small Optimal strategies at $t=8$. }
    \end{subfigure}
			\caption{\small Comparing different optimal strategies and optimal terminal wealth for defaultable contract with $\delta=0.3$ and $\alpha=0.4$. }
			\label{Fig:compared1}
  		 \end{figure}

Let us now compare the unconstrained investment strategy with the constrained one under a VaR or a PI constraint. To serve this aim, the minimum guarantee in the PI problem and the default probability are chosen as $l=0.2L_T$ and $\beta=0.025$. The comparison of different strategies is presented in Figure \ref{Fig:compared1} from which we can observe that the exposure to risk when the wealth process is small (the market condition gets worse) will be reduced by a VaR constraint. In particular, the region where the terminal wealth is zero shrinks. Furthermore, we observe that the Solvency II VaR-type constraints lead to more prudent investment than investments which are not regulated (i.e., come from unconstrained optimization problems). This result is contrary to the situation where the insurer maximizes the utility of the total wealth of the company (without distinguishing between equity holders and policyholders), in which case a VaR constraint may induce the insurer to take excessive risk leading to higher losses than in the case of no regulation, see \cite{BasakShapiro01,CNS17,kraft2013,CNSmult,wei2018}.
These effects are more significant if a PI regulation is used. We also obtain similar effects on the optimal strategies for the case of full protected contracts.


\begin{figure}[h]
       \centering\includegraphics[width=0.8\columnwidth,height=7cm]{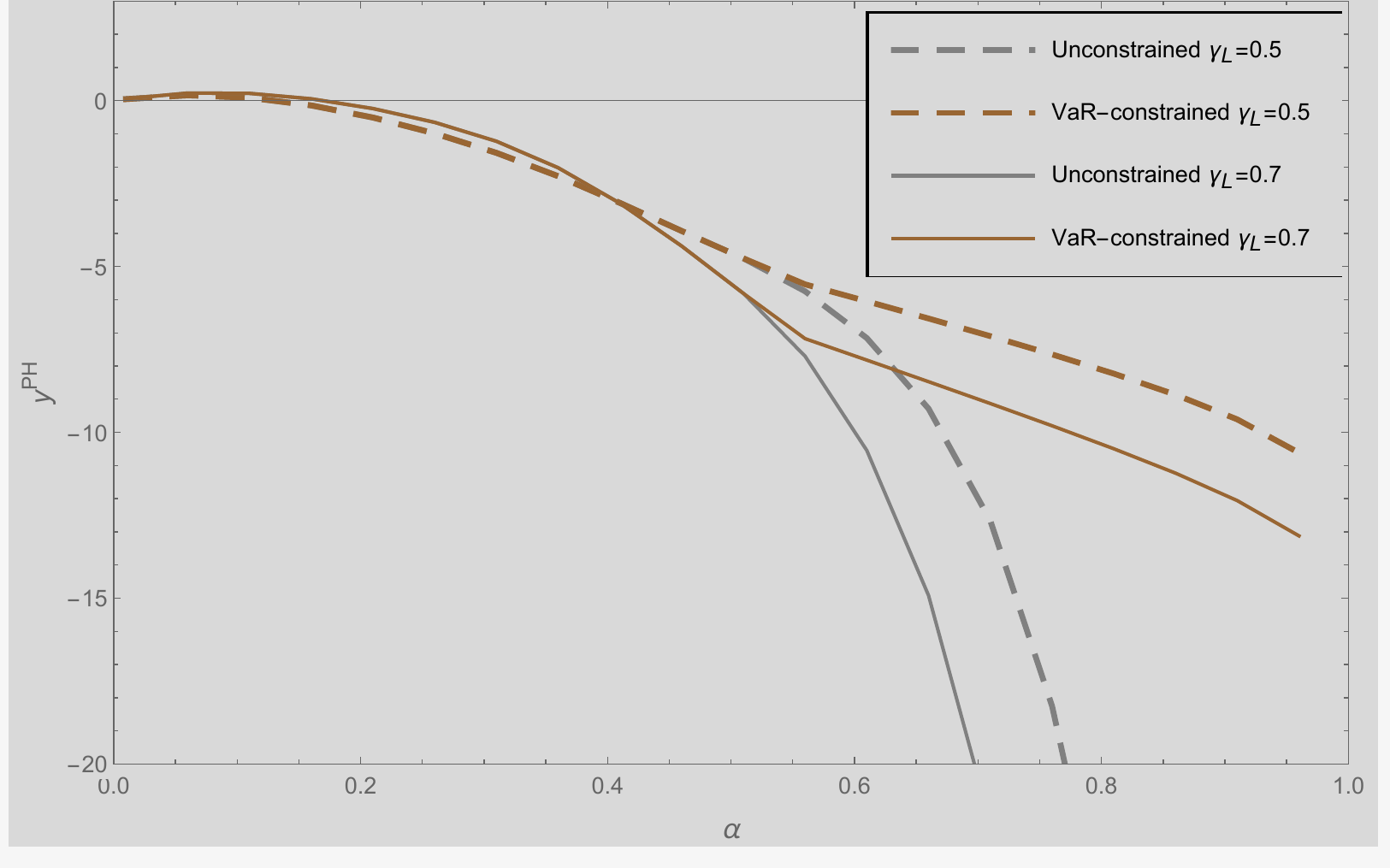}%
				\caption{\small Effect of the risk aversion and VaR constraint in $y^{PH}$ for defaultable contracts with $\delta=0.7$.  }
			\label{Fig:DeltaPH}
  		 \end{figure}

In Figure \ref{Fig:DeltaPH}, we report the effect of the participation rate in the policyholder's well-being $y^{PH}$ which is measured by the difference between the certainty equivalent of the unconstrained/VaR solution with the individual (unconstrained) Merton solution using $L_0=\alpha X_0$ as the initial capital. It can be observed that while the policyholder overall may suffer losses for high participation rates, she can even profit from participating in the contract for low participation rates. However, compared to the unregulated solution, losses are reduced by the use of a VaR constraint, which means that the policyholder will prefer a VaR-constrained portfolio to an unregulated portfolio. Moreover, the greater the risk aversion the more pronounced this effect, and different bonus rates overall do not change this conclusion. 

In Table \ref{Tab1} we look at the certainty equivalents of both sides of a defaultable contract under a VaR or a PI regulation. We assume that the single policyholder also  evaluates her terminal payoff by a power utility with the same risk aversion 
$\gamma_L=0.5$ as the equity holder. To see the effect of the (VaR or PI) regulation (compared to the unregulated problem), we compute $\Delta^{PH}$ (resp. $\Delta^{EH}$) as {\em the difference between the certainty equivalent} of the VaR-solution/PI-solution and the unconstrained solution for the policyholder (resp. the equity holder). Some interesting conclusions can be derived from Table \ref{Tab1}. Firstly, while both VaR and PI regulations imply a smaller expected utility for the insurer, it is in favor of the policyholder, i.e., she receives more expected utility from a VaR or a PI-constrained portfolio than from an unregulated portfolio. For both types of regulation, the gains in terms of expected utility are greater for higher participation rates, while being smaller for higher bonus rates. This observation is consistent with Figure \ref{Fig:DeltaPH}. Secondly, a stricter regulation (a smaller allowed default probability in the VaR problem or a higher level $l$ in the $PI$ problem) will enhance the benefit of the policyholder but deteriorate that of the insurer.

\begin{table}[h]%
\begin{center}
{\footnotesize
\begin{tabular}{|l|l|ll|ll|ll|ll|}
\hline
 \multicolumn{2}{|c|}{} & \multicolumn{2}{c|}{$\delta=0.3$} &\multicolumn{2}{c|}{$\delta=0.5$} &\multicolumn{2}{c|}{$\delta=0.7$}&\multicolumn{2}{c|}{$\delta=0.9$} \\
\hline
\multicolumn{2}{|r|}{$\alpha$} & $\Delta ^{PH}$ & $\Delta ^{EH}$& $\Delta ^{PH}$ & $\Delta^{EH}$& $\Delta ^{PH}$ & $\Delta^{EH}$& $\Delta ^{PH}$ & $\Delta ^{EH}$\\
\hline
  \multirow{5}{*}{\rotatebox[origin=c]{90}{VaR, $\beta=0.025$}}
 &0.1 & 0& 0 & 0 & 0 & 0 & 0 & 0& 0 \\
 &0.3 & 0 & 0 & 0 &0& 0 & 0 & 0 & 0 \\
 &0.5 & 5.3812 & -0.62107 & 1.59776 & -0.10904 & 0. & 0 & 0 & 0. \\
 &0.7 & 27.2518 & -5.54112 & 16.7027 & -2.37756 & 4.85723 & -0.41025 & 0 & 0 \\
 &0.9 & 55.9921 & -16.2817 & 47.4525 & -10.4539 & 35.6956 & -4.83685 & 8.81385 & -0.35244 \\
\hline
  \multirow{5}{*}{\rotatebox[origin=c]{90}{VaR, $\beta=0.05$}}
 &0.1 & 0 & 0 & 0. & 0. & 0 & 0 & 0& 0 \\
 &0.3 & 0 & 0 & 0& 0& 0 & 0 & 0 & 0 \\
 &0.5 & 2.93424 & -0.176109 & 0 & 0 & 0 & 0 & 0 & 0 \\
 &0.7 & 24.004 & -4.14662 & 14.0463 & -1.53169 & 2.78551 & -0.11097 & 0 & 0\\
 &0.9 & 51.7909 & -13.9376 & 43.9733 & -8.84345 & 32.9618 & -3.94988 & 6.86441 & -0.17468 \\
\hline
\multirow{5}{*}{\rotatebox[origin=c]{90}{PI, $l=0.2 L_T$}}
 &0.1 & \text{3.12}$*{10}^{-7}$ & -\text{3.75}$*{10}^{-8}$ & \text{2.20}$*{10}^{-7}$ & -\text{2.66}$*{10}^{-8}$ & \text{1.50}$*{10}^{-7}$& -\text{1.86}$*{10}^{-8}$ & \text{1.00}$*{10}^{-7}$ & -\text{1.28}$*{10}^{-8}$ \\
 &0.3 & 0.25411 & -0.02999 & 0.12641 & -0.01546 & 0.05253 & -0.00695 & 0.01686 & -0.00261 \\
 &0.5 & 5.78032 & -0.98316 & 2.99816 & -0.49857 & 0.98081 & -0.17662 & 0.10744 & -0.03192 \\
 &0.7 & 18.8767 & -3.35867 & 14.8676 & -2.04577 & 6.31463 & -0.76591 & 0.48069 & -0.07828 \\
 &0.9 & 30.1992 & -5.43378 & 29.6519 & -3.98221 & 27.0092 & -2.45201 & 10.5917 & -0.50173 \\
\hline
\multirow{5}{*}{\rotatebox[origin=c]{90}{PI, $l=0.5 L_T$}}
 &0.1 & \text{3.23}$*{10}^{-7}$ & -\text{3.89}$*{10}^{-8}$& \text{2.24}$*{10}^{-7}$& -\text{2.75}$*{10}^{-8}$  & \text{1.53}$*{10}^{-7}$& -\text{1.92}$*{10}^{-8}$ & \text{1.02}$*{10}^{-7}$ & -\text{1.32}$*{10}^{-8}$ \\
 &0.3 & 0.28132 & -0.03518 & 0.13706 & -0.01783 & 0.05567 & -0.00789 & 0.01737 & -0.00291\\
 &0.5 & 7.81038 & -1.54023 & 3.61826 & -0.71464 & 1.02386 & -0.22852 & 0.08654 & -0.03771 \\
 &0.7 & 28.8837 & -6.73458 & 19.1421 & -3.55756 & 6.63452 & -1.10254 & 0.38439 & -0.08946 \\
 &0.9 & 48.5809 & -13.1827 & 45.0528 & -9.4583 & 37.6635 & -5.35405 & 10.1804 & -0.67902 \\
\hline
\multirow{5}{*}{\rotatebox[origin=c]{90}{ PI, $l=0.9 L_T$}}
 &0.1 & \text{3.23}$*{10}^{-7}$ & -\text{3.89}$*{10}^{-8}$  & \text{2.24}$*{10}^{-7}$ & -\text{2.75}$*{10}^{-8}$ & \text{1.53}$*{10}^{-7}$  & -\text{1.92}$*{10}^{-8}$  & \text{1.02}$*{10}^{-7}$ & -\text{1.32}$*{10}^{-8}$  \\
 &0.3 & 0.28147 & -0.03522 & 0.1371 & -0.01785 & 0.05567 & -0.00790 & 0.01737 & -0.00292 \\
 &0.5 & 7.90473 & -1.56879 & 3.62667 & -0.72132 & 1.02165 & -0.22938 & 0.08613 & -0.03776 \\
 &0.7 & 30.5322 & -7.50628 & 19.1683 & -3.72849 & 6.56806 & -1.11328 & 0.38303 & -0.08952 \\
 &0.9 & 60.1018 & -19.1469 & 50.6672 & -12.4667 & 37.9901 & -6.00355 & 10.1406 & -0.68012 \\

\hline
\end{tabular}
}
\end{center}
\caption{\small Gain/Loss in terms of certainty equivalent for defaultable contracts under regulations.}
\label{Tab1}
\end{table}

\section{A two-period optimization framework}\label{sec:Twoperiod}
In this section we assume that the insurer is able to update the intertemporal information at an intermediate time point $T_1$, with $0<T_1<T$. More precisely, we assume that $n$ policyholders participate in the pool at time zero and the death event of policyholder $i$ in the first period $(0,T_1]$ is modelled by $\{d_i^{(I)}=1\}$, where $d_i^{(I)}$ is a binomial random variable. Let $N_1:=\sum_{i=1}^n{\bf 1}_\zs{d_i^{(I)}=1}$ be the number of deaths in the first time interval $(0,T_1]$. At time $T_1$, dividends are paid out to the survivors and death benefits are paid out to the designated beneficiaries of the persons deceased. We suppose that $d_1^{(I)},\cdots, d_n^{(I)}$ are mutually independent and independent of the assets $S$. 

Given that policyholder $i$ has survived in the first period, i.e., $d_i^{(I)}=0$, we model by $\{d_i^{(II)}=1\}$ the death event of policyholder $i$ in the second period $(T_1,T]$. To be consistent with the results in the previous sections, we assume that conditional on $d_i^{(I)}=0$, $d_i^{(II)}$ is a binomial random variable, and $d_i^{(II)}=0$ a.s. if $d_i^{(I)}=1$. Again, we suppose that $d_1^{(II)},\cdots, d_n^{(II)}$ are mutually independent and independent of the financial assets. 

For such a two-period framework, we now have to work on an enlarged probability space $(\Omega,\cG, \P)$, where $\bbg=(\cG_t)_\zs{t\in[0,T]}$ is the enlargement of the financial filtration $\bbf$ that contains the information generated by the death variables. $\bbg$ represents the structure of the global information available to the insurer over the time horizon $[0,T]$. Specifically, let $\cD^{(m)}:=\sigma(d_i^{(m)},i=1,\cdots,n)$, $m\in \{I,II\}$ be the sigma algebra generated by the death variables in the first and second period. From the model setting, we can choose the enlargement as follows 
$$
\cG_t=
\begin{cases}
\cF_t,&\quad \mbox{if}\quad 0\le t<T_1,\\
\cF_{t}\vee\cD^{(I)}&\quad \mbox{if}\quad T_1\le t<T,\\
\cF_{T}\vee\cD^{(I)} \vee \cD^{(II)}& \quad\mbox{if} \quad t=T.
\end{cases}
$$
Our admissible strategy is defined as 
$\nu_t=\nu_t^{(I)}{\bf 1}_\zs{t\in[0,T_1)}+\nu_t^{(II)}{\bf 1}_\zs{t\in[T_1,T]},$
 where $\nu_t^{(I)}$ is an $(\cF_t)_\zs{t\in[0,T_1)}$-adapted process and $\nu_t^{(II)}$ is an $(\cF_{t}\vee\cD^{(I)})_\zs{t\in[T_1,T]}$-adapted process. Note that the investment strategy in the last period $[T_1,T]$ depends on the death realizations at time $T_1$. Due to the death benefit and dividend payments, the portfolio process $X_t^{\nu,X_0}$ admits a downside jump at $T_1$ whose size is equal to the total death payment in the first period $(0,T_1]$. The policy of death and dividend payments will be specified below. We note that such a policy needs to guarantee the admissibility in the interval $(T_1,T]$ under a VaR regulation at maturity $T$.

Let $N_2=\sum_{i=1}^{n-N_1}{\bf 1}_\zs{d_i^{(II)}=1}$ be the number of deaths in $(T_1,T]$. The number of policyholders alive at maturity is given by $n-N_1-{N}_2$. We assume that conditional on $\cG_{T_1}=\cF_{T_1}\vee \cD^{(1)}$, the death variables in $(T_1,T]$, $(d_i^{(II)})_\zs{i=1,\cdots,n}$, are independent. Put $\wh{\kappa}_2:=\sum_{i=1}^{n-N_1} k_i {\bf 1}_\zs{d_i^{(II)}=1}$, which economically represents the total proportion of the guaranteed payment $\min (L_T,X_T)$ that the insurer has to pay to those who have died between $T_1$ and $T$. General death benefit and dividend payments will be specified below. 

We assume that the insurer has to satisfy a VaR regulation imposed at $T$. To simplify the presentation we consider the case of a defaultable contract in what follows. The case of fully protected contracts can be treated in a similar way. The insurer has to solve the following VaR optimization problem: 
\begin{align} \label{eq:2VaR.1}
 \sup_{ (\nu_t)_\zs{t\in[0,T]}} \E\left[{U}^{\cS}( V_E^{p,d}(X_T^{\nu,X_0}))\right], 
 \quad\text{s.t.} \quad \P(X_T^{\nu,X_0}<L_T\vert \cG_\zs{T_1})<\beta,
\end{align}
for some probability default level $0\le \beta<1$. We remark that for any admissible portfolio $X_T^{\nu,X_0}$ of Problem \eqref{eq:2VaR.1} the unconditional VaR constraint $\P(X_T^{\nu,X_0}<L_T)<\beta$ always holds.

For simplicity below we drop the dependency of the portfolio on the strategy and initial wealth. Due to the information updated at the intertemporal time $T_1$, Problem \eqref{eq:2VaR.1} can be solved recursively. More precisely, given the death information and the wealth level $X_{T_1}$ at time $T_1$ (after death benefit and dividend payments are made), from time $T_1$ on, an optimal strategy has to attain the supremum in 
\begin{align} \label{eq:2VaR.11}
 \esssup_\zs{ (\nu_t)_\zs{t\in[0,T]}}\E\left[{U}^{\cS}( V_E^{p,d}(X_T))\vert X_{T_1}\right],
 \quad\text{s.t.} \quad \P(X_T<L_T\vert X_{T_1})<\beta,
\end{align}
However, Problem \eqref{eq:2VaR.11} can be transformed into an equivalent optimization in a fully complete market as in Section \ref{sec:mod}. To this end, let $\Xi\subset [0,1]$ be the finite set of all possible outcomes of $\wh{\kappa}_2$. Using the independence of $\wh{\kappa}_2$ and $X_T$ we define
	\begin{align}
	U_{\wh{\kappa}_2}(x):=\E_{N_1}[U(\wt{f}(x,\wh{\kappa}_2))]=\sum_{\nu\in\Xi} {U}\big(\wt{f}(x,\nu)\big)\P(\wh{\kappa}_2=\nu\vert N_1),\nonumber
		\label{eq:Uepsg}
	\end{align}
where $	\E_{N_1}$ stands for the conditional expectation operator with respect to the distribution of $\wh{\kappa}_2$ given $N_1$. As shown in Lemma \ref{Le:Ukappa} the function $U_{\wh{\kappa}_2}$ is strictly increasing and concave on $[\wt{L}_T,\infty)$ with derivative	given by a similar form as in \eqref{eq:UepsgUprime}.

 Problem \eqref{eq:2VaR.1} on the interval $[T_1,T]$ is equivalent to the following optimization in a fully complete market: 
\begin{equation}
 \esssup_{ X_T\in\cX}\E\left[\wt{U}^{\cS,p,\wh{\kappa}_2}(X_T)\vert X_{T_1}\right], 
 \quad\text{s.t.} \quad \P(X_T<L_T\vert X_{T_1})<\beta, \label{eq:2VaR.110}
\end{equation}
and the budget constraint $
\E[\xi_T\xi^{-1}_\zs{T_1} X_T \vert X_{T_1}]\le X_{T_1}. 
$
Note that the VaR constraint will only be fulfilled if the wealth value after death payment at time $T_1$ is greater or equal to the time-$T_1$ value of the (stochastic) level $L_T {\bf 1}_\zs{\xi_T\le \bar{\xi}}$, where $\bar{\xi}$ is implicitly determined by $\P(\xi_\zs{T}>\bar{\xi}\vert \cG_{T_1})=\beta$, that is,
\begin{equation}
X_{T_1}\ge
\E[L_T \xi_T \xi_\zs{T_1}^{-1} {\bf 1}_\zs{\xi_T\le \bar{\xi}}\vert \cG_\zs{T_1}]:=l_{T_1}.
\label{eq:wtx0}
\end{equation}
The optimal terminal portfolio is given by Theorem \ref{Th:const}. 
For simplicity, we assume that the first case holds, i.e., $\Upsilon^{1,q_j}(\wt{L}_T)>0$ and the optimal terminal wealth is given by
	\begin{equation}
X^{VaR,p,\wh{\kappa}_2,*}_T=\cWD_\zs{[\xi_\zs{\wt{L}}^{\wh{\kappa}_2},\xi_\zs{\wh{L}},\bar{\xi}]} 
^{[I_{\wh{\kappa}_2},\wt{L},I+L,0]} {\bf 1}_\zs{ \bar{\xi}\ge \wh{\xi}^{1,q_j}}
+\cWD_\zs{[\xi_\zs{\wt{L}}^{\wh{\kappa}_2},\xi_\zs{\wh{L}},\wh{\xi}^{1,q_p}]}^{[I_{\wh{\kappa}_2},\wt{L},I+L,0]}{\bf 1}_\zs{ \bar{\xi}<\wh{\xi}^{1,q_p}},
		\end{equation}
where the second term indicates the case where the VaR constraint is not active. The other cases can be treated similarly. Note that the reference points $\xi_\zs{\wt{L}}^{\wh{\kappa}_2},\xi_\zs{\wh{L}},\wh{\xi}^{1,q_j}$ are defined by \eqref{eq:xi11} where the multiplier $\lambda$ is characterized by the budget equation at time $T_1$: 
	\begin{equation}
	\E[\xi_T \xi_\zs{T_1}^{-1} X^{VaR,p,\wh{\kappa}_2,*}_T\vert X^{VaR,p,\wh{\kappa}_2,*}_{T_1}= X_{T_1}]=X_{T_1}.
	\label{eq:}
	\end{equation}
Hence, $\lambda$ should be considered as a function of $X_{T_1}$, the wealth level after death benefit payment at time $T_1$.

\vspace{2mm}
\noindent{\bf Indirect value function:} Assume $x\ge l_{T_1}$. Consider the indirect value function of the VaR constrained problem
which will serve as the objective function in the next backward period.
\begin{align} \label{eq:2VaR.12}
 V_\zs{(N_1)}(x):=\E\left[\wt{U}^{\cS,p,\wh{\kappa}_2}(X^{VaR,j,\wh{\kappa}_2,*}_T)\vert \cG_{T_1},X_{T_1}=x\right].
\end{align}

For simplicity in the rest of this section we assume a simple Black-Schole's model in which the wealth process starting with an initial wealth $X_0>0$ related to the strategy $\nu_t=\nu_t^{(I)}{\bf 1}_\zs{t\in[0,T_1)}+\nu_t^{(II)}{\bf 1}_\zs{t\in[T_1,T]}$(where $\nu_t^{(I)}$ is an $(\cF_t)_\zs{t\in[0,T_1)}$-adapted process and $\nu_t^{(II)}$ is an $(\cF_{t}\vee\cD^{(I)})_\zs{t\in[T_1,T]}$-adapted process) can be expressed as $X_t^{\nu}=X_t^{\nu^{(I)}}{\bf 1}_\zs{t\in[0,T_1)}+X_t^{\nu^{(II)}}{\bf 1}_\zs{t\in[T_1,T]}$, where
\begin{align}
\d X_t^{\nu^{(I)}} =  (r_t X_t^{\nu^{(I)}}+ \nu_t^{(I)}(\mu_t-r_t))  \d t + \nu_t^{(I)}\sigma_t \d W_t,\quad t \in[0,T_1), 
\end{align}
and $X_{T_1}^{\nu^{(II)}}=X_{T_1}^{\nu^{(I)}}$ minus the death benefit and dividend payments at time $T_1$, and 
\begin{align}
\d X_t^{\nu^{(II)}} =  (r_t X_t^{\nu^{(II)}}+ \nu_t^{(II)}(\mu_t-r_t))  \d t + \nu_t^{(II)}\sigma_t \d W_t,\quad t \in[T_1,T], 
\end{align}
where $r,\mu,\sigma$ are positive and deterministic processes. We denote by $\theta_t=(\mu_t-r_t)/\sigma_t$ the market price of risk. Recall that the investment strategy in the last period $[T_1,T]$ depends on the death realizations at time $T_1$. Due to death benefit and dividend payments, the portfolio process $X_t^{\nu,X_0}$ admits a downside jump at $T_1$ whose size is equal to the total death and dividend payment in the first period $(0,T_1]$, i.e., the starting capital for the last period is defined by $X_{T_1}^{\nu^{(II)}}=X_{T_1^-}^{\nu^{(I)}}$ subtracted by the total death and dividend payment. The policy of death payment will be specified below. We note that such a policy needs to guarantee the admissibility in the interval $(T_1,T]$ under a VaR regulation at maturity $T$.

We follow the technique in a multiple-VaR framework in \cite{CNSmult}. Note that our current framework \eqref{eq:2VaR.1} is more general due to the non-concavity of $\wt{U}^{\cS,p,\wh{\kappa}_2}$ and the presence of mortality. To make the presentation self-contained we will present the argument in detail. To this end, let us introduce some notation.

\vspace{2mm}

\noindent{\bf Notation:}\label{Nota}
Below, $\Vert f\Vert_\zs{s,t}$ stands for the Lebesgue norm of a deterministic function $f$ on $[s,t]$, i.e.
$\Vert f\Vert_\zs{s,t}:=\sqrt{\int_s^t f^2_u \d u}.
$
Note that the subscript $\{s,t\}$ is used to emphasize the considered interval time $[s,t]$ for  $s<t$. This notation will frequently appear in the sequel. 
For convenience, we will make use of the following functions
\begin{equation}
R_\zs{s,t}=\int_s^t r_u\d u \quad \mbox{and} \quad\xi_\zs{s,t}(z): =\exp\{-R_\zs{s,t}-\frac{1}{2}\Vert \theta\Vert_\zs{s,t}^2-z\Vert \theta\Vert_\zs{s,t}\}.
\label{eq:Indirect0}
\end{equation}
Hence, $R_\zs{s,t}$ is the accumulated interest rate in the interval $[s,t]$. The following simple property is frequently used in the analysis below. 
\begin{remark}\label{Re:0}
 For any $0\le s\le t\le T$ and with the notation above, we can write the state price density $\xi_t$ as
\begin{equation}
\xi_t=\xi_\zs{s}\xi_\zs{s,t}(Z),
\label{eq:Z}
\end{equation}
where $Z=\Vert \theta\Vert_\zs{s,t}^{-1}\int_s^t \theta_s \d W_s$ is a standard normal variable independent of ${\cal F}_\zs{s}$.
\end{remark}
Now, it is straightforward to see that
\begin{equation}
\bar{\xi}=\xi_\zs{T_\zs{1}}\exp\left\{-R_\zs{T_1,T}-\frac{1}{2}\Vert \theta\Vert_\zs{T_1,T}^2-q_\zs{\beta}\Vert \theta\Vert_\zs{T_1,T}\right\}=\xi_\zs{T_\zs{1}}\xi_\zs{t_1,T}(q_\zs{\beta}),
\label{eq:xibar.0}
\end{equation}
where $q_\beta$ is the lower $\beta$-quantile of the standard normal law.
Let
\begin{equation}
d_\zs{s,t}(\eta,\xi)=-\frac{\ln(\eta/\xi)+R_\zs{s,t}+\frac{1}{2}\Vert \theta\Vert_\zs{s,t}^2}{\Vert \theta\Vert_\zs{s,t}}\,\quad \mbox{and} \quad \wh{d}_\zs{s,t}(\eta,\xi):=d_\zs{s,t}(\eta,\xi)+\Vert \theta\Vert_\zs{s,t}.
\label{eq:Indirect1}
\end{equation}
The following result is useful for our analysis.
\begin{lemma}\label{Le:bas}
For any $a>0$ and $0\le s\le t\le T$, we have
\begin{align}
\E [{\bf 1}_\zs{\xi_t\le a}\vert \cF_s]=1-\Phi(d_\zs{s,t}(a,\xi_s)),\quad
\E [\xi_t \xi_s^{-1}{\bf 1}_\zs{\xi_t\le a}\vert \cF_s]=e^{-R_\zs{s,t}}(1-\Phi(\wh{d}_\zs{s,t}(a,\xi_s))).
\label{eq:Q0}
\end{align}
\end{lemma}
\proof As in Remark \ref{Re:0}, observe that
$\{\xi_t\le a\}=\{Z \ge d_\zs{s,t}(a,\xi_s)\}$ and the conclusion follows from a straightforward calculation. \endproof

\vspace{3mm}
For later use we define
\begin{align}
\wh{Q}_\zs{s,t}(a,b,\xi_s):=e^{R_\zs{s,t}}\E [\xi_t \xi_s^{-1}{\bf 1}_\zs{a\le \xi_t< b}\vert \cF_s]=\Phi(\wh{d}_\zs{s,t}(a,\xi_s))-\Phi(\wh{d}_\zs{s,t}(b,\xi_s)),
\label{eq:Q}
\end{align}
where we have applied Lemma \ref{Le:bas}. Economically, $\wh{Q}_\zs{s,t}$ gives us the market value at time $s$ of \$1 paid at time $t$ in case of an intermediate market scenario where $a\le \xi_t< b$. 

The optimal wealth process is given by the following lemma:
\begin{lemma}\label{Le:n2.0}
Given a realized wealth level at time $T_1$, the optimal wealth process on $[T_1,T)$ is given by ${X}_t^{VaR}=\X_\zs{t,T}^{VaR}(\lambda,\xi_{t})$, where
\begin{align}
\X_\zs{t,T}^{VaR}(\lambda,\xi_{t})&:=\int_\zs{z<\xi_\zs{\wt{L}}^{\wh{\kappa}_2}/\xi_{t}} zI_{\wh{\kappa}_2}(\lambda\xi_t\,z) \d F_{\xi_\zs{t,T}}(z)
+ \wt{L}_T e^{-R_\zs{t,T}}  \wh{Q}_\zs{t,T}(\xi_\zs{\wt{L}}^{\wh{\kappa}_2},\xi_\zs{\wh{L}},\xi_t)\notag\\
&\quad+
\int_\zs{\xi_\zs{\wh{L}}/\xi_{t}\le z<\bar{\xi}/\xi_{t}}z I(\lambda\xi_t\,z) \d F_{\xi_\zs{t,T}}(z)
+ L_T e^{-R_\zs{t,T}} \wh{Q}_\zs{t,T}(\xi_\zs{\wh{L}},\bar{\xi},\xi_t),
\label{eq:Indirect3}
\end{align}
where $F_{\xi_\zs{t,T}}$ is CDF of the log-normal variable $\xi_\zs{t,T}(Z)$ and $\lambda$ satisfies the budget constraint at time $T_1$.
\end{lemma}

Note that the dependency on $X_{T_1}$ results from the reference points $\xi_\zs{\wt{L}}^{\wh{\kappa}_2},\xi_\zs{\wh{L}},\wh{\xi}^{1,q_p}$ which are defined by \eqref{eq:xi11} where the multiplier $\lambda$ is characterized by the budget equation at time $T_1$. Importantly, taking \eqref{eq:xi11} into account we can represent the wealth process as $\X_\zs{t,T}^{VaR}(\lambda\xi_{t})$, a functional of the product $\lambda \xi_{t}$.

Recall that the VaR constraint at time $T$ may not be active if the wealth level at time $T_1$ is large enough. In this case the optimal terminal wealth is given by the unconstrained solution whose optimal portfolio is given by 
\begin{align}
\X_\zs{t,T}^{unco}(\lambda,\xi_{t})&:=\int_\zs{z<\xi_\zs{\wt{L}}^{\wh{\kappa}_2}/\xi_{t}} zI_{\wh{\kappa}_2}(\lambda\xi_t\,z) \d F_{\xi_\zs{t,T}}(z)
+ \wt{L}_T e^{-R_\zs{t,T}}  \wh{Q}_\zs{t,T}(\xi_\zs{\wt{L}}^{\wh{\kappa}_2},\xi_\zs{\wh{L}},\xi_t)\notag\\
&\quad+
\int_\zs{\xi_\zs{\wh{L}}/\xi_{t}\le z<\wh{\xi}^{1,q_p}/\xi_{t}}z I(\lambda\xi_t\,z) \d F_{\xi_\zs{t,T}}(z)
+ L_T e^{-R_\zs{t,T}} \wh{Q}_\zs{t,T}(\xi_\zs{\wh{L}},\wh{\xi}^{1,q_p},\xi_t),
\label{eq:Indirect31}
\end{align}
which results from replacing $\bar{\xi}$ with $\wh{\xi}^{1,q_p}$ in \eqref{eq:Indirect3}. This critical wealth level can be explicitly determined from setting $\bar{\xi}=\wh{\xi}^{1,q_p}$, 
or equivalently,
$$
\lambda \xi_{T_1} =U'_{\wh{\kappa}_2}(\wh{\xi}^{1,q_p})/\xi_{T_1,T}(q_{\beta}):=\lambda^{bind},$$ which is independent of $\xi_{T_1}$. As before, we can consider the unbinding wealth process as $\X_\zs{t,T}^{unco}(\lambda\xi_{t})$, a functional of the product $\lambda \xi_{t}$. 
Using the monotonicity of the inverse marginal utility function, one deduces that the terminal VaR constraint is not binding if and only if $\lambda \xi_{T_1}\le \lambda^{bind}$ or equivalently, the wealth level at $T_1$ is greater or equal to
\begin{equation}
x^{bind}_{T_1}:=\X_\zs{T_1,T}^{unco}(\lambda^{bind}).
\label{eq:Xbind}
\end{equation}
The value function at time $T_1$ for the case of a binding VaR constraint can be computed similarly as
\begin{align}
 V^{bind}_\zs{(N_1)}(x)&=\E\left[\wt{U}^{\cS,p,\wh{\kappa}_2}(X^{VaR,p,\wh{\kappa}_2,*}_T)\vert \cF_{T_1}\vee N_1,X_{T_1}=x\right]\notag\\
&=\int_\zs{z<\xi_\zs{\wt{L}}^{\wh{\kappa}_2}/\xi_{T_1}} U_{\wh{\kappa}_2}(I_{\wh{\kappa}_2}(\lambda\xi_{T_1}\,z)) \d F_{\xi_\zs{{T_1},T}}(z)
+ U(\wh{L}_T) {Q}_\zs{{T_1},T}(\xi_\zs{\wt{L}}^{\wh{\kappa}_2},\xi_\zs{\wh{L}},\xi_{T_1})\notag\\
&\quad+
\int_\zs{\xi_\zs{\wh{L}}/\xi_{T_1}\le z<\bar{\xi}/\xi_{T_1}} U( I(\lambda \xi_{T_1}\,z)) \d F_{\xi_\zs{{T_1},T}}(z).
\label{eq:IndirectVbind}
\end{align}
Analogously, the indirect value function for the unbinding terminal VaR constraint can be computed by taking $\bar{\xi}=\wh{\xi}^{1,q_p}$ in \eqref{eq:IndirectVbind} to obtain
\begin{align} 
 V^{unco}_\zs{(N_1)}(x)&=\int_\zs{z<\xi_\zs{\wt{L}}^{\wh{\kappa}_2}/\xi_{T_1}} U_{\wh{\kappa}_2}(I_{\wh{\kappa}_2}(\lambda^{unco}\xi_{T_1}\,z)) \d F_{\xi_\zs{{T_1},T}}(z)
+ U(\wh{L}_T) {Q}_\zs{{T_1},T}(\xi_\zs{\wt{L}}^{\wh{\kappa}_2},\xi_\zs{\wh{L}},\xi_{T_1})\notag\\
&\quad+
\int_\zs{\xi_\zs{\wh{L}}/\xi_{T_1}\le z<\wh{\xi}^{1,q_p}/\xi_{T_1}} U( I(\lambda^{unco} \xi_{T_1}\,z)) \d F_{\xi_\zs{{T_1},T}}(z).
\label{eq:Indirect301}
\end{align}
Therefore, the indirect value function at time $T_1$ is given by
\begin{equation}
{V}_\zs{(N_1)}(x)={V}^{bind}_\zs{(N_1)}(x){\bf 1}_\zs{x< x^{bind}_{T_1}}+{V}^{unco}_\zs{(N_1)}(x){\bf 1}_\zs{x\ge x^{bind}_{T_1}},\quad x\in (l_{T_1},\infty),
\label{eq:VN1}
\end{equation}
where $x^{bind}_{T_1}$ is the natural switching point at which the terminal wealth ``jumps'' from the binding VaR case to the unbinding VaR case. It can be checked directly that ${V}_\zs{(N_1)}$ is strictly increasing and $x^{bind}_{T_1}$ is a continuous point of ${V}_\zs{(N_1)}$. Moreover, we observe that both value functions $V_\zs{(N_1)}$ and ${V}^{unco}_\zs{(N_1)}$, can be seen as functions of the product $\lambda \xi_\zs{T_1}$, and $\lambda^{bind}$ is a ``smooth switching point'' at which ${V}_\zs{(N_1)}$ is continuously differentiable. Due to the budget constraint at time $T_1$, the multiplier $\lambda$ can be viewed as a function of the wealth level at time $T_1$, $\lambda=\lambda(x)$. 
\begin{lemma}\label{Le:Indirect1}
The indirect value function ${V}_\zs{(N_1)}$ is a globally strictly concave function satisfying Inada's condition on $[l_{T_1},\infty)$  and its two first derivatives are given by
\begin{equation}
({V}_\zs{(N_1)})'(x)=\lambda(x)\xi_\zs{T_\zs{1}}\quad \mbox{and} \quad (V_\zs{(N_1)})^{''}(x)=\lambda'(x) \xi_\zs{T_\zs{1}}.
\label{eq:Indirect4}
\end{equation}
Moreover, the inverse of $(V_\zs{(N_1)})'$ is given by $I_\zs{(N_1)}$ which is defined by
\begin{equation}
I_\zs{(N_1)}(z)=I^{bind}_\zs{(N_1)}(z){\bf 1}_\zs{z> \lambda^{bind}}+I^{unco}_\zs{(N_1)}(z){\bf 1}_\zs{z\le \lambda^{bind}},
\label{eq:I2VaR}
\end{equation}
where
\begin{align}
I^{bind}_\zs{(N_1)}(u)&:=
\int_\zs{z<U_\zs{\wh{\kappa}_2}'(\wt{L}_T)/u} zI_{\wh{\kappa}_2}(u\,z) \d F_{\xi_\zs{t,T}}(z)
+ \wt{L}_T e^{-R_\zs{t,T}}  \wh{Q}_\zs{t,T}(U_\zs{\wh{\kappa}_2}'(\wt{L}_T),U'(\wh{L}_T),u)\notag\\
&\quad+
\int_\zs{U'(\wh{L}_T)/u\le z<\xi_{T_1,T}(q_{\beta})/u}z I(u\,z) \d F_{\xi_\zs{t,T}}(z)
+ L_T e^{-R_\zs{t,T}} \wh{Q}_\zs{t,T}(\xi_\zs{\wh{L}},\xi_{T_1,T}(q_{\beta}),u),
\label{eq:Indirect8}
\end{align}
and 
\begin{align}
I^{unco}_\zs{(N_1)}(u)&:=
\int_\zs{z<U_\zs{\wh{\kappa}_2}'(\wt{L}_T)/u} zI_{\wh{\kappa}_2}(u\,z) \d F_{\xi_\zs{t,T}}(z)
+ \wt{L}_T e^{-R_\zs{t,T}}  \wh{Q}_\zs{t,T}(U_\zs{\wh{\kappa}_2}'(\wt{L}_T),U'(\wh{L}_T),u)\notag\\
&\quad+
\int_\zs{U'(\wh{L}_T)/u\le z<U_{\wh{\kappa}_2}'(\wh{y}^{\wh{\kappa}_2,q_p})/u}z I(u\,z) \d F_{\xi_\zs{t,T}}(z)
+ L_T e^{-R_\zs{t,T}} \wh{Q}_\zs{t,T}(\xi_\zs{\wh{L}},U_{\wh{\kappa}_2}'(\wh{y}^{\wh{\kappa}_2,q_p}),u).
\label{eq:Indirect8unc}
\end{align}

\end{lemma}
\proof Following the same steps as in Lemma 5 in \cite{CNSmult} and noting $x=I^{bind}_\zs{(N_1)}(\lambda \xi_{T_1})$ it can be checked directly that 
$$({V}^{bind}_\zs{(N_1)})'(x)=\lambda \xi_{T_1},\quad x\in[l_{T_1},x^{bind}_\zs{T_1}].
$$ 
Similarly, for $x\in[x^{bind}_\zs{T_1},\infty)$ we have $({V}^{unco}_\zs{(N_1)})'(x)=\lambda^{unco} \xi_{T_1}$ and $x=I^{{unco}}_\zs{(N_1)}(\lambda^{unco} \xi_{T_1})$ . \endproof

\vspace{1cm}

\noindent{\bf Death benefit and dividend payment}: 
Below we assume that at time $T_1$ the death and dividend payments are made only up to the level that the portfolio value after such payments $X_{T_1}$ is sufficient to satisfy the VaR regulation at maturity. In particular, let $D\ge 0 $ be a constant which represents the payment size for each death in the first period $(0,T_1]$. Hence, the total death payment is equal to $DN_1$. We assume that each surviving policyholder receives a dividend payment equal to $\rho (X_{T_1^-}-DN_1-X_0)^+/(n-N_1)$ for some constant $\rho\in[0,1)$. We observe that the dividend payments are proportional to the difference of the portfolio value $X_{T_1^-}$ and the sum of the total death payment $DN_1$ and the initial capital $X_0$. This dividend policy takes an option-like form on the portfolio level with strike price equal to the sum of the minimal capital for the VaR constraint and the initial investment.  These full death benefit and dividend payments are only possible if $X_{T_1^-}-DN_1 -\rho(X_{T_1^-}-DN_1-X_0)^+>l_{T_1}$, i.e., if the portfolio value at time $T_1$ is sufficient to cover the VaR constraint at maturity. 

If $X_{T_1^-}-DN_1 -\rho(X_{T_1^-}-DN_1-X_0)^+\le l_{T_1}$, the payments are made up to the level $l_{T_1}$, the minimal capital level that is needed to satisfy the VaR regulation at maturity. The total death and dividend payment for this case is measured by $X_{T_1^-}-l_{T_1}$.\footnote{From the insurer's perspective, further specifications on the payments are not needed for the optimization problem \eqref{eq:deathdivi}.} 


Thus, due to death benefit and dividend payments, the portfolio value at time $T_1$  admits a downside jump which is characterized by
\begin{equation}
\begin{cases} DN_1+\rho (X_{T_1^-}-DN_1-X_0)^+&\mbox{if}\quad X_{T_1^-}> l_{T_1}+\rho(X_{T_1^-}-DN_1-X_0)^+,\\
X_{T_1^-}-l_{T_1} &\mbox{else}.
\end{cases}
\label{eq:}
\end{equation}
Let 
$$
\wh{l}^{\rho}:=\frac{\rho}{1-\rho}X_0+\frac{1}{1-\rho}l_{T_1}.
$$
The portfolio value after death benefit and dividend (if applicable) payments is given by
\begin{align}
X_{T_1}=(X_{T_1^-}-DN_1-\rho (X_{T_1^-}-DN_1-X_0)){\bf 1}_\zs{X_{T_1^-}>DN_1+\wh{l}^{\rho}}+
l_{T_1}{\bf 1}_\zs{X_{T_1^-}\le DN_1+\wh{l}^{\rho}}.
\label{eq:}
\end{align}
Recall that the insurer's value function at time $T_1$ is given by ${V}_\zs{(N_1)}$, defined by \eqref{eq:VN1}. Due to the dynamic programming principle, the insurer therefore has to solve the following optimization problem
\begin{align} 
 \sup_{ (\nu_t)_\zs{t\in[0,T_1]}} \E\left[ {V}_\zs{(N_1)}( X_{T_1})\right].\label{eq:deathdivi}
\end{align}
Note that the condition $X_{T_1^-}\ge l_{T_1}$ has been implicitly included in the use of the indirect value ${V}^{VaR}_\zs{(N_1)}$ which is only defined on $[l_{T_1},\infty)$. Now, let us introduce $\varepsilon_k:=\P(N_1=k)$ for $k\in\{0,1,\cdots,n\}$. By conditioning on the financial filtration $\cF_{T_1}$ and using the assumption that $X_{T_1^-}$ is independent of $N_1$ we can represent the optimization problem \eqref{eq:deathdivi} as
\begin{align} 
 \sup_{ (\nu_t)_\zs{t\in[0,T_1]}} \E\bigg[\sum_{k=0}^n \varepsilon_k\bigg({V}_\zs{(k)}( (1-\rho)X_{T_1^-}-(1-\rho)Dk+\rho X_0){\bf 1}_\zs{X_{T_1^-}>Dk+\wh{l}^{\rho}}
+
{V}_\zs{(k)}( l_{T_1}){\bf 1}_\zs{X_{T_1^-}\le Dk+\wh{l}^{\rho}}
 \bigg)\bigg].
\label{eq:T1optdivi}
\end{align} 
Note that \eqref{eq:T1optdivi} is expressed as a non-concave optimization problem in a complete market. To solve it we look at its static version
\begin{align} 
 \sup_{\cF_\zs{T_1}\ni X\ge l_\zs{T_1} }\E[\wh{V}(X)]:= 
\E\bigg[\sum_{k=0}^n \varepsilon_k\bigg(&{V}_\zs{(k)}( (1-\rho)X-(1-\rho)Dk+\rho X_0){\bf 1}_\zs{X>Dk+\wh{l}^{\rho}}
+
{V}_\zs{(k)}( l_{T_1}){\bf 1}_\zs{X\le Dk+\wh{l}^{\rho}}
 \bigg)\bigg]
\label{eq:T1optdivi2}
\end{align} 
subject to the usual budget constraint $\E[\xi_{T_1}X]\le X_0$. For convenience, we introduce the sequence $(x_m^{\rho})_\zs{m\in\{-1,0,1\cdots,n\}}$ defined by
\begin{equation}
x_m^{\rho}:=\wh{l}^{\rho}+mD,\quad m\in\{0,1\cdots,n\},
\label{eq:}
\end{equation}
with $x_{-1}^{\rho}:=l_{T_1}$ and $x_{n+1}^{\rho}:=\infty$ by convention. For the presentation of the solution we introduce the following sequence of increasing and concave functions $\wh{U}_\zs{(m)}$ defined in $[x_m^{\rho},x_{m+1}^{\rho})$ 
\begin{equation}
\wh{U}_\zs{(m)}  	(x):=\sum_{k=0}^m \varepsilon_k {V}_\zs{(k)}( (1-\rho)X-(1-\rho)Dk+\rho X_0)
+\sum_{k=m+1}^n \varepsilon_k{V}_\zs{(k)}( l_{T_1}),\quad m\in\{-1,0\cdots,n\}.
\label{eq:U}
\end{equation}
Note that by construction, $\wh{U}_\zs{(-1)}(x) =\sum_{k=0}^n \varepsilon_k{V}_\zs{(k)}( l_{T_1})$, which is a constant. Now, it can be observed that the objective function $\wh{V}$ is identical to $\wh{U}_\zs{(m)}$ in the interval $[x_m^{\rho},x_{m+1}^{\rho})$ for each $m\in\{-1,0\cdots,n\}$, or more precisely, 
\begin{equation}
\wh{V}(x)=\sum_{m=-1}^n\wh{U}_\zs{(m)}  	(x){\bf 1}_\zs{[x_m^{\rho},x_{m+1}^{\rho})}(x).
\label{eq:T1opt2divi}
\end{equation}
It is clear that $\wh{V}$ is a continuous, increasing and piecewise concave function in $[l_{T_1},\infty)$. Note that $\wh{V}$ is not differentiable at $\{x_m^{\rho},\,m=0,\cdots,n\}$ because
\begin{equation}
\wh{U}_\zs{(m-1)}'(x_m^{\rho})<\wh{U}'_\zs{(m)}(x_m^{\rho}), \quad m=0,\cdots,n.
\label{eq:}
\end{equation}

The solution of the non-concave optimization \eqref{eq:T1optdivi2} is given by looking at the concavified function $\wh{V}^{**}$ which is the double conjugate of $\wh{V}$. The convex conjugate of $\wh{V}$ itself can be related to a family of upper-half hyperplanes whose intersection equals the region below the graph of $\wh{V}$. In particular, each member of this family is the smallest affine function of the form $xy+c$ which dominates $\wh{V}$, i.e.,
$$
\wh{V}(x)\le xy +c,\quad \forall x\in[l_{T_1},\infty).
$$
Then, for a given slope $y$, the corresponding conjugate of $\wh{V}$ is the smallest constant  being determined by
$$
\wh{V}^*(y):=\sup_{x\ge l_{T_1}}(\wh{V}(x)-xy).
$$
Since $\wh{V}$ is not differentiable at the utility changing points $x_m^{\rho}$, we have to first compute the above supremum on each interval $[x_m^{\rho},x_{m+1}^{\rho})$. The convex conjugate $\wh{V}^*$ follows from a comparison among these local maximums. In addition, $\wh{V}$ is concave on each of these intervals and the corresponding supremum defines a convex and decreasing function in $y$. Therefore, $\wh{V}^*(y)$ can be seen as a decreasing, convex and differentiable function except at a finite number of points (which are explicitly determined below as the tangency points depending on our parameters). Classical results from convex analysis ensure that the double conjugate $\wh{V}^{**}$ is the smallest concave function which dominates $\wh{V}$, and on an interval where $\wh{V}^{**}\neq \wh{V}$, $\wh{V}^{**}$ is linear \cite{carassus2009,reichlin2013}. The optimal terminal wealth is then given\footnote{$f'_+$ stands for right derivative of $f$} by 
$- (\wh{V}^*)_+'(\wh{\lambda} \xi_{T_1})$ for some Lagrangian multiplier $\wh{\lambda}$ determined via the budget constraint. For more details, see  \cite{carpenter2000, karoui2005, carassus2009,reichlin2013,bichuch2014}. 

\begin{theorem}
Assume that $X_0\ge \E[l_{T_1}\xi_\zs{t_1}] =e^{-rT}L_T\xi_\zs{0,T}(q_\zs{\beta}).$ The optimal wealth before death benefit and dividend payments at time $T_1$ is then given by $X^*_{T_1^-}=- (\wh{V}^*)_+'(\wh{\lambda} \xi_{T_1})$, where $\wh{\lambda}$ is determined via $\E[\xi_{T_1} X^*_{T_1^-}]=X_0$, the budget constraint at time zero.
\end{theorem} 
\proof This essentially follows from the results in Section 3 and 4 of \cite{reichlin2013}. It remains to check the integrability for the value function $\wh{V}$. Indeed, for any admissible portfolio value $X_{T_1}\ge l_{T_1}$ at time ${T_1}$, we observe from \eqref{eq:T1optdivi2} that  
$$
\E[\wh{V}(X_{T_1})]\le 
\E\left[\sum_{k=0}^n \varepsilon_k {V}_\zs{(k)}( X_{T_1})\right]\le\E\left[\wt{U}^{\cS,p,\kappa}( X_T^{p,\kappa,*})\right]<\infty,
$$
where $X_T^{p,\kappa,*}$ is the unconstrained optimal terminal wealth for the case of defaultable contracts. 
\endproof

We conclude this section by looking at the case where there is one policyholder in the pool, i.e., $n=1$ and only the death payment is made at time $T_1$, i.e., $\rho=0$. In this case, the solution can be given explicitly. First, noting that $\wh{l}^{0}=l_{T_1}$ the value function defined by \eqref{eq:T1opt2divi} is given by  
$$
\wh{V}(x)=\wh{U}_\zs{(0)}  	(x){\bf 1}_\zs{\wh{l}^{\rho}\le x< \wh{l}^{\rho}+D}+\wh{U}_\zs{(1)}  	(x){\bf 1}_\zs{x\ge \wh{l}^{\rho}+D}.$$
In this situation, the concavified function of $\wh{V}$ can be determined explicitly leading to an explicit solution. To this end, let $\wh{I}_\zs{(m)}=(\wh{U}_\zs{(m)}')^{-1}$ be the inverse function of the first derivative $\wh{U}_\zs{(m)}'$,  $ m=0,\cdots,n$. We are now able to present the solution of the optimization problem \eqref{eq:T1optdivi2}.
\begin{theorem}
Assume that $n=1$ and $X_0\ge \E[l_{T_1}\xi_\zs{t_1}] =e^{-rT}L_T\xi_\zs{0,T}(q_\zs{\beta}).$  The optimal portfolio of Problem \eqref{eq:T1optdivi2} is given by
\begin{equation}
X^*_\zs{T_1^{-}}=\wh{I}_\zs{(1)}(\wh{\lambda}\xi_\zs{T_1}){\bf 1}_\zs{\xi_\zs{T_1}<\wh{\xi}}+\wh{I}_\zs{(0)}(\wh{\lambda}\xi_\zs{T_1}){\bf 1}_\zs{\xi_\zs{T_1}\ge\wh{\xi}},
\label{eq:}
\end{equation}
where $\wh{\xi}$ is defined by
\begin{equation}
\wh{U}_\zs{(1)}(\wh{I}_\zs{(1)}(\wh{\lambda}\wh{\xi}))-
\wh{U}_\zs{(0)}(\wh{I}_\zs{(0)}(\wh{\lambda}\wh{\xi}))
=\wh{\lambda}\wh{\xi} \bigg(\wh{I}_\zs{(1)}(\wh{\lambda}\wh{\xi})-\wh{I}_\zs{(0)}(\wh{\lambda}\wh{\xi})\bigg),
\label{eq:commonline}
\end{equation}
and $\wh{\lambda}$ is determined such that the budget constraint $\E[\xi_\zs{T_1}X^*_\zs{T_1^{-}}]=X_0$ is satisfied.
\end{theorem}
Before presenting the proof let us remark that \eqref{eq:commonline} defines the linear line that is jointly tangent to the curve $\wh{U}_\zs{(0)}$ (defined on $[l_{T_1}, l_{T_1}+D)]$) at a point (say $\wh{y}_\zs{(0)}<l_{T_1}+D$), and $\wh{U}_\zs{(1)}$ (defined on $[l_{T_1}+D,\infty)$) at another point (say $\wh{y}_\zs{(1)}>l_{T_1}+D$). The concavified function of $\wh{V}$ is then defined by $\wh{U}_\zs{(0)}$ on $[l_{T_1}, \wh{y}_\zs{(0)})$, the joint tangency line on $[\wh{y}_\zs{(0)},\wh{y}_\zs{(1)})$, and $\wh{U}_\zs{(1)}$ on $[\wh{y}_\zs{(1)},\infty)$.

\proof
For $\wh{\lambda}>0$ and $\xi>0$, consider the Lagrangian 
 \begin{equation}
\Psi(x):=\wh{V}(x)-\wh{\lambda} \xi x=\wh{U}_\zs{(0)}  	(x){\bf 1}_\zs{l_{T_1}\le x< l_{T_1}+D}+\wh{U}_\zs{(1)}  	(x){\bf 1}_\zs{x\ge l_{T_1}+D}-\wh{\lambda} \xi x.
\label{eq:}
\end{equation}
Note first that $\wh{U}_\zs{(i)}$ attains maximum at $\wh{I}_\zs{(i)}({\wh{\lambda}}{\xi})$, $i=0,1$. Furthermore, it follows from \eqref{eq:U} that $\wh{I}_\zs{(1)}({\wh{\lambda}}{\xi})>\wh{I}_\zs{(0)}({\wh{\lambda}}{\xi})$ and $\wh{I}_\zs{(1)}({\wh{\lambda}}{\xi})>l_{T_1}+D$ for all $\wh{\lambda}>0$ and $\xi>0$. Let $\xi_0:=\frac{\wh{U}'_\zs{(0)}(l_{T_1}+D) }{\wh{\lambda}}$. We observe that $\Psi$ is increasing in $[l_{T_1},\wh{I}_\zs{(1)}({\wh{\lambda}}{\xi}))$ and decreasing in $[\wh{I}_\zs{(1)}({\wh{\lambda}}{\xi}),\infty)$ if $\xi<\xi_0$. This implies that $\wh{I}_\zs{(1)}({\wh{\lambda}}{\xi})$ is the global maximizer of $\Psi$ for $\xi<\xi_0$. Assume next $\xi\ge\xi_0$. The global optimality of $\Psi$ results from the comparison of $\Psi(\wh{I}_\zs{(1)}({\wh{\lambda}}{\xi}))$ and $\Psi(\wh{I}_\zs{(0)}({\wh{\lambda}}{\xi}))$. To this end, consider the function 
$
f(\xi):=\Psi(\wh{I}_\zs{(1)}({\wh{\lambda}}{\xi}))-\Psi(\wh{I}_\zs{(0)}({\wh{\lambda}}{\xi})). 
$
Obviously $f'(\xi)=-\wh{\lambda}(\wh{I}_\zs{(1)}(\wh{{\lambda}}{\xi})-\wh{I}_\zs{(0)}(\wh{{\lambda}}{\xi}))<0$, which implies that $f$ is decreasing in $\xi\in[\xi_0,\infty)$. Furthermore, since $\wh{U}'_\zs{(1)}(l_{T_1}+D)=\infty$ and $\wh{U}'_\zs{(0)}(l_{T_1})=\infty$ we obtain
$\lim_{\xi\to\infty}f(\xi)=-\infty$. On the other hand, noting that $\wh{U}_\zs{(1)}(l_{T_1}+D)=\wh{U}_\zs{(0)}(l_{T_1}+D)$ we obtain
$$
f(\xi_0)\ge
\wh{U}_\zs{(1)}(\wh{I}_\zs{(1)}({\wh{\lambda}}{\xi}_0))
-\wh{U}_\zs{(1)}(l_{T_1}+D)-\wh{U}'_\zs{(1)}(\wh{I}_\zs{(1)}({\wh{\lambda}}{\xi}_0))(\wh{I}_\zs{(1)}({\wh{\lambda}}{\xi}_0)-(l_{T_1}+D))>0,
$$
because $\wh{U}_\zs{(1)}$ is strictly concave in $[l_{T_1}+D,\infty)$. Therefore, there exists $\wh{\xi}>\xi_0$ such that $f(\wh{\xi})=0$ which gives the concavification equation \eqref{eq:commonline}. Note that $f$ is strictly positive in $[\xi_0,\wh{\xi}]$ and strictly negative in $[\wh{\xi},\infty)$. The global maximizer of $\Psi$ is then given by $\wh{I}_\zs{(1)}({\wh{\lambda}}{\xi})$ if $\xi<\wh{\xi}$ or $\wh{I}_\zs{(0)}({\wh{\lambda}}{\xi})$ if $\xi\ge\wh{\xi}$. The sufficient verification can be done similarly as in the proof of Lemma \ref{Le:unconst.1}.\endproof

\vspace{2mm}
\begin{remark}
It is important to note that the modelling with one terminal constraint does not fully reflect the optimization problem of financial institutions in practice. More precisely, the assumption that the financial institutions never reevaluate their VaR portfolios after the initiation is quite unrealistic, especially for long-term portfolios. In practice, VaR constraints are imposed over a yearly, monthly or weekly time horizon.\footnote{For instance, according to Solvency II, the solvency capital requirement for the insurance companies is based on an annual VaR-measure calibrated to a 99.5\% confidence level (c.f. Article 100 of Solvency II directive).} 
Note that the regulation horizon is typically chosen by regulators, and financial institutions in general have to periodically report their VaR estimates for a prefixed shorter holding period\footnote{For example, as noted in \cite{Alex2012}, regulatory VaR requires that VaR estimates be made for a holding period of 10 working days at a significance level of 1\%. Regulators use the number of violations as a proxy for the quality of the VaR modelling and an internal model is only accepted by the regulator if the number of violations is smaller than or equal to 10.}, leading to a multiple-VaR constrained optimization problem. Such a multiple-VaR regulation has been considered in the literature 
by many authors e.g. \cite{Cuoco2006,Schyns2010,kraft2013,Xu2017,ShiWerker12,CNSmult} and we refer the reader to these works for further discussions. 
In particular, it is shown in e.g. \cite{CNSmult} that for a concave utility framework under a multiple VaR constraint, by taking intertemporal time instances into account, the regulator can better control the risk of intertemporal bankruptcy \emph{before} the terminal time period is reached, which is ignored if only the terminal VaR constraint is considered. Furthermore, a dynamic regulatory framework allows the financial institution to consistently update its \emph{overall} strategy based on new information at the end of each period. 

The two-period optimization framework in this section essentially shows that the utility maximization considered in this work can be extended to  frameworks where there are multiple VaR constraints imposed within the regulation horizon and death benefit and dividend payments are made at a finite number of intermediate time points. This means that our analysis can be implemented according to the Solvency II capital requirement for insurance companies. Note that our analysis can be considered as an extension of the setting in \cite{CNSmult} where the utility function is strictly concave and the entire market is assumed to be complete. 

\end{remark}
\noindent{\bf Fair pricing constraint}: 
The above unconstrained optimal strategy maximizes the equity holders' utility, leaving the question whether the policyholders are willing to participate in such a contract. To make the contract more attractive the insurer can choose the investment strategy that results from an optimization under a so-called fair pricing or fair participating constraint (see, e.g., \cite{pitacco2011,Chen2017}),  meaning that the initial value of each policyholder's payoff is equal to her initial contribution. 
Note however that in this paper we mainly focus on regulatory aspects of VaR constraints motivated from the Solvency II regulation rather than on the problem of optimal product design. On the other hand, our framework allows for a pool of multiple contracts of heterogeneous policyholders with mortality risk for which the individual fairness is not the main point from the insurer's perspective as it might be too challenging to take it into account for every policyholder. It seems in our setting more reasonable to assume that only those policyholders enter the contract in the first place, for whom a fair pricing constraint holds without the need for the insurer to readjust its strategy. In addition, it seems very challenging to specify even for one policyholder the individual fair participating constraint in the two-period framework with death and dividend payments studied above due to the market incompleteness\footnote{To incorporate fair pricing constraints in an incomplete market setting, one might use ``good deal bounds'' (initiated by \cite{cochrane2000} and extended by \cite{bjork2006}) to narrow the set of pricing measures and discuss possible ways of defining the fair pricing constraint.} and possible intermediate money withdrawals.



Below we briefly explain how the optimization developed here can be applied to the case of one policyholder without mortality or dividend payments. Mathematically, we now consider the following optimization problem (for $j\in\{p,np\}$) 
$$ \sup_{ X_T\in\cX}\E\left[\wt{U}^{\cS,j,\kappa}(X_T)\right]$$ under the fair pricing constraint $\E[\xi_T V_L^j(X_T)]\ge L_0$,
where the objective utility $\wt{U}^{\cS,j,\kappa}$ is defined by \eqref{Ueps}. 

We remark first that if $\mathbb{E}_\mathbb{P}\big[ \xi_T V_L^{j}(X_T^{j,\kappa,*})\big]> {L}_0$ (i.e., the fair pricing constraint is not binding), the policyholder is willing to participate in such a contract because her contract value is higher than her initial investment $L_0$. The solution to this problem is identical to the unconstrained solution given in Theorem \ref{Th:unconstPTeps}. We exclude these cases in the following by assuming that the contract parameters $(\alpha,\delta,X_0,L_0,L_T,T)$ satisfy $\mathbb{E}_\mathbb{P}\big[ \xi_T V_L^{j}(X_T^{j,\kappa,*})\big]\le {L}_0$. 
The optimization problem now reads
\begin{align}\label{eq:Fair1}
 \sup_{ X_T\in\cX}\E\left[\wt{U}^{\cS,j,\kappa}(X_T)\right],\quad s.t. \quad \E[\xi_T V_L^j(X_T)]=L_0,
\end{align}

To solve the fair-pricing constrained optimization problem \eqref{eq:Fair1} we can rely on a Lagrangian approach with the usual concavification technique
\begin{align*}
\sup_{X_T} \E\big[ \wt{U}^{\cS,j,\kappa}(X_T)\big) + \lambda_1 \big( \nu - \xi_TX_T \big) + \lambda_2 \big( L_0 -  \xi_T V_L^{j}(X_T) \big)  \big]
\end{align*}
for multipliers $\lambda_1> 0$ and $\lambda_2>0$. For each type of payoffs, we consider the convex conjugate
\begin{align}  \label{conv_2}
{\chi}^{j}(y,z) :=\operatorname{argsup}_{x> 0}  \big[ \wt{U}^{\cS,j,\kappa}(x) - y\cdot x -z\cdot V^j(x)\big]\,,\quad j\in\{p,np\}.
\end{align}
This leads to $\wh{X}^{j}(\lambda_1,\lambda_2,\xi_T): = \chi^{j}(\lambda_1 \xi_T, \lambda_2 \xi_T)$ as a candidate for the optimal terminal wealth. The verification step is analogous to Lemma \ref{Le:unconst.1} and the proof of existence of the Lagrangian multipliers requires some additional analysis of indifference curves which can be done along the lines of Appendix E in \cite{Chen2017}.

The analysis can be extended to the case where a VaR constraint at maturity is included. Obviously, such a study requires a more thorough analysis. However, as mentioned above it seems very challenging to treat all  individual fair participating constraints for a pool of individual contracts in the two-period framework with death and dividend payments and a regulatory VaR constraint.

\section{Conclusion}
We solve a non-concave utility maximization problem for the equity holders of a participating life insurance contract under a VaR-type regulatory constraint imposed at maturity. We obtain a closed-form solution extending the martingale approach to constrained non-concave utility maximization problems. Our theoretical and numerical results show that for the case of a defaultable contract, the risk exposure in bad market conditions will be reduced by a VaR constraint. This result is contrary to most other results derived for VaR risk management e.g. \cite{BasakShapiro01,CNS17,kraft2013,CNSmult,wei2018} where the introduction of a VaR constraint increases the losses in the bad states as the risk manager has an incentive to push significant parts of the remaining risk into those parts of the tails not captured by VaR. The effects of the parameters of the contract are also discussed numerically. It is shown that the policyholders indeed benefit from a Solvency II type VaR-regulation in the sense of having greater expected utility. The prudent investment behavior for participating insurance contracts is more pronounced if a VaR-type regulation is replaced by a PI-type regulation. Furthermore, a stricter regulation (a smaller allowed default probability in the VaR problem or a higher minimum guarantee level in the $PI$ problem) enhances the benefit of the policyholder but deteriorates that of the insurer. For both types of regulation, the gains in terms of expected utility are greater for higher participation rates, while being smaller for higher bonus rates. However, our results also show that in certain situations, for instance in the cases of relatively high participation rates and relatively low bonus rates, policyholders overall may in terms of expected utility prefer a VaR regulation, over a PI regulation with a relatively low minimum guarantee level. These results can be extended to multiple periods.



\setcounter{section}{0}
\renewcommand{\thesection}{\Alph{section}}
\section{Proofs}
 \subsection{Proof of Theorem \ref{Th:unconstPTeps} (the unconstrained case)} \label{sec:unpro}

This section is devoted to the detailed proof for the uncontrained problem \eqref{eq:TPmort}. This problem can be solved by considering the static optimization $\max_\zs{X\ge 0} \Psi_j(X)$ of the Lagrangian 
\begin{equation}
\Psi_j(X)=\wt{U}^{\cS,j,\kappa}(X)-\lambda \xi_T X.
\label{eq:f}
\end{equation}
Notice that $\Psi_j$ is not concave, which makes the problem more challenging to solve. We make use of a modified dual approach which shows that the global maximal value of $\Psi_j$ can be attained at the local maximizers or at the utility changing points $L$ and $\wt{L}$, or even at the boundary point $0$. Note that $\Psi_j$ is not differentiable at $L_T$ and $\wt{L}_T$.

\begin{lemma}\label{Le:unconst.1}
For $\lambda>0$ and $\xi>0$, the unique solution of the problem $\max_\zs{X\ge 0} \Psi_j(X)$, for $j\in\{p,np\}$,
is given by
	\begin{equation*}\label{eq:uopteplemma}
	X^{j,\kappa,*}(\lambda,\xi)=\cWD_\zs{[\xi_\zs{\wt{L}}^{\kappa},\xi_\zs{\wh{L}},\wh{\xi}^{1,q_{j}}]}^{[I_\kappa,\wt{L},I+L,0]}{\bf 1}_\zs{\Upsilon^{1,q_{j}}(\wt{L}_T)>0}+ \cWD_\zs{[\xi_\zs{\wt{L}}^{\kappa},\xi_\zs{U}^{q_{j}}]}^{[I_\kappa,\wt{L},0]} {\bf 1}_\zs{\Upsilon^{\kappa,q_{j}}(\wt{L}_T)\ge0\ge\Upsilon^{1,q_{j}}(\wt{L}_T)}+ \cWD_\zs{[\wh{\xi}^{\kappa,q_{j}}]}^{[I_\kappa,0]}	{\bf 1}_\zs{\Upsilon^{\kappa,q_{j}}(\wt{L}_T)<0}.
	\end{equation*}

\end{lemma}
\proof Note first that we are looking for the global maximizer $X^*$ of a three-part function $\Psi_j(X)=Q_1(X){\bf 1}_\zs{X< L}+Q_2(X){\bf 1}_\zs{L\le X\le  \wt{L}}+Q_3(X){\bf 1}_\zs{X>\wt{L}}$, for $j\in\{p,np\}$ where
\begin{align*}
Q_1(X):=-U_{lo}(\epsilon_j (L-X))-\lambda\xi X,\quad Q_2(X):= U(X-L)-\lambda\xi X, \quad Q_3(X):= U_\kappa(X)-\lambda\xi X.
\end{align*}
Clearly, $Q_2,Q_3$ are concave functions having local maximizers $X_2:=I(\lambda\xi)+L$ and $X_3:=I_\kappa(\lambda \xi)$ respectively. Note that since $Q_1$ is convex, its maximum can be attained at the boundary points zero or $L$. Due to the continuity at $L$, i.e. $\lim_{x\uparrow L}Q_1(x)=Q_2(L)\le Q_{2,\max}:=\max_\zs{L\le X\le  \wt{L}}Q_2(X)$, the global optimum is obtained by comparing $Q_{1}(0)=- U_{lo}(\epsilon_jL)=-q_j$ with $Q_{2,\max}$ and $Q_{3,\max}:=\max_{X> \wt{L}}Q_3(X)$.

From \eqref{eq:xi11} we observe first that $X_2\in [L,\wt{L}]$ iff $\xi\ge \xi_{\wh{L}}$ and $X_3>\wt{L}$ iff $\xi< \xi_\zs{\wt{L}}^{\kappa}$. Therefore, we study the Lagrangian on the following subintervals of values of $\xi$:
\begin{enumerate}
	\item[(a)] For $\xi< \xi_\zs{\wt{L}}^{\kappa}$, we have $X_2\ge \wt{L}$ and $X_3> \wt{L}$ so the Lagrangian $\Psi_j$ is increasing from $L$ to $X_3$ and decreasing again in $(X_3,\infty)$. So global optimality can be attained at $0$ or $X_3$. To conclude, we consider 
	$$
	Q_{3,\max}-Q_{1}(0)=U_\kappa(I_\kappa(\lambda\xi))-\lambda\xi I_\kappa(\lambda\xi)+q_j:=\Delta_a(\xi),
$$
which is a decreasing function in $\xi$. It follows that $Q_{3,\max}-Q_{1}(0)\ge \Delta_a(\xi_\zs{\wt{L}}^{\kappa})=\Upsilon^{\kappa,q_j}(\wt{L}_T)$. If $\Upsilon^{\kappa,q_j}(\wt{L}_T)>0$ then $Q_{3,\max}-Q_{1}(0)>0$ and the global maximizer is $X_3$. When $\Upsilon^{\kappa,q_j}(\wt{L}_T)\le 0$, by Lemma \ref{Le:conf} there exists $\wh{\xi}^{\kappa,q_j}\le \xi_\zs{\wt{L}}^{\kappa}$ such that $\Delta_a(\wh{\xi}^{\kappa,q_j})=0$. In this case, the global maximizer can be chosen as $X_3$ for $\xi< \wh{\xi}^{\kappa,q_j}$ or zero for $\xi\in[\wh{\xi}^{\kappa,q_j},\xi_\zs{\wt{L}}^{\kappa})$. 

\item [(b)] For $\xi_\zs{\wt{L}}^{\kappa}\le \xi<\xi_{\wh{L}}$, $X_2\ge \wt{L}$ but $X_3\le\wt{L}$ so the Lagrangian is increasing from $L$ to $\wt{L}$ and decreasing again in $(\wt{L},\infty)$. So global optimality can be attained at $0$ or $\wt{L}$. To conclude, we consider $
Q_{2,\max}-Q_{1}(0)=Q_{2}(\wt{L})+q_j=U(\wt{L}-L)-\lambda\xi \wt{L}+q_j:=\Delta_b(\xi)$, which is a decreasing function in $\xi$. It follows that $\Delta_b(\xi)\in(\Upsilon^{1,q_j}(\wt{L}_T),\Upsilon^{\kappa,q_j}(\wt{L}_T)]$. If $\Upsilon^{1,q_j}(\wt{L}_T)>0$ then $\Delta_b({\xi})>0$ and the global maximizer is $\wt{L}$. When $\Upsilon^{\kappa,q_j}(\wt{L}_T)< 0$, the global maximizer is zero. It remains to consider the case $\Upsilon^{1,q_j}(\wt{L}_T)\le 0\le\Upsilon^{\kappa,q_j}(\wt{L}_T)$. By Lemma \ref{Le:conf}, $\Delta_b({\xi}_U^{q_j})=0$ and $\xi_{\wt{L}}^{\kappa}\le {\xi}_U^{q_j}<\xi_\zs{\wh{L}}$. This implies that the global maximizer is $\wt{L}$ for $\xi_\zs{\wt{L}}^{\kappa}\le \xi<\xi_U^{q_j}$ or is equal to zero for $\xi_U^{q_j}\le\xi<\xi_\zs{\wh{L}}$. 
\item [(c)] For $\xi\ge \xi_{\wh{L}}$, $X_2\le\wt{L}$ and $X_3< \wt{L}$ so the Lagrangian is increasing from $L$ to $X_2$ and decreasing again in $(X_2,\infty)$. So global optimality can be attained at $0$ or $X_2$. We need to study
$Q_{2,\max}-Q_{1}(0)=Q_{2}(X_2)-q_j=U(I(\lambda\xi))-\lambda\xi (I(\lambda\xi)+L)+q_j:=\Delta_c(\xi),$ which is a decreasing function in $\xi$. Thus, $\Delta_c(\xi)\le \Delta_c(\xi_{\wh{L}})=\Upsilon^{1,q_j}(\wt{L}_T)$. Therefore, zero is the maximizer if $\Upsilon^{1,q_j}(\wt{L}_T)\le 0.$ Suppose now that $\Upsilon^{1,q_j}(\wt{L}_T)>0$. By Lemma \ref{Le:conf}, there exists $\wh{\xi}^{1,q_j}> \xi_{\wh{L}}$ such that $\Delta_c(\wh{\xi}^{1,q_j})=0$ and the global optimality is attained at $X_2$ for $\wh{\xi}^{1,q_j}> \xi\ge \xi_{\wh{L}}$ and at zero for ${\xi}\ge \wh{\xi}^{1,q_j}$, respectively.
 	\end{enumerate}
Note that $\Delta_a(\wh{\xi}^{\kappa,q_j})=0$ is equivalent to \eqref{eq:yqPT} with $\wh{y}^{\kappa,q_j}=I_\kappa(\lambda\wh{\xi}^{\kappa,q_j})$, whereas $\Delta_c(\wh{\xi}^{1,q_j})=0$ is equivalent to \eqref{eq:yPT} with $\wh{y}^{1,q_j}=I(\lambda\wh{\xi}^{1,q_j})+L$. \endproof

We now prove Theorem \ref{Th:unconstPTeps} and then Corollaries \ref{Th:unconst} and \ref{Th:unconstPT}. Let us prove that $ X^{j,\kappa,*}(\lambda,\xi_T)$ defined by \eqref{eq:uopteplemma} is the solution to Problem \eqref{eq:TPmort2}. For any terminal wealth ${X}_T\ge 0$ satisfying the budget constraint we have
\begin{align}
\E [\wt{U}^{\cS,j,\kappa}(X_T))]&\le \E[\wt{U}^{\cS,j,\kappa}({X}_T))+\lambda(X_0-\xi_T X_T)]\label{V2}\\
&\le
\E \left[\sup_\zs{X\ge 0}(\wt{U}^{\cS,j,\kappa}({X})-{\lambda} \xi_T X)\right]+X_0\lambda\notag\\
&=\E[\wt{U}^{\cS,j,\kappa}(X^{j,\kappa,*}(\lambda,\xi_T))]+\lambda(X_0-\E[\xi_T X^{j,\kappa,*}(\lambda,\xi_T)])\notag\\
&=\E[\wt{U}^{\cS,j,\kappa}(X^{j,\kappa,*}(\lambda,\xi_T))]. \notag
\end{align}
The last equality folows from $\E[\xi_T X^{j,\kappa,*}(\lambda,\xi_T)]=X_0$. Hence, $X^{j,\kappa,*}$ is optimal. The existence of the Lagrangian multiplier $\lambda$ follows from the lemma below.
\begin{lemma}\label{Le:suff}
For any $\lambda\in (0,\infty)$, we have
\begin{enumerate}
\item  $ \E [U(I(\lambda\xi_T))]<\infty\Longrightarrow \E[\wt{U}^{\cS,j,\kappa}(X^{j,\kappa,*}(\lambda,\xi_T))]<\infty $;
	\item $\E [\xi_TI(\lambda\xi_T)]<\infty \Longrightarrow \E [\xi_TX^{j,\kappa,*}(\lambda,\xi_T)]<\infty$;
	\item Furthermore, if $\E[\xi_T I(\lambda\xi_T)]<\infty$ for any $\lambda\in(0,\infty)$, the mapping $\psi:\lambda\longmapsto \E[\xi_TX^{j,\kappa,*}(\lambda,\xi_T)]$ is strictly decreasing, continuous and surjective from $(0,\infty)$ to $(0,\infty)$.
\end{enumerate}
\end{lemma}
\proof The first two conclusions follow from the observation that 
$$
X^{j,\kappa,*}(\lambda,\xi_T)\le \wt{L}_T+(1-\wt{\delta})^{-1}I((1-\wt{\delta})^{-1}\lambda\xi_T),
$$
using Lemma \ref{Le:Ukappa} and \eqref{eq:comkappa}. For the last one, it suffices to notice that $I$ is strictly decreasing, so for a.s. all $\omega\in\Omega$,  $\lambda\longmapsto X^{j,\kappa,*}(\lambda,\xi_T(\omega))$ is decreasing, which implies that $\psi$ is also decreasing. Now, if $\E[\xi_T I(\lambda\xi_T)]<\infty$ then $\psi$ is well defined. Noting that the price density $\xi_T$ has no atom, (i.e., for all $a\in\bbr$, $\P(\xi_T=a)=0$), we deduce that $\psi$ is continuous on $(0,\infty)$. 
Moreover, for a.s. all $\omega$ we have $\lim_\zs{\lambda\to 0}X^{j,\kappa,*}(\lambda,\xi_T(\omega))=\infty$ and $\lim_\zs{\lambda\to \infty}X^{j,\kappa,*}(\lambda,\xi_T(\omega))=0$ due to the Inada condition and the monotone convergence theorem, which implies that $\psi$ is surjective.\endproof
\

 \subsection{Proof of Theorem \ref{Th:const} (the VaR-constrained case)} \label{sec:pro}
To study the constrained problem \eqref{eq:const.1}, we again consider the static optimization of the Lagrangian 
\begin{equation}
\max_\zs{X\ge 0}\overline{\Psi}_j(X):=\wt{U}^{\cS,j,\kappa}(X)-\lambda \xi X-\lambda_2{\bf 1}_\zs{X<L},
\label{eq:fVaR}
\end{equation}
for given $\lambda>0,\lambda_2\ge 0$. Again, the subscript $T$ is dropped for simplicity. Note first that $\overline{\Psi}_j$ is not concave and Problem \eqref{eq:fVaR} is more challenging than the unconstrained case due to the presence of the additional indicator function. Below, we show that the optimality can be obtained within a modified dual 
approach. 


The key point is to choose the second multiplier $\lambda_2$ as a function of the first one. This is because we need to check the ``bindingness'' of the VaR constraint by comparing the unconstrained solution with the VaR-threshold $L_T$. Let $\bar{\xi}$ be defined by $\P(\xi_T>\bar{\xi})=\beta$. From Theorem \ref{eq:uopteps} we observe that 
\begin{equation}\label{eq:prob}
\P(X^{j,\kappa,*}<L_T)=
\begin{cases}
\P(\xi_T>\wh{\xi}^{1,q_j}) \quad \mbox{if}\quad \Upsilon^{1,q_j}(\wt{L}_T)>0,\\
\P(\xi_T>{\xi}_U^{q_j}) \quad \mbox{if}\quad \Upsilon^{\kappa,q_j}(\wt{L}_T)\ge0\ge\Upsilon^{1,q_j}(\wt{L}_T),
\\P(\xi_T>\wh{\xi}^{1,q_j}_\delta) \quad \mbox{if}\quad \Upsilon^{\kappa,q_j}(\wt{L}_T)<0,
\end{cases}
\end{equation}
for $j\in\{p,np\}$. Therefore, to incorporate the bindingness of the VaR constraint, $\lambda_2$ should be chosen such that it is zero when the above probability is smaller than the insolvency level $\beta$. For example, if $\Upsilon^{1,q_j}(\wt{L}_T)>0$ then $\lambda_2>0$ if $\wh{\xi}^{1,q_j}<\overline{\xi}$ and $\lambda_2=0$ if $\wh{\xi}^{1,q_j}\ge\overline{\xi}$. The other cases can be seen similarly. On the other hand, $\lambda_2$ needs to reflect the corresponding comparison of local maxima in the proof of Lemma \ref{Le:unconst.1}. Then, a closer inspection suggests defining 
\begin{equation}\label{eq:lambda2}
\hspace{-1mm}\lambda_2=
\begin{cases}
-\Delta_c(\overline{\xi}){\bf 1}_\zs{\wh{\xi}^{1,q_j}<\overline{\xi}}, \quad \mbox{if}\quad \Upsilon^{1,q_j}(\wt{L}_T)>0,\\
-\Delta_b(\overline{\xi}){\bf 1}_\zs{{\xi}_U^{q_j}<\overline{\xi}\le \xi_{\wh{L}}} -\Delta_c(\overline{\xi}){\bf 1}_\zs{\overline{\xi}\ge \xi_{\wh{L}}},\quad \mbox{if}\quad \Upsilon^{\kappa,q_j}(\wt{L}_T))\ge0\ge\Upsilon^{1,q_j}(\wt{L}_T),
\\
-\Delta_a(\overline{\xi}){\bf 1}_\zs{\wh{\xi}^{1,q_j}<\overline{\xi}\le \xi_\zs{\wt{L}}^{\kappa}}-
\Delta_b(\overline{\xi}){\bf 1}_\zs{\xi_\zs{\wt{L}}^{\kappa}<\overline{\xi}\le \xi_{\wh{L}}}
-\Delta_c(\overline{\xi}){\bf 1}_\zs{\xi_{\wh{L}}<\overline{\xi}},\quad \mbox{if}\quad \Upsilon^{\kappa,q_j}(\wt{L}_T)<0.
\end{cases}
\end{equation}
\begin{lemma}\label{Le:const.1}
For $\lambda>0$ and $\xi>0$, the unique solution $X^{VaR,j,\kappa,*}(\lambda,\xi):=\mbox{argmax}_\zs{X\ge 0} \overline{\Psi}_j(X)$ is given by:
\begin{itemize}
	\item If $\Upsilon^{1,q_j}(\wt{L}_T)>0$ then
	\begin{equation*}
X^{VaR,j,\kappa,*}_T=\cWD_\zs{[\xi_\zs{\wt{L}}^{\kappa},\xi_\zs{\wh{L}},\bar{\xi}]}^{[I_\kappa,\wt{L},I+L,0]}{\bf 1}_\zs{ \bar{\xi}\ge \wh{\xi}^{1,q_j}}+\cWD_\zs{[\xi_\zs{\wt{L}}^{\kappa},\xi_\zs{\wh{L}},\wh{\xi}^{1,q_{j}}]}^{[I_\kappa,\wt{L},I+L,0]}{\bf 1}_\zs{ \bar{\xi}< \wh{\xi}^{1,q_{j}}}.
		\end{equation*}
		\item If $\Upsilon^{\kappa,q_j}(\wt{L}_T)\ge0\ge\Upsilon^{1,q_j}(\wt{L}_T)$ then
\begin{equation*}
X^{VaR,j,\kappa,*}_T=\cWD_\zs{[\xi_\zs{\wt{L}}^{\kappa},\xi_\zs{\wh{L}},\bar{\xi}]}^{[I_\kappa,\wt{L},I+L,0]}{\bf 1}_\zs{ \bar{\xi}\ge \xi_\zs{\wh{L}}}+\cWD_\zs{[\xi_\zs{\wt{L}}^{\kappa},\bar{\xi}]}^{[I_\kappa,\wt{L},0]}{\bf 1}_\zs{ \xi_\zs{U}^{q_{j}}\le\bar{\xi}< \xi_\zs{\wh{L}}}
+\cWD_\zs{[\xi_\zs{\wt{L}}^{\kappa},\xi_\zs{U}^{q_{j}}]}^{[I_\kappa,\wt{L},0]}{\bf 1}_\zs{ \bar{\xi}< \xi_\zs{U}^{q_{j}}}.
		\end{equation*}
		\item If $\Upsilon^{\kappa,q_j}(\wt{L}_T)<0$ then
\begin{align*}
X^{VaR,j,\kappa,*}_T=\cWD_\zs{[\xi_\zs{\wt{L}}^{\kappa},\xi_\zs{\wh{L}},\bar{\xi}]}^{[I_\kappa,\wt{L},I+L,0]}{\bf 1}_\zs{ \bar{\xi}\ge \xi_\zs{\wh{L}}}+\cWD_\zs{[\xi_\zs{\wt{L}}^{\kappa},\bar{\xi}]}^{[I_\kappa,\wt{L},0]}{\bf 1}_\zs{ \xi_\zs{\wt{L}}^{\kappa}\le \bar{\xi}< \xi_\zs{\wh{L}}}
+
\cWD_\zs{[\bar{\xi}]}^{[I_\kappa,0]}{\bf 1}_\zs{ \wh{\xi}^{\kappa,q_j}\le \bar{\xi}<\xi_\zs{\wt{L}}^{\kappa}}
+
\cWD_\zs{[ \wh{\xi}^{\kappa,q_j}]}^{[I_\kappa,0]}{\bf 1}_\zs{\bar{\xi} <\wh{\xi}^{\kappa,q_j}}.
\end{align*}
\end{itemize}
\end{lemma}
\proof We keep the notation from the proof of Lemma \ref{Le:unconst.1}. The only difference is that $Q_1(X):=-U_{lo}(\epsilon_j (L-X))-\lambda_2$, where $\lambda_2$ is defined by \eqref{eq:lambda2}. Therefore, we take $Q_{1}(0)=-q_j-\lambda_2$. Note that $Q_{2}$ and $Q_{3}$ remain the same as before. As in \eqref{eq:delta_l} we have
\begin{equation}
\Upsilon^{1,q_j}(\wt{L}_T)=\Delta_b(\xi_\zs{\wh{L}})=\Delta_c(\xi_\zs{\wh{L}})\quad \mbox{and} \quad \Upsilon^{\kappa,q_j}(\wt{L}_T)=\Delta_b(\xi_\zs{\wt{L}}^{\kappa})=\Delta_a(\xi_\zs{\wt{L}}^{\kappa}).
\label{eq:delta}
\end{equation}We consider the following cases:

\noindent{\bf Case 1}:	If $\Upsilon^{1,q_j}(\wt{L}_T)>0$ then $\lambda_2=-\Delta_c(\bar{\xi}){\bf 1}_\zs{\wh{\xi}^{1,q_j}<\overline{\xi}}$. It suffices to suppose that $\wh{\xi}^{1,q_j}<\overline{\xi}$, i.e., to consider the case where the VaR constraint is binding.
\begin{enumerate}
	\item[(a)] For $\xi<\xi_\zs{\wt{L}}^{\kappa}$, we consider $\Delta:=Q_{3,\max}-Q_{1}(0)=\Delta_a(\xi)+\lambda_2\ge \Delta_a(\xi_\zs{\wt{L}}^{\kappa})+\lambda_2=\Upsilon^{\kappa,q_j}(\wt{L}_T)+\lambda_2>0$. So $X_3$ is the global optimizer of the constrained problem for $\xi< \xi_\zs{\wt{L}}^{\kappa}$.
	
	\item [(b)]For $\xi_\zs{\wt{L}}^{\kappa}\le\xi< \xi_{\wh{L}}$, global optimality can be attained at $0$ or $\wt{L}$. In this case we have $\Delta=Q_{2,\max}-Q_{1}(0)=\Delta_b(\xi)+\lambda_2$. Because $\Upsilon^{1,q_j}(\wt{L}_T)>0$ we have $\Delta_b({\xi})>0$ as in the proof of Lemma \ref{Le:unconst.1} (b) and the global maximizer is $\wt{L}$. 
	
	\item [(c)] For $\xi\ge \xi_{\wh{L}}$, global optimality can be attained at $0$ or $X_2$. We need to study 
$\Delta=Q_{2,\max}-Q_{1}(0)=\Delta_c(\xi)+\lambda_2=\Delta_c(\xi)-\Delta_c(\bar{\xi})$. Recall that $\Delta_c$ is decreasing. Therefore, $\Delta\ge 0$ if $\xi_{\wh{L}}\le {\xi}<\bar{\xi}$ and the global optimality is attained at $X_2$. Similarly, $\Delta<0$ for ${\xi}\ge \bar{\xi}$, which means that zero is the optimizer.
\end{enumerate}	
 
\noindent{\bf Case 2}: Suppose $\Upsilon^{\kappa,q_j}(\wt{L}_T)\ge0\ge\Upsilon^{1,q_j}(\wt{L}_T)$. It suffices to assume that ${\xi}_U^{q_j}\le \overline{\xi}$, i.e. the VaR constraint is binding. For $\xi< \xi_\zs{\wt{L}}^{\kappa}$, $\Delta=Q_{3,\max}-Q_{1}(0)=\Delta_a(\xi)+\lambda_2\ge \Upsilon^{\kappa,q_j}(\wt{L}_T)+\lambda_2>0$ because $\Upsilon^{\kappa,q_j}(\wt{L}_T)>0$ and $\lambda_2>0$ are strictly positive as in the first case. Hence, $X_3$ is the optimizer. 
		
Assume next that $\xi_\zs{\wt{L}}^{\kappa}\le\xi< \xi_{\wh{L}}$. Global optimality can be attained at $0$ or $\wt{L}$.
\begin{itemize}
	\item Assume that ${\xi}_U^{q_j}\le \overline{\xi}< \xi_{\wh{L}}$ then $\lambda_2=-\Delta_b(\bar{\xi})$. In this case we have $\Delta=Q_{2,\max}-Q_{1}(0)=\Delta_b(\xi)-\Delta_b(\bar{\xi})$. As $\Delta_b(\xi)$ decreases in $\xi$ we conclude that $\Delta>0$ for $\xi_\zs{\wt{L}}^{\kappa}\le \xi<\overline{\xi} $, which means $\wt{L}$ is the global maximizer. For $\overline{\xi} \le \xi<\xi_{\wh{L}}$, $\Delta<0$ and optimality is attained at zero. 
	\item If $\overline{\xi}\ge\xi_{\wh{L}}$ then $\lambda_2=-\Delta_c(\bar{\xi})$. From \eqref{eq:delta}, we have $\Delta=Q_{2,\max}-Q_{1}(0)=\Delta_b(\xi)-\Delta_c(\bar{\xi})\ge \Delta_b(\xi)-\Delta_c(\xi_{\wh{L}})=\Delta_b(\xi)-\Delta_b(\xi_{\wh{L}}) \ge 0$ since $\xi_\zs{\wt{L}}^{\kappa}\le \xi<\xi_{\wh{L}}$, which means that optimality is attained at $\wt{L}$ for $\xi_\zs{\wt{L}}^{\kappa}\le\xi< \xi_{\wh{L}}$ . 
	\end{itemize}
	It remains to consider the case $\xi\ge \xi_{\wh{L}}$. As before, global optimality can be attained at $0$ or $X_2$. We need to study 
$\Delta=Q_{2,\max}-Q_{1}(0)=\Delta_c(\xi)+\lambda_2$.
\begin{itemize}
	\item If ${\xi}_U\le \overline{\xi}< \xi_{\wh{L}}$ then $\lambda_2=-\Delta_b(\bar{\xi})$ which implies that $\Delta=\Delta_c(\xi)-\Delta_b(\bar{\xi}).$ Observe that $\Delta_b(\bar{\xi})\ge\Delta_b(\xi_{\wh{L}})=\Delta_c(\xi_{\wh{L}})$ since $\bar{\xi}< \xi_{\wh{L}}$. This leads to $\Delta\le \Delta_c(\xi)-\Delta_c(\xi_{\wh{L}})$. Because $\Delta_c$ is decreasing we conclude that $\Delta\le 0$ for ${\xi}\ge\xi_{\wh{L}}$, which means that zero is the optimizer.
	\item If $\overline{\xi}\ge\xi_{\wh{L}}$ then $\lambda_2=-\Delta_c(\bar{\xi})$ and $\Delta=\Delta_c(\xi)-\Delta_c(\bar{\xi})$. As $\Delta$ is decreasing, it is positive for $\xi_{\wh{L}}\le \xi<\bar{\xi}$, hence, $X_2$ is the optimizer, and negative for $\xi\ge \bar{\xi}$, which means that optimality is attained at zero.
	\end{itemize}

\noindent{\bf Case 3}: Suppose $\Upsilon^{\kappa,q_j}(\wt{L}_T)<0$. We only consider the case $\wh{\xi}^{1,q_j}_\delta\le\overline{\xi}$, i.e., the VaR constraint is binding. Let us first consider the case $\xi< \xi_\zs{\wt{L}}^{\kappa}$. Again, global optimality can be attained at $0$ or $X_3$. 	To conclude, we consider $\Delta=Q_{3,\max}-Q_{1}(0)=\Delta_a(\xi)+\lambda_2$.
	\begin{itemize}
		\item If $\wh{\xi}^{1,q_j}_\delta\le\overline{\xi}<\xi_\zs{\wt{L}}^{\kappa}$ then $\lambda_2=-\Delta_a(\overline{\xi})$ and hence $\Delta=\Delta_a(\xi)- \Delta_a(\overline{\xi})$. So the global optimizer is $X_3$ for $\xi< \bar{\xi}$ and is zero for $\bar{\xi}\le \xi< \xi_\zs{\wt{L}}^{\kappa}$.
			\item If $\xi_\zs{\wt{L}}^{\kappa}\le\overline{\xi}< \xi_{\wh{L}}$ then $\lambda_2=-\Delta_b(\overline{\xi})$ and hence $\Delta=\Delta_a(\xi)- \Delta_b(\overline{\xi})\ge \Delta_a(\xi)- \Delta_a(\xi_\zs{\wt{L}}^{\kappa})\ge 0 $, which implies that $X_3$ is the global optimizer.
				\item If $\overline{\xi}\ge \xi_{\wh{L}}$ then $\lambda_2=-\Delta_c(\overline{\xi})$ and hence $\Delta=\Delta_a(\xi)- \Delta_c(\overline{\xi})\ge \Delta_a(\xi_\zs{\wt{L}}^{\kappa})- \Delta_c(\xi_{\wh{L}})= \Delta_b(\xi_\zs{\wt{L}}^{\kappa})- \Delta_b(\xi_{\wh{L}})\ge 0 $, which implies that $X_3$ is the global optimizer.
	\end{itemize}

Assume now that $\xi_\zs{\wt{L}}^{\kappa}\le\xi< \xi_{\wh{L}}$. The global optimality can be attained at $0$ or $\wt{L}$. In this case we have $\Delta=Q_{2,\max}-Q_{1}(0)=\Delta_b(\xi)+\lambda_2$.
	\begin{itemize}
		\item If $\wh{\xi}^{1,q_j}_\delta\le\overline{\xi}<\xi_\zs{\wt{L}}^{\kappa}$ then $\lambda_2=-\Delta_a(\overline{\xi})$ and hence $\Delta=\Delta_b(\xi)- \Delta_a(\overline{\xi})<\Delta_b(\xi)- \Delta_a(\xi_\zs{\wt{L}}^{\kappa})=\Delta_b(\xi)- \Delta_b(\xi_\zs{\wt{L}}^{\kappa})<0$. Thus the global optimizer is zero for $\overline{\xi}< \xi_\zs{\wt{L}}^{\kappa}\le \xi<\xi_{\wh{L}}$.
			\item If $\xi_\zs{\wt{L}}^{\kappa}\le \overline{\xi}< \xi_{\wh{L}}$ then $\lambda_2=-\Delta_b(\overline{\xi})$ and hence $\Delta=\Delta_b(\xi)- \Delta_b(\overline{\xi})$. The global optimizer will be $\wt{L}$ if $\xi_\zs{\wt{L}}^{\kappa}\le\xi<\overline{\xi} <\xi_{\wh{L}}$ and equal to zero if $\overline{\xi}\le \xi<\xi_{\wh{L}}$ .
				\item If $\overline{\xi}\ge \xi_{\wh{L}}$ then $\lambda_2=-\Delta_c(\overline{\xi})$ and hence $\Delta=\Delta_b(\xi)- \Delta_c(\overline{\xi})\ge \Delta_a(\xi_\zs{\wt{L}}^{\kappa})- \Delta_c(\xi_{\wh{L}})= \Delta_b(\xi_\zs{\wt{L}}^{\kappa})- \Delta_b(\xi_{\wh{L}})\ge 0 $, which implies that $\wt{L}$ is the global optimizer.
	\end{itemize}
	Next, we assume that $\xi\ge \xi_{\wh{L}}$. Thus, global optimality can be attained at $0$ or $X_2$ depending on the sign of 
$\Delta=Q_{2,\max}-Q_{1}(0)=\Delta_c(\xi)+\lambda_2.$ We consider the following subcases:
	\begin{itemize}
		\item If $\wh{\xi}^{1,q_j}_\delta\le\overline{\xi}< \xi_\zs{\wt{L}}^{\kappa}$ then $\lambda_2=-\Delta_a(\overline{\xi})$ and hence $\Delta=\Delta_c(\xi)- \Delta_a(\overline{\xi})\le \Delta_b(\xi_{\wh{L}})- \Delta_b(\xi_\zs{\wt{L}}^{\kappa})\le 0$. So the global optimizer is zero.
			\item If $\xi_\zs{\wt{L}}^{\kappa}\le\overline{\xi}< \xi_{\wh{L}}$ then $\lambda_2=-\Delta_b(\overline{\xi})$ and hence $\Delta=\Delta_c(\xi)- \Delta_b(\overline{\xi})\le \Delta_c(\xi_{\wh{L}})-  \Delta_b(\overline{\xi})=\Delta_b(\xi_{\wh{L}})-  \Delta_b(\overline{\xi})\le 0 $, which implies that $0$ is the global optimizer.
				\item If $\overline{\xi}\ge \xi_{\wh{L}}$ then $\lambda_2=-\Delta_c(\overline{\xi})$ and hence $\Delta=\Delta_c(\xi)- \Delta_c(\overline{\xi})$, which implies that $X_2$ is the global optimizer in $\xi_{\wh{L}}\le \xi<\overline{\xi}$, and zero is the optimizer for $\xi\ge\overline{\xi}$. 
	\end{itemize}
		
Thus, the lemma is proved.\endproof

Using Lemma \ref{Le:const.1}, it is straightforward to show that $X^{VaR,j,\kappa,*}(\lambda,\xi)$ is an optimal solution to constrained problem \eqref{eq:const.1}, see the arguments in  \eqref{V2}. The existence of the Lagrangian multiplier $\lambda$ can be seen from Lemma \ref{Le:suff}.\endproof

\subsection{Proof of Theorem \ref{Th:l}  (the PI-constraint case)}\label{se:profPI}
The proof follows from a modification of Lemma \ref{Le:unconst.1}. In fact, due to the PI constraint, the problem boils down to consider the global optimality on the interval $[l,\infty)$  of $\Psi_j(X)$ defined in \eqref{eq:f}. 
\begin{lemma}\label{Le:unconst.11}
For $\lambda>0$ and $\xi>0$, the problem $\max_\zs{X\ge l} \Psi_j(X)$, for $j\in\{p,np\}$, has the solution 
	\begin{equation*}
		X_T^{PI,j,\kappa,*}=\cWD_\zs{[\xi_\zs{\wt{L}}^{\kappa},\xi_\zs{\wh{L}},\wh{\xi}^{1,q_{j}^l}_l]}^{[I_\kappa,\wt{L},I+L,l]}{\bf 1}_\zs{\Upsilon^{1,q_{j}^l}_l(\wt{L}_T)>0}+ \cWD_\zs{[\xi_\zs{\wt{L}}^{\kappa},\xi_\zs{U}^l]}^{[I_\kappa,\wt{L},l]} {\bf 1}_\zs{\Upsilon^{\kappa,q_{j}^l}_l(\wt{L}_T)\ge0\ge\Upsilon^{1,q_{j}^l}_l(\wt{L}_T)}+ \cWD_\zs{[\wh{\xi}^{\kappa,q_{j}^l}_l]}^{[I_\kappa,l]}	{\bf 1}_\zs{\Upsilon^{\kappa,q_{j}^l}_l(\wt{L}_T)<0}, \quad j\in\{p,np\},
	\label{eq:XPI}
	\end{equation*}
\end{lemma}

\proof Global optimality on $[l,\infty)$ of $\Psi_j$ can be attained at $X_1=l$, $X_2=L+I(\lambda \xi)$, $X_3=I_\kappa(\lambda \xi)$ or at the utility changing point $\wt{L}$. In particular, we need to consider the  $Q_\zs{1}(l)=-q_j^l-\lambda \xi l$ with $Q_\zs{2,max}$ and $Q_\zs{3,max}$ which remain unchanged. As a result, the analysis needs to be modified when we compare the local maximizers $X_2=L+I(\lambda \xi)$, $X_3=I_\kappa(\lambda \xi)$ and $\wt{L}$ to $X_1=l$. For example, for the comparison between $X_3$ and $X_1=l$, we study the difference 
$$
Q_{3,max}-Q_\zs{1}(l)=\Delta_a(\xi)+\lambda \xi l=U_\kappa(I_\kappa(\lambda\xi))-\lambda\xi I_\kappa(\lambda\xi)+\lambda \xi l+q_j^l:=\Delta_a^l(\xi).
$$
In the same way, we also have $\Delta_b^l(\xi):=\Delta_b(\xi)+\lambda \xi l$ and $\Delta_c^l(\xi):=\Delta_c(\xi)+\lambda \xi l$. Note that 
\begin{equation}
\Upsilon^{1,q_{j}^l}_l(\wt{L})=\Delta_b^l(\xi_\zs{\wh{L}})=\Delta_c^l(\xi_\zs{\wh{L}})\quad \mbox{and} \quad \Upsilon^{\kappa,q_{j}^l}_l(\wt{L})=\Delta_b^l(\xi_\zs{\wt{L}}^{\kappa})=\Delta_a^l(\xi_\zs{\wt{L}}^{\kappa}).
\label{eq:delta_l}
\end{equation}
Furthermore, $\wh{\xi}^{1,q_j^l}_l,\wh{\xi}^{\kappa,q_j^l}_l$ and $\xi_U^l$ (in case they exist) are zero points of  $\Delta_c^l$, $\Delta_a^l$ and $\Delta_b^l$ respectively. 

The remainder of the proof follows the same lines in Lemma \ref{Le:unconst.1}.\endproof
%

\bibliography{VaRequity_references}
\bibliographystyle{siamplain}

\end{document}